\pdfoutput=1
\documentclass[traditabstract]{aa} 
\usepackage{natbib}
\usepackage{fixltx2e} 
\usepackage{graphicx}
\usepackage{txfonts}
\begin{document}

\title{Optical and Radio variability of the Northern VHE gamma-ray emitting BL Lac objects}

\author{
E.J.~Lindfors\inst{1,}\inst{2}\and
T.~Hovatta\inst{3,}\inst{4}\and
K.~Nilsson\inst{2}\and
R.~Reinthal\inst{1}\and
V.~Fallah Ramazani\inst{1}\and
V.~Pavlidou\inst{5,}\inst{6}\and
W.~Max-Moerbeck\inst{7}\and
J.~Richards\inst{8}\and
A.~Berdyugin\inst{1}\and
L.~Takalo\inst{1}\and
A.~Sillanp\"a\"a\inst{1}\and
A. C. S.~Readhead\inst{8}
}
\institute {
Tuorla Observatory, Department of Physics and Astronomy, University of Turku Finland
\and Finnish Centre for Astronomy with ESO (FINCA), University of Turku, Finland
\and Aalto University Mets\"ahovi Radio Observatory, Mets\"ahovintie 114, 02540, Kylm\"al\"a, Finland
\and Aalto University Department of Radio Science and Engineering,P.O. BOX 13000, FI-00076 AALTO, Finland
\and Foundation for Research and Technology - Hellas, IESL, Voutes, GR-7110 Heraklion, Greece
\and Department of Physics and Institute for Plasma Physics, University of Crete, GR-71003 Heraklion, Greece
\and Max-Planck-Institut für Radioastronomie, Auf dem Hügel 69, 53121, Bonn, Germany
\and Cahill Center for Astronomy \& Astrophysics, Caltech, 1200 E. California Blvd, Pasadena, CA, 91125, U.S.A.
}
\date{Received XX / Accepted XX}

\abstract{We compare the variability properties of very high energy
  gamma-ray emitting BL Lac objects in the optical and radio
  bands. We use the variability information to distinguish multiple
  emission components in the jet, to be used as a guidance for
  spectral energy distribution modelling. Our sample includes 32
  objects in the Northern sky that have data for at least 2 years in
  both bands. We use optical R-band data
  from the Tuorla blazar monitoring program and 15\,GHz radio data
  from the Owens Valley Radio Observatory blazar monitoring
  program. We estimate the variability amplitudes using the intrinsic
  modulation index, and study the time-domain connection by
  cross-correlating the optical and radio light curves assuming power law power spectral density. Our sample objects are in
general more variable in the optical than radio. We find
correlated flares in about half of the objects, and correlated
long-term trends in more than 40\% of the objects. In these objects we
estimate that at least 10\%-50\% of the optical emission originates in the
same emission region as the radio, while the other half is due to
faster variations not seen in the radio. This implies that simple
single-zone spectral energy distribution models are not adequate for
many of these objects.}

\keywords{galaxies: active, BL Lacertae objects:general, galaxies: jets}

\maketitle


\section{Introduction}
Today 67 extragalactic very high energy (VHE) $\gamma$-ray
sources are known{\footnote{As of January 2016, http://tevcat.uchicago.edu}}. The vast majority of these sources are Active
Galactic Nuclei (AGN) of blazar type. In blazars the relativistic jet, where
electrons travel with the speed close to the speed of light, points close
to our line of sight. 
The blazar group consists of
flat-spectrum radio quasars and BL Lac objects. 
Of the VHE $\gamma$-ray emitting blazars, the 55 BL Lac objects form the majority.

The spectral energy distribution (SED) of the blazars shows two peaks;
one in the infra-red to X-ray range and the second in the X-ray to
$\gamma$-rays. The first peak is synchrotron emission while the second is
most commonly attributed to inverse Compton emission. Based on the
location of the first peak, the BL Lac objects are traditionally
divided into three classes: low, intermediate and high energy
synchrotron peaking (LSP, ISP and HSP, \citealt{1LAC}). The high energy synchrotron peaking objects have their synchrotron
peak in the UV to X-ray range and have therefore been considered as
best candidates to emit VHE $\gamma$-ray energies \citep[e.g.][]{cg02}. 
Indeed within the known VHE $\gamma$-ray blazars
they are the most numerous, which could also in part be an observational
bias as the pointed observations focus on best candidates, and no
full-sky survey exists. However, all BL Lac object sub-classes 
are present in the VHE $\gamma$-ray emitting blazar class. It should also be
noted that in many BL Lac objects the synchrotron
peak moves to higher energies during flares \citep[e.g.][]{pian98}, 
and therefore the division between the different classes is not well
defined. Additionally, there seems to exist a class of extreme BL Lac
objects that show very hard spectra in X-ray and VHE $\gamma$-ray regime \citep[e.g.][]{costamante01}. 

Blazars in general show variability in all wavelengths from
radio to $\gamma$-rays. Many VHE $\gamma$-ray blazars show fast, large
amplitude variability in VHE $\gamma$-rays \citep[e.g. PKS~2155-304;][]{aha07} \citep[Mrk~501;][]{Alb07a}, while for some no variability has been detected \citep[e.g. 1ES~0414+009;][]{Aliu12a}. In MeV-GeV $\gamma$-rays the BL Lac objects are generally less variable than the FSRQs \citep{1LAC}. The
variability in X-rays often shows correlation with the VHE $\gamma$-rays
\citep[e.g.][]{fossati08}. There is also a connection between optical outbursts and emission of the VHE $\gamma$-rays, witnessed by the success of optically triggered target of opportunity observations in detecting new sources as well as high flux states in VHE $\gamma$-rays \citep[][and references therein]{reinthal,Alek15b}.
In radio bands the HSPs are in general weak and less
variable \citep[e.g.][]{nieppola07}, while LSPs are bright and show
frequent large amplitude outbursts. Due to their radio faintness, the
HSPs have not been well represented in large VLBI programs, but there is
evidence that high energy synchrotron peaking BL Lac objects show lower Doppler factors in the $\gamma$-ray loud AGN class \citep{lister11}. 

In general, the SEDs of VHE $\gamma$-ray BL Lacs are modelled with a
single-zone synchrotron self-Compton models, where the emission region
is located close to the central black hole, and is therefore opaque to
radio emission \citep[e.g.][]{tavecchio10}. The radio emission is
assumed to originate in the parsec scale jet and is therefore
typically excluded from the modelling. However, the multiwavelength
campaigns are typically short in duration (weeks to $1-2$ months) or
include only sparse radio observations and therefore it has not been
well studied if this assumption is justified.

The connection between optical and radio
outbursts in blazars have been investigated in several works \citep[e.g.][]{tornikoski94,hanski} finding that the connection is not straight
forward as sometimes there is a correlation between these two bands
while other times there is not. However, these works largely
concentrate on FSRQs and low frequency BL Lac objects that are bright
in radio frequencies. For VHE $\gamma$-ray BL~Lacs, the radio-optical
connection has not been studied in detail. However, recently for two sources, connection between optical and radio was found \citep{Alek14a, Alek14b}. In this paper we study the optical and radio
variability properties of the VHE $\gamma$-ray emitting BL Lacs using the
long term monitoring data from Tuorla blazar monitoring program and
Owens Valley Radio Observatory (OVRO) monitoring program. The study is the first
of its kind for a sample of VHE $\gamma$-ray blazars. We compare the optical and
radio variability behavior of the sources and investigate if the assumptions 
used in the modelling of the SEDs are justified.

\section{Observations and Data analysis} 

\subsection{Tuorla Blazar Monitoring Program} 

The optical R-band observations have been performed as a part of the
Tuorla blazar monitoring
program{\footnote{http://users.utu.fi/kani/1m}}. The observations are
done using the 35cm telescope attached to the 60 cm Kungliga Vetenskapsakademi (KVA) telescope (and can be used with the latter simultaneously) at La
Palma and Tuorla 1.03 meter telescope in Finland. The monitoring program is concentrated on blazars with $\delta>-20^o$. KVA is remotely operated from
Finland. The observations are coordinated with the MAGIC Telescope
and while the monitoring observations are typically performed
two to three times a week (weather permitting), during MAGIC
observations the sources are observed every night.

The data are analyzed using the standard procedures with the
semi-automatic pipeline developed in Tuorla (Nilsson et al. in prep.). The
magnitudes are measured using the differential photometry and
comparison star magnitudes from \cite{nilsson07, smith91, smith98, monet, villata98, fiorucci96, fiorucci98}. For five sources VER~0521+211, VER~0648+151, RGB~0847+115, MAGIC~J2001+435 and B3~2247+381 the comparison stars were calibrated by us using the observations of sources with known comparison star magnitudes from same night (see Appendix~B). 
The magnitudes are converted into Janskys using the standard formula
$S=3080\times10^{-(mag/2.5)}$. 

For many sources the contribution of the host galaxy to the
total flux density is significant. Therefore, it has been
subtracted using the host galaxy fluxes from \cite{nilsson07}
or host galaxy magnitudes from \cite{scarpa00, meisner10, nilsson03, Nil08, Alek14b}. 
For the three sources VER~0521+211, VER~0648+151 and RGB~0847+115 neither of these were available, so the host galaxy was assumed to be a
standard elliptical with M$_R=-22.8$ and effective radius of 8\,kpc
and its contribution to the measured flux density within our aperture (5'')
was estimated using the standard formulae. Host galaxy values for these three sources are given in Appendix~B.

Finally, the measured fluxes were corrected for the galactic absorption using the values from
NED{\footnote{http://ned.ipac.caltech.edu}}. 

\subsection{OVRO}
Regular 15\,GHz observations of the sources were carried out as part
of the blazar monitoring program at OVRO \citep{richards11,richards14}. The program includes all the Fermi detected blazars with $\delta>-20^o$ from 1FGL and 2FGL and the candidate gamma-ray emitters from \cite{healey}. 

The OVRO 40-m telescope uses off-axis dual-beam optics and a cryogenic high electron mobility transistor (HEMT) low-noise amplifier with a
15.0~GHz center frequency and 3~GHz bandwidth. The two sky beams are Dicke switched using the off-source beam as a reference, and the source is alternated between the two beams in an ON-ON fashion to remove atmospheric and ground contamination. Calibration is achieved
using a temperature-stable diode noise source to remove receiver gain
drifts and the flux density scale is derived from observations of
3C~286 assuming the \cite{baars77} value of 3.44\,Jy at
15.0\,GHz. 

A noise level of approximately 3--4\,mJy in quadrature
with about 2\% additional uncertainty, mostly due to pointing errors,
is achieved in a 70~s observation period.
The systematic uncertainty of about 5\% in the flux density
scale is not included in the error bars.  
Complete details of the
reduction and calibration procedure are found in \cite{richards11}.

\section{Sample}

\begin{table*}
\caption{VHE BL Lac sample}              
\label{table:1}      
\centering                                    
\begin{tabular}{lcccccccccc}          
\hline\hline                       
Source name	&RA		&DEC		&data\tablefootmark{a} &$F(>200GeV)/10^{-11}$\tablefootmark{b}	&Ref.\tablefootmark{c}	&z			&Ref.\tablefootmark{d}		&SED Type\tablefootmark{e}\\   
		&		&	&[years]	&[$\rm Ph/cm^2/s$]	&	&			&		&	 \\ 
\hline  
                                
1ES~0033+595	&0:35:53	&59:50:05&5	&0.32	&Alek15a	&$>0.24$\tablefootmark{f}	
&Sbaru05		&HSP	\\
RGB~0136+391	&1:36:33	&39:06:03&5	&?	&			&$>0.41$\tablefootmark{f}	&Nil12	&HSP	\\
3C 66A		&2:22:40	&43:02:08&6	&2.50	&Abdo11	&$>0.335$\tablefootmark{g}	&Furn13A	&ISP	\\
1ES~0229+200	&2:32:49	&20:17:18&6	&0.46	&Aha07	&0.139			&FK99	&(HSP)	\\
HB89~0317+185	&3:19:52	&18:45:34&5	&0.24	&Aliu12b		&0.190			&2LAC	&HSP	\\
1ES~0414+009	&4:16:52	&1:05:24&5	&0.19	&Abram12		&0.287			&2LAC	&HSP	\\
1ES~0502+675	&5:07:56	&67:37:24&5	&2.55	&Majum11		&0.340	&Shaw13	&HSP	\\
VER~J0521+211	&5:21:55	&21:11:24&2.1	&7.08	&Archam13	&0.108			&Shaw13	&ISP	\\
VER~J0648+151	&6:48:48	&15:16:25&2.1	&0.75	&Aliu11		&0.179			&Aliu11	&HSP	\\
1ES~0647+250	&6:50:46	&25:03:00&6	&?	&			&0.410			&Kot11	&HSP	\\
S5~0716+714	&7:21:53	&71:20:36&6	&4.10	&Ander09		&0.31			&Nil08	&ISP	\\
1ES~0806+524	&8:09:49	&52:18:58&6	&2.45	&Alek15b		&0.137			&2LAC	&HSP	\\
RGB~0847+115	&8:47:13	&11:33:50&3.2	&0.57	&Mirz14a		&0.198			&Plot10	&HSP	\\
1ES~1011+496	&10:15:04	&49:26:01&5.4	&23.0	&Ahnen16a		&0.212			&Alb07b	&HSP	\\
Mkn~421	&11:04:27	&38:12:32&6	&232.65	&Cort13		&0.031			&2LAC	&HSP	\\
Mkn~180	&11:36:26	&70:09:27&6	&2.25	&Alb06a	&0.046			&2LAC	&HSP	\\
RGB~1136+676	&11:36:30	&67:37:04&3.8	&0.34	&Mirz14c	&0.134			&Plot10	&HSP	\\
ON~325		&12:17:52	&30:07:01&6	&0.77	&Alek12a	&0.130			&2LAC	&HSP	\\
1ES~1218+304	&12:21:22	&30:10:37&5	&5.52	&Accia10 	&0.184			&2LAC	&HSP	\\
ON~231		&12:21:32	&28:13:59&6	&6.22	&Accia09		&0.103			&2LAC	&ISP	\\
PG~1424+240	&14:27:00	&23:48:00&5	&0.53	&Alek14a	&$>0.604$\tablefootmark{g}	&Furn13b	&HSP	\\
1ES~1426+428	&14:28:33	&42:40:21&6	&66.21	&Horan02		&0.129			&Laur98	&HSP	\\
PG~1553+113	&15:55:43	&11:11:24&5	&4.44	&Aliu15	&$>0.395$\tablefootmark{h}	&Danf10	&HSP	\\
Mkn~501	&16:53:52	&39:45:37&5	&98.06	&Aha99	&0.034			&2LAC	&HSP	\\
H~1722+119	&17:25:04	&11:52:15&6	&0.33	&Ahnen16b	&$>0.17$\tablefootmark{f}	&Sbaru06	&HSP	\\
1ES~1727+502	&17:28:19	&50:13:10&6	&0.26	&Archam15 &0.055			&oke	&HSP	\\
1ES~1741+196	&17:43:58	&19:35:09&6	&0.19	&Berg11		&0.083			&heidt	&HSP	\\
1ES~1959+650	&20:00:00	&65:08:55&6	&48.89	&Krawc04		&0.047			&2LAC	&HSP	\\
MAGIC~J2001+439	&20:01:14	&43:53:03&3.4	&1.80	&Alek14b&0.190			&Alek14b	&ISP	\\
BL~Lacertae	&22:02:43	&42:16:40&5	&34.00	&Arlen13	&0.069				&2LAC	&ISP$^{*}$	\\
B3~2247+381	&22:50:06	&38:24:37&5	&0.50	&Alek12b	&0.119			&Laur98	&HSP	\\
1ES~2344+514	&23:47:05	&51:42:18&6	&13.91	&Accia11	&0.044			&2LAC	&HSP	\\
\hline                                            
\end{tabular}
\tablefoot{
\tablefoottext{a}{The length of the period of optical and radio data used for the analysis}\\
\tablefoottext{b}{Highest flux reported in the literature, the fluxes have been converted to $>$200 GeV for easier comparison.}\\
\tablefoottext{c}{References:\cite{Alek15a};\cite{Abdo11}; \cite{Aharonian07}; \cite{Aliu12b}; \cite{Abram12}, \cite{majumdar11}, \cite{Archam13}, \cite{Aliu11}, \cite{Ander09}, \cite{Alek15b}, \cite{Mirz14a}, \cite{Ahnen16a}, \cite{Cort13}, \cite{Alb06}, \cite{Mirz14c}, \cite{Alek12a}, \cite{Accia10}, \cite{Accia09}, \cite{Alek14a}, \cite{Horan02}, \cite{Aliu15} \cite{Aharonian99}, \cite{Ahnen16b} 
\cite{Archam15}, \cite{Berg11}, \cite{Krawc04}, \cite{Alek14b}, \cite{Arlen13}, \cite{Alek12b}, \cite{Accia11}}\\
\tablefoottext{d}{References:(Sbaru05) \cite{Sbaru05}; (Nil12) \cite{Nil12}; (Furn13A) \cite{Furn13A}; (FK99) \cite{FK99}; 2LAC \cite{2LAC}; (Shaw13) \cite{Shaw13}; (Aliu11) \cite{Aliu11}; (Kot11) \cite{Kot11}; (Nil08) \cite{Nil08}; (Plot10) \cite{Plot10}; (Alb07b) \cite{Alb07b}; (Furn13b) \cite{Furn13b}; (Laur98) \cite{Laur98}; (Danf10) \cite{Danf10}; (Sbaru06) \cite{Sbaru06}; (oke) \cite{oke}; (heidt) \cite{heidt}; (Alek14b) \cite{Alek14b}}\\
\tablefoottext{e}{HSP=High synchrotron peak frequency source, ISP=Intermediate synchrotron peak frequency source. From \cite{2LAC} except for 1ES~0229+200, which is not included in 2LAC.}\\
\tablefoottext{f}{Lower limit based on non-detection of the host}\\
\tablefoottext{g}{Lower limit based on $Ly\alpha$}\\
\tablefoottext{h}{Lower limit based on a confirmed $Ly\alpha+O_{vi}$ absorber}\\
\tablefoottext{*}{In many other catalogues classified as LSP=Low synchrotron peak frequency source.}\\
\tablefoottext{?}{The flux density has not been reported even if the detection has been announced.}\\
}
\end{table*}

The number of known VHE $\gamma$-ray emitting BL Lacs is 55 (as of January 2016){\footnote{This number includes IC~310 and HESS~J1943+213, both of which have multiple classifications}}. The redshift range is
from 0.03 to $\sim0.6$, although some sources still have uncertain or unknown
redshift. Most of the sources have high synchrotron peak frequency and
are classified as HSPs (log $\nu_{peak}>15.0$), while only 4 intermediate and 2
low synchrotron peaking sources are known. The VHE $\gamma$-ray fluxes of
the sources range from very weak ($<1\%$ of Crab Nebula flux at 200\,
GeV) to very bright ($>5$ Crab Nebula flux) and many sources are
variable.

For 39 of these sources we could find a radio measurement from the
literature and the 5 GHz flux densities range from $\sim$0.01\,Jy to
$\sim$3.5\,Jy, the faintest being 1ES~0347$-$121 and the brightest
one BL~Lacertae. For all sources archival optical data from R-band is
available and the observed flux density range is from 0.1\,mJy  (1ES~0229+200, host galaxy subtracted) to 25\,mJy (Mrk~421, host galaxy subtracted).

The Tuorla blazar monitoring sample consists of a core sample of 24
TeV candidate BL Lac objects from \cite{cg02}
with $\delta>20^o$ (to be observable from Tuorla). These blazars have been
monitored since the fall of 2002. Later many sources have been added to the
monitoring and it now includes most of the VHE $\gamma$-ray emitting
AGN with $\delta>-20^o$. The declination limit excludes 11 VHE
$\gamma$-ray blazars. The monitoring program does not include IC~310 and HESS J1943+243. Additionally there are 10 VHE $\gamma$-ray emitting
blazars that are not part of the program: SHBL~J001355.9$-$185406, KUV~00311$-$1938, S2~0109+22, 1ES~0152+017, 1ES~0347$-$121, RGB~0710+591, MS~1221.8+2452, S3~1227+25, 1ES~1440+122 and RGB~2243+203. This gives us a sample of 32 VHE $\gamma$-ray emitting BL Lac objects with optical light curves with at least two years of data (see Table 1).

The OVRO blazar monitoring program started in 2008 including all the sources from the Candidate Gamma-Ray Blazar Sample in \cite{healey} with $\delta>-20^o$. All Fermi-detected sources from 1FGL and 2FGL catalogues have been subsequently added to the monitoring.
For all the 32 sources for which we have
long enough optical light curves, there exists more than two years of OVRO
data. The source sample, and the time range of the data for each
source, is presented in Table~1. The sample represents well the known population of VHE $\gamma$-ray emitting BL Lacs in redshift range,
classification, VHE $\gamma$-ray fluxes and range of optical and radio
flux densities found in literature. The majority of the sources are HSPs, all of the intermediate objects are included and one of the two known LSPs is
included (although BL Lac is classified as ISP in \cite{2LAC}). In redshifts, the population is mostly concentrated to z$<0.2$. Only four BL Lac sources are known at z$>0.4$, three of which are in our sample. 

\begin{figure}
\includegraphics[width=0.3\textwidth, angle=-90]{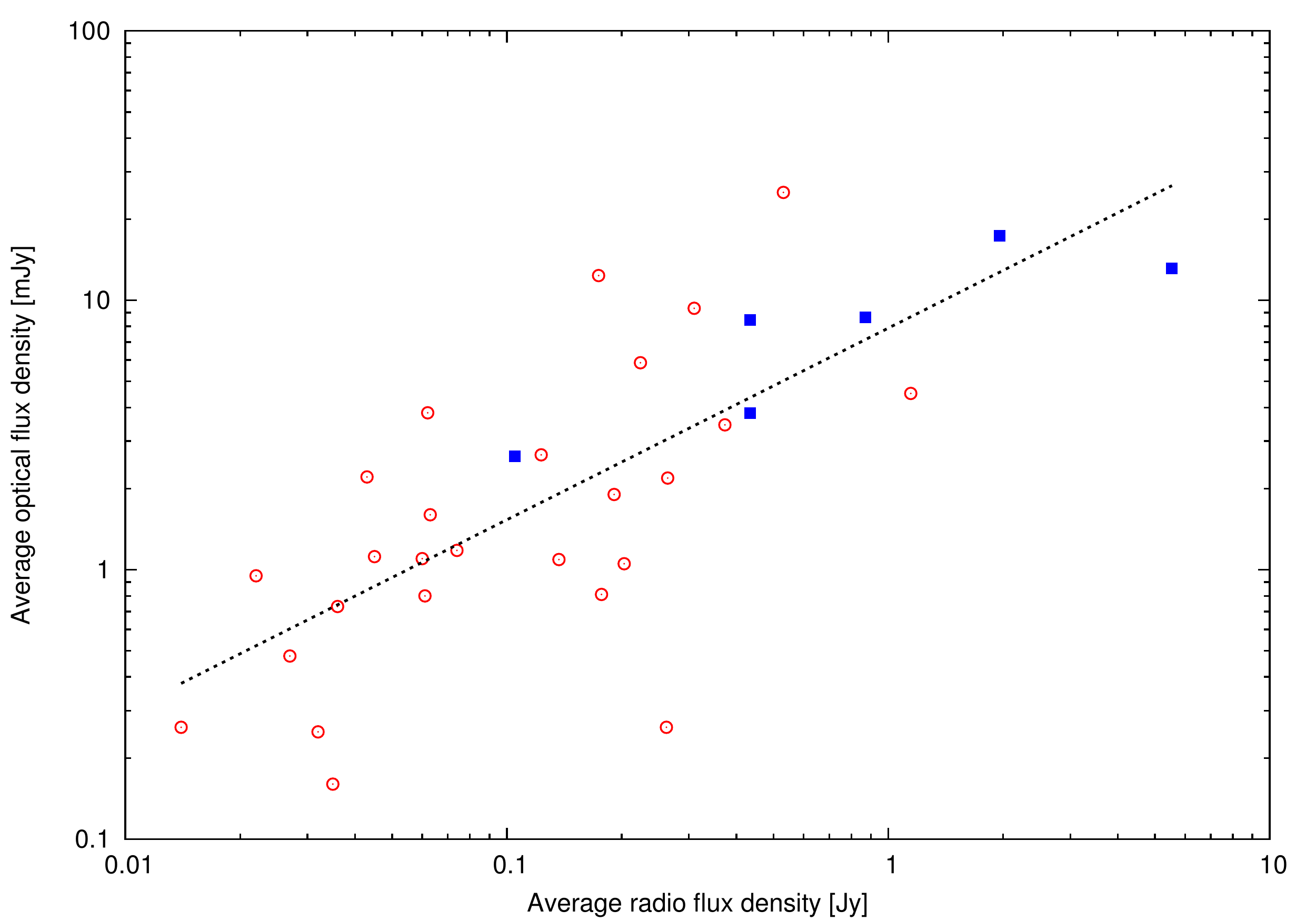}
\caption{The average radio flux density versus the average optical
  flux density. The average flux densities in these two bands show
  strong correlation ($> 4\sigma$), the dotted line shows the best-fit for the correlation. The different symbols denote different source
  classes (HSPs, red circles; ISPs, blue rectangles). As expected, the ISP's seem to be brighter in radio and optical bands.}
  \label{fig:aveflux}
\end{figure}

The average radio and optical flux densities for the sample range in radio from 0.014\,Jy to 5.544\,Jy and
in optical from 0.16\,mJy to 25.13\,mJy (Tables 2 and 3), in the period indicated in Table 1. The median
for average radio flux density is 0.17\,Jy and for optical flux density
2.19\,mJy. The average flux densities are shown in Figure~\ref{fig:aveflux} and Tables 2 and 3. 
There is a clear correlation
between the average flux density in radio and optical with the Spearman's $\rho=0.75$. We estimate the
significance of the correlation using simulated samples in the
luminosity space as proposed by 
\cite{pavlidou12}, in order to account for the common redshift in the
two wavebands. For the calculation of the luminosity we assume a flat
spectral index of 0 in the radio band and $-1.1$, $-1.3$, and $-1.5$
in the optical band for the HSP, ISP, and LSP sources, respectively \citep{fiorucci04}.
By simulating $10^7$ uncorrelated samples, we obtain
a significance of $p=1.9\times10^{-5}$ ($> 4\sigma$) for the correlation.

As the sample studied in this work is VHE $\gamma$-ray selected, we
also checked if average flux densities correlate with the VHE $\gamma$-ray flux given in Table~1. We found no significant correlation, but it should be noted that the VHE flux densities present the highest observed flux density, not the average one, and are typically non-simultaneous to the optical and radio data. 

\begin{table*}
\begin{center}
\caption{Analysis results of the optical R-band light curves}
\begin{tabular}{lccccccc}
\hline
\hline
name&nr of obs& avg flux density\tablefootmark{a}& mod ind\tablefootmark{b}
&$\rho$\tablefootmark{c}&$\rho_{boot}$\tablefootmark{d}& p\tablefootmark{e}\\
\hline
1ES~0033+595& 237& 1.18&$0.230_{-0.014}^{+0.015}$&$-$0.041&$-0.054$&0.26\\
RGB~0136+391& 225&  2.21&$0.163_{-0.008}^{+0.008}$&$-$0.481&$-$0.475&$<10^{-8}$\\ 
3C~66A&362&8.64&$0.368_{-0.015}^{+0.016}$&$-$0.294&$-$0.297&$10^{-8}$\\
1ES~0229+200& 126&0.16 &$0.00{*}$&0.284&0.282&$6.0\cdot10^{-4}$\\\
HB89~0317+185&91&0.26&$0.252_{0.023}^{0.027}$&$-$0.794&$-$0.781&$<10^{-8}$\\
1ES~0414+009&120&1.10&$0.234_{-0.016}^{+0.018}$&0.135&0.133&0.0707\\
1ES~0502+675&191&0.95&$0.256_{-0.014}^{+0.015}$&$-$0.367&$-$0.346&$8.0\cdot10^{-
8}$\\
VER~J0521+211&59&8.44&$0.375_{-0.037}^{+0.044}$&0.744&0.741&$<10^{-8}$\\
VER~J0648+152&44&0.73&$0.319_{-0.036}^{+0.044}$&0.838&0.829&$<10^{-8}$\\
1ES~0647+250&218&1.60&$0.183_{-0.009}^{+0.010}$&0.594&0.597&$<10^{-8}$\\
S5~0716+716&355&17.44&$0.507_{-0.023}^{+0.027}$&$-$0.283&$-$0.278&$3.0\cdot10^{-
8}$\\
1ES~0806+524&245&2.67&$0.383_{-0.019}^{+0.021}$&0.186&0.187&$1.7\cdot10^{-3}$\\
RGB~0847+115&41&0.26&$0.188_{-0.026}^{+0.031}$&$-$0.557&$-$0.546&$7.0\cdot10^{-
5}$\\
1ES~1011+496&239&2.19&$0.299_{-0.014}^{+0.016}$&$-$0.720&$-$0.724&$<10^{-8}$\\
Mkn~421&449&25.13&$0.455_{-0.017}^{+0.019}$&0.576&0.576&$<10^{-8}$\\
Mkn~180&295&1.90&$0.359_{-0.016}^{+0.018}$&0.871&0.868&$<10^{-8}$\\
RGB~1136+676&102&0.25&$0.122_{-0.019}^{+0.020}$&$-$0.162&$-$0.168&0.053\\
ON~325&206&3.45&$0.221_{-0.011}^{+0.012}$&$-$0.109&$-$0.096&0.059\\
1ES~1218+304&151&1.12&$0.367_{-0.023}^{+0.026}$&$-$0.132&$-$0.118&0.0526\\
ON~231& 214&3.80&$0.360_{-0.019}^{+0.021}$&$-$0.793&$-$0.788&$<10^{-8}$\\
PG~1424+240& 177&9.34&$0.120_{0.006}^{0.007}$&0.362&0.353&$3.6\cdot10^{-
8}$\\
1ES~1426+428&165&0.48&$0.150_{-0.016}^{+0.017}$&$-$0.182&$-$0.177&$9.7\cdot10^{-
3}$\\
PG~1553+113&344&12.36&$0.238_{-0.009}^{+0.010}$&$-$0.014&$-$0.002&0.401\\
Mkn~501&447&4.51&$0.091\pm0.004$&$-$0.393&$-$0.395&$<10^{-8}$\\
H~1722+119&327& 3.82&$0.278_{-0.011}^{+0.012}$&0.721&0.724&$<10^{-8}$\\
1ES~1727+502&289&1.09&$0.226_{-0.010}^{+0.011}$&0.607&0.605&$<10^{-8}$\\
1ES~1741+196&212&  1.05&$0.039\pm0.007$&$-$0.181&$-$0.176&0.0042\\
1ES~1959+650&516&5.86&$0.338_{-0.011}^{+0.012}$&0.319&0.324&$<10^{-8}$\\
MAGIC~J2001+439&144&2.63&$0.468_{-0.031}^{+0.036}$&$-$0.750&$-$0.739&$<10^{-8}$\\
BL~Lac&404&13.13&$0.600_{-0.027}^{+0.029}$&0.537&0.537&$<10^{-8}$\\
B3~2247+381&232&0.80&$0.339_{-0.017}^{+0.019}$&$-$0.374&$-$0.373&$<10^{-8}$\\
1ES~2344+514&271&0.81&$0.304_{-0.016}^{+0.017}$&0.379&0.380&$<10^{-8}$\\ 
\hline
\hline
\end{tabular}
\vskip 0.4 true cm
\label{opt}
\tablefoot{\tablefoottext{a}{Average flux density in mJy}\\
\tablefoottext{b}{Modulation index} \\
\tablefoottext{c}{Spearman $\rho$ for the 2D linear regression} \\
\tablefoottext{d}{Spearman $\rho$ with bootstrapping for the 2D linear regression}\\ 
\tablefoottext{e}{p-value for the null hypothesis of no correlation, 5$\sigma$ limit is $3\cdot10^{-7}$}\\
\tablefoottext{*}{The source was too faint for estimating the modulation index}\\
}
\end{center}
\end{table*}

\begin{table*}
\begin{center}
\caption{Analysis results of the 15\,GHz radio light curves}
\begin{tabular}{lccccccccc}
\hline
\hline
name&nr of obs& avg flux density\tablefootmark{a}& mod ind\tablefootmark{b}&$\rho$\tablefootmark{c} &$\rho_{boot}$\tablefootmark{d}&p\tablefootmark{e}\\
\hline
1ES~0033+595& 369&  0.074& $0.094_{-0.007}^{+0.007}$& $-$0.151& $-$0.155&0.0019\\
RGB~0136+391& 304&  0.043& $0.00^{*}$&0.081&0.087&0.0798\\ 
3C~66A      & 466&  0.870&$0.133_{-0.004}^{+0.005}$&$-$0.678&$-$0.676&$<10^{-8}$\\
1ES~0229+200& 341&  0.035& $0.076\pm0.011$&0.244&0.243 &$2.6\cdot10^{-6}$\\
HB89~0317+185&283& 0.262& $0.045\pm0.003$&0.665&0.656&$<10^{-8}$\\
1ES~0414+009& 269&  0.060& $0.087_{-0.007}^{+0.008}$&0.151&0.135&0.0065\\
1ES~0502+675& 359&  0.022& $0.00^{*}$&$-$0.054&$-$0.057&0.1519\\
VER~J0521+211& 104&0.434&$0.074_{-0.006}^{+0.007}$&0.492&0.487&$2.0\cdot10^{-8}$\\
VER~J0648+152&  94&  0.036& $0.00^{*}$& 0.595& 0.592& $<10^{-8}$\\
1ES~0647+250& 261&  0.062& $0.088_{-0.006}^{+0.007}$& 0.387& 0.383 &$<10^{-8}$\\
S5~0716+714& 452&  1.956& $0.351_{-0.013}^{+0.014}$&0.209&0.206&$3.6\cdot10^{-6}$\\
1ES~0806+524& 411&  0.122& $0.124\pm0.005$& 0.620& 0.619 &$<10^{-8}$\\
RGB~0847+115& 153&  0.014& $0.00^{*}$&$-$0.045&$-$0.038&0.29\\
1ES~1011+496& 289&  0.265& $0.063\pm0.003$& $-$0.724& $-$0.720 &$<10^{-8}$\\
Mkn~421     & 561&  0.531& $0.217\pm0.007$& 0.698&0.694&$<10^{-8}$\\
Mkn~180     & 335&  0.191& $0.128_{-0.005}^{+0.006}$& 0.874& 0.872&$<10^{-8}$\\
RGB~1136+676& 189&  0.032& $0.086\pm0.015$&$-$0.033&$-$0.020&0.323\\
ON~325      & 355&  0.373& $0.090\pm0.004$&$-$0.605&$-$0.603&$<10^{-8}$\\ 
1ES~1218+304& 363&  0.045& $0.149_{-0.009}^{+0.010}$&0.215& 0.208&$1.8\cdot10^{-5}$\\
ON~231      & 443&  0.434& $0.125_{-0.004}^{+0.005}$&0.291&0.294&$<10^{-8}$\\
PG~1424+240 & 245&  0.311& $0.154_{-0.007}^{+0.008}$&0.957&0.955&$<10^{-8}$\\
1ES~1426+428& 292&  0.028& $0.150_{-0.016}^{+0.017}$&$-$0.111&$-$0.115&0.0292\\
PG~1553+113 & 313&  0.173& $0.088_{-0.004}^{+0.005}$& 0.269 &0.266 &$6.7\cdot10^{-7}$\\  
Mkn~501     & 335&  1.145&$0.041\pm0.002$&$-$0.536&$-$0.537&$<10^{-8}$\\
H~1722+119  & 347& 0.061&$0.150_{-0.011}^{+0.012}$&$-$0.059&$-$0.060&0.138\\
1ES~1727+502& 363&  0.045& $0.060\pm0.003$ & 0.215& 0.211&$1.8\cdot10^{-5}$\\
1ES~1741+196& 252&  0.203& $0.042\pm0.003$& 0.357&0.351&$<10^{-8}$\\
1ES~1959+650& 457&  0.223& $0.137\pm0.005$ & 0.775&0.773&$<10^{-8}$\\
MAGIC~J2001+439&398&0.105&$0.292_{-0.012}^{+0.013}$&$-$0.771&$-$0.768&$<10^{-8}$\\
BL~Lac     & 311&  5.544& $0.319_{-0.014}^{+0.015}$&0.736&0.739&$<10^{-8}$\\ 
B3~2247+381 & 284& 0.061& $0.060\pm0.008$&$-$0.155&$-$0.155&0.0044\\ 
1ES~2344+514& 402&  0.177& $0.111\pm0.005$&0.558&0.570&$<10^{-8}$\\
\hline
\hline
\end{tabular}
\vskip 0.4 true cm
\label{radio}
\tablefoot{\tablefoottext{a}{Average flux density in Jy}\\
\tablefoottext{b}{Modulation index} \\
\tablefoottext{c}{Spearman $\rho$ for the 2D linear regression} \\
\tablefoottext{d}{Spearman $\rho$ with bootstrapping for the 2D linear regression}\\ 
\tablefoottext{e}{p-value for the null hypothesis of no correlation, 5$\sigma$ limit is $3\cdot10^{-7}$} \\
\tablefoottext{*}{The source was too faint for estimating the modulation index}\\
}
\end{center}
\end{table*}

\section{Variability analysis}

\subsection{Variability amplitudes}\label{sect:modindex}
We determine the variability amplitudes of the sources in the optical
and radio bands using the intrinsic modulation index
\citep{richards11}, defined as 
\begin{equation}
\centering
\overline{m} = \frac{\sigma_0}{S_0},
\end{equation}
where $\sigma_0$ is the intrinsic standard deviation and $S_0$ is the
intrinsic mean flux density of the source. Here the term intrinsic
denotes values that would be obtained if the observational uncertainties were
zero and we would have infinite number of samples. 
The intrinsic values are calculated using a likelihood approach, which
assumes the observed flux densities to follow a normal distribution
with Gaussian errors. The measurement errors are accounted for in the
calculation of the joint likelihood for $S_0$ and $\overline{m}$. For
the full derivation of the likelihoods see \cite{richards11}. 
The main advantage of this method over other variability estimates
is that it provides an uncertainty estimate for the variability, which increases when the flux uncertainty is larger or the number of points in the light curves is small.

\begin{figure}
\includegraphics[scale=0.8]{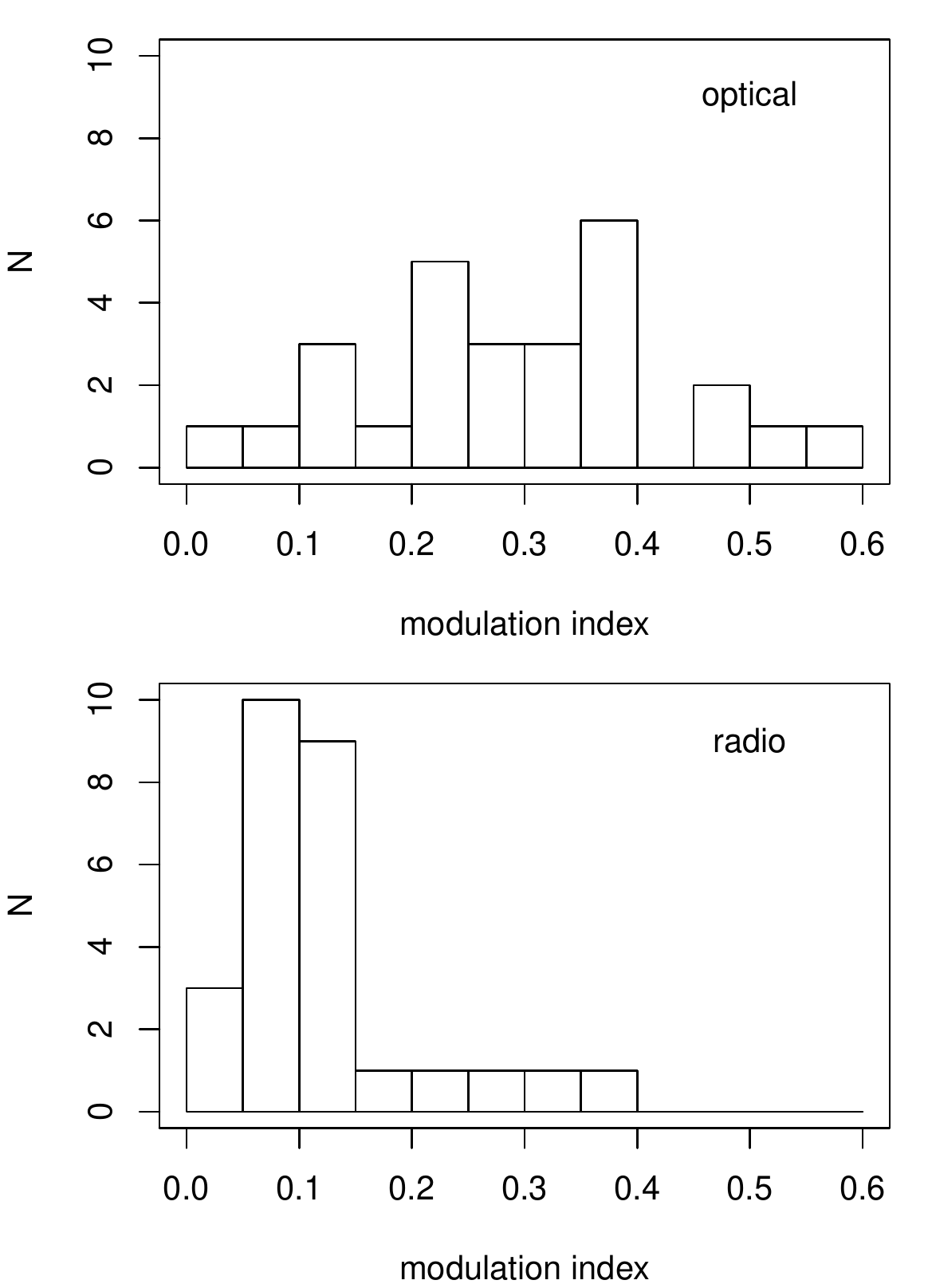}
\caption{Histogram of the intrinsic modulation indices $\overline{m}$
  for the optical (top) and radio (bottom) light curves. The uncertainty of the intrinsic modulation indices (typically 0.01, largest value 0.044) are smaller than the bin size. According to a K-S test, the probability that optical and radio modulation index distributions come from the same parent population is very low ($p=1.3\times10^{-5}$).}\label{fig:modindex}
\end{figure}
The modulation indices and uncertainties for each source are shown in
Table~\ref{opt} and Table~\ref{radio}. In four cases (RGB~0136+391, 1ES~0502+675, RGB~0847+115, VER~J0648+152) in the radio and in one case (1ES~0229+200) in the optical there were too many low signal-to-noise points for estimating the modulation index. 
All the remaining sources were variable at a more than
$3\sigma$ level.

The mean value of modulation index for the
optical and radio light curves is $\overline{m}_{\rm{opt}} = 0.29$ and $\overline{m}_{\rm{rad}} = 0.13$, respectively. The uncertainty is typically $\sim0.01$, largest value being 0.044. The distributions
  of the modulation indices are shown in Fig.~\ref{fig:modindex} and are clearly different.
According to a non-parametric Kolmogorov-Smirnov (K-S) test, the
optical and radio modulation indices come from the same population
with a probability $p=1.3\times10^{-5}$.

\begin{figure}
\includegraphics[width=0.35\textwidth, angle=270]{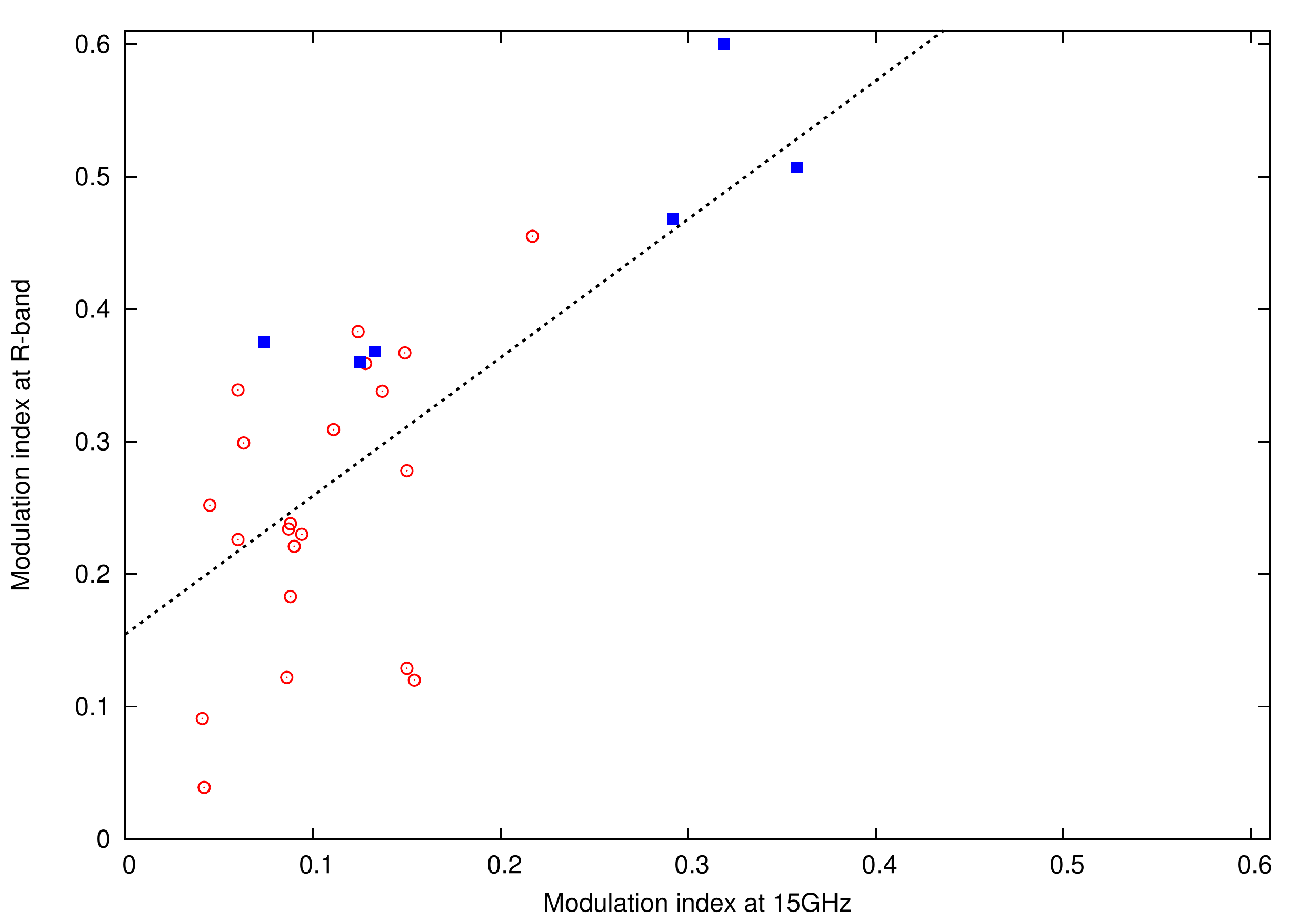}
\caption{The modulation index at 15\,GHz versus the modulation index
  at optical R-band for the sample sources (HSPs red circles, ISPs blue rectangles). The sources for which one
  of the modulation indices could not be determined were excluded from
  the plot. As expected the ISPs in general have larger modulation indices in optical and radio than HSPs. Therefore the indices between the two band show significant ($>3\sigma$) correlation (see text). The dotted line shows the best linear fit to the data.}\label{fig:modindex_corr}
\end{figure}

Fig.~\ref{fig:modindex_corr} shows the modulation indices at 15\,GHz versus the modulation indices at R-band. The two show significant correlation with Spearman's $\rho=0.58$ corresponding to significance of $>3\sigma$. However, this correlation is largely due to ISPs showing larger modulation index values in general and for HSPs only the Spearman's $\rho=0.32$, i.e. the correlation is not significant.  

\subsection{Trends}\label{sect:trends}

Blazars have been long known to show variability in time scales of
years, in addition to the fast variability, which is typically
described as flares \citep{smith93}. \cite{smith95}
determined that for BL Lacs in optical band the observed time scale of
such slow variations is typically $5-7.5$ years, with the average at 7.2
years. In the simplest case such slow variability would show up as increasing or decreasing trend in our data as the studied light curves have duration
less than the average time scale of these variations. In the case of PKS~1424+240, such trend is also present in the 15\,GHz data \citep{Alek14a}.  

To look for such trends in our optical and radio light curves, we simply test 
if there is a significant correlation between time and flux density.
We use the five methods for obtaining the linear regression fits from \cite{isobe}. 
In Table 2 and 3 we report the Spearman $\rho$ value for these fits for optical and radio light curves, respectively. As the analysis does
not take into account the uncertainties of the flux density measurements, we also test the significance using bootstrapping analysis for re-sampling the light curves. Also these values are given in the Tables as well as p-values for null hypothesis of the no trend. If the null hypothesis can be excluded at 5$\sigma$ level we conclude that there is a significant trend in the data.

We find that 21 of our optical and 22 of our radio light curves{\footnote{from this count we
exclude the source VER~0648+121 for which a modulation index
cannot be determined}} show significant
increasing or decreasing trend. In 13 sources\footnote{{14 if VER 0648+121 was
not excluded}} the trends in optical and radio bands are
to the same direction.

To investigate the probability of a random occurrence of this result in presence of red noise, we perform simulations. We first simulate 1000 optical and 1000 radio light curves for each source assuming that the light curves are red noise with a power law slope of $-1.5$ (optical, Nilsson et al. in prep) and $-1.7$ (radio, derived for this sample using the methods in \cite{max-moerbeck14}){\footnote{For BL Lac these slopes are $-0.8$ and $-1.95$, respectively}}. 
The light curves are then interpolated to have the same rms scatter and sampling as the
observed light curve. Finally, we perform the same linear regression
analysis to these simulated light curves. The results are summarized
in Table 4. In the simulations we find on average 21.9 sources with a trend in optical and 17.3 sources with A trend in radio, so also the sample statistics are in acceptable agreement with what we see in the real data, even if there is on average slightly less trends in the simulated radio light curves than in the real radio data. However, for some of the weak
(F$<$0.1 Jy) radio sources, the simulated light curves do not show any
trends or occur only in $1-2$ of 1000 light curves. We suggest that
this is related to the size of the uncertainties of the actual radio light
curves for these sources. The simulation results are in general
agreement with the real data, i.e. for these sources we find no
significant trends in the observed light curves either. There is only one
exception, VER~0648+121, for which our trend test suggests significant
trend, but light curve analysis cannot determine a modulation index. Therefore, we
exclude this source from statistics of the whole sample. We find that
the observed fraction of trends in the same direction ($\geq13/31$) is found only in
5 cases out of 1000 in the simulations (p=0.005).
 Therefore, we conclude that the result indicates true physical connection, meaning that the slowly variable optical component has common origin with the 15\,GHz radio emission.

\begin{table*}
\caption{Analysis results of the connection between optical and radio
  light curves}
\begin{tabular}{lccccccc}
\hline
\hline
name&optical trend\tablefootmark{a}& radio trend\tablefootmark{a}& sum trend\tablefootmark{b}&sim opt trend\tablefootmark{c}& sim radio trend\tablefootmark{c}& sim sum trend\tablefootmark{d}& $t_{lag}$[d]\tablefootmark{e}\\ \hline
1ES~0033+595& 0&0&0&750&1&0&610\\
RGB~0136+391&$-$&0&0&785&2&0&NA\\ 
3C~66A&$-$&$-$&1&791&842&316&860\\
1ES~0229+200&0&+&0&295&2&1&NA\\
HB89~0317+185&$-$&+&0&571&721&192&NA\\
1ES~0414+009&0&0&0&649&1&0&20\\
1ES~0502+675&$-$&0&0&748&1&0&NA\\
VER~J0521+211&+&+&1&560&712&181&$-$110\\
1ES~0647+250&+&+&1&746&570&204&$-$1030\\
VER~J0648+152&+&+&1$^{*}$&623&0&0&NA\\
S5~0716+716&$-$&+&0&800&879&370&NA\\
1ES~0806+524&0&+&0&740&843&323&$-$110\\
RGB~0847+115&$-$&0&0&469&0&0&NA\\
1ES~1011+496&$-$&$-$&1&759&816&294&$-$370\\
Mkn~421&+&+&1&844&917&376&$-$60\\
Mkn~180&+&+&1&803&836&345&$-$10\\
RGB~1136+676&0&0&0&541&9&3&NA\\
ON~325&0&$-$&0&735&815&311&NA\\
1ES~1218+304&0&+&0&706&772&288&$-$50\\
ON~231&$-$&+&0&734&858&321&NA\\
PG~1424+240&+&+&1&790&788&293&NA\\
1ES~1426+428&0&0&0&748&781&307&NA\\
PG~1553+113&0&+&0&769&823&312&$-$200\\
Mkn~501&$-$&$-$&1&840&834&363&NA\\
H~1722+119&+&0&0&848&0&0&$-$190\\
1ES~1727+502&+&+&1&787&799&311&$-$50\\
1ES~1741+196&0&+&0&523&381&108&NA\\
1ES~1959+650&+&+&1&839&839&383&NA\\
MAGIC~J2001+439&$-$&$-$&1&680&808&273&70\\
BL~Lac&+&+&1&511&879&223&$-$560\\
B3~2247+381&$-$&0&0&766&0&0&$-$90\\
1ES~2344+514&+&+&1&747&820&282&$-$70\\ 
\hline
\hline
\end{tabular}
\vskip 0.4 true cm
\label{modelparam}
\tablefoot{
\tablefoottext{a}{In the optical trend and radio trend columns, 0 indicates that the linear regression analysis gave p>0.0005 for null hypothesis (no trend), $-$ a negative trend (with p<0.0005 for null hypothesis of no trend) and + a positive trend (with p<0.0005 for null hypothesis of no trend).}\\
\tablefoottext{b}{0 if trend in optical and radio have different signs or no significant trend was found, 1 if the trend was in same direction.}\\
\tablefoottext{c}{Number of simulated light curves (out of 1000) for which no trend was excluded with p>0.0005}\\
\tablefoottext{d}{Number of simulations (out of 1000) in which case the trend in optical and radio light curves is in same direction.}\\
\tablefoottext{e}{Time lag for the most significant peak of the DCF, - means that optical is leading radio, NA that there was no peaks in DCF with significance of 3$\sigma$}\\
\tablefoottext{*}{Excluded from the sample statistic, see text.}
}  
\end{table*}

\subsection{Cross-correlation of light curves}
We use the discrete correlation function (DCF) \citep{edelson88} with
local normalization \citep[LCCF;][]{welsh99} to study the correlation between
the optical and radio light curves. In the calculation of the LCCF, we use time binning of 10 days and also require that each LCCF bin has at least 10 elements.
We include all sources that are
variable according to the modulation indices estimated in
Sect.~\ref{sect:modindex}. For these sources the cross correlation
functions are shown in the bottom panel of Figs.~\ref{Fig:lc1},~\ref{Fig:lc3},~\ref{Fig:lc5},~\ref{Fig:lc6},~\ref{Fig:lc8},~\ref{Fig:lc10}-\ref{Fig:lc32}.

The significance of the cross-correlation is estimated using simulated
light curves, similarly as in \cite{max-moerbeck14}. We take 1000
simulated uncorrelated light curves of each source (we use the best-fitting power-law index values, as determined in
Sect.~\ref{sect:trends}) and cross-correlate them similarly as the
real data. This way we can estimate the occurrence of false
correlations due to random fluctuations in the data. The significance
levels are also shown in the Figs.~\ref{Fig:lc1},~\ref{Fig:lc3},~\ref{Fig:lc5},~\ref{Fig:lc6},~\ref{Fig:lc8},~\ref{Fig:lc10}-\ref{Fig:lc32}.
In Table~4, we list the most significant time lag between the optical and
radio light curves for sources showing significant correlations at a
$3\sigma$ level. We only list time lags that are shorter than half the
length of the light curves to discard time lags due to single events.

We note that the significance estimates depend strongly on the slope
of the power spectral density used to simulate the light curves
\citep{max-moerbeck14}. Furthermore, as shown by
\cite{emmanoulopoulos13}, if the flux density distributions are
non-Gaussian, it will also have a large effect on the estimated
significances. Therefore it is likely that the significance of our
correlations is over estimated in some cases where the obtained time
lag is close to the $3\sigma$ limit. This seems to be the case for sources such as 1ES~0033+595 where the most significant delays seem to be produced by a few outlier points with small uncertainties in the radio light curve. In cases like this, when the majority of the radio data points have fairly large uncertainties, the simulations do not produce such outliers into the light curves. This results in an overestimation of the significance of the correlation, as the simulated light curves do not reproduce the observed ones perfectly. In other sources, such as VER~0521+211, the significant correlation is most likely produced by a common linear trend in the data. We discuss the individual correlations in Appendix~A.

\begin{figure}
\includegraphics[scale=0.7]{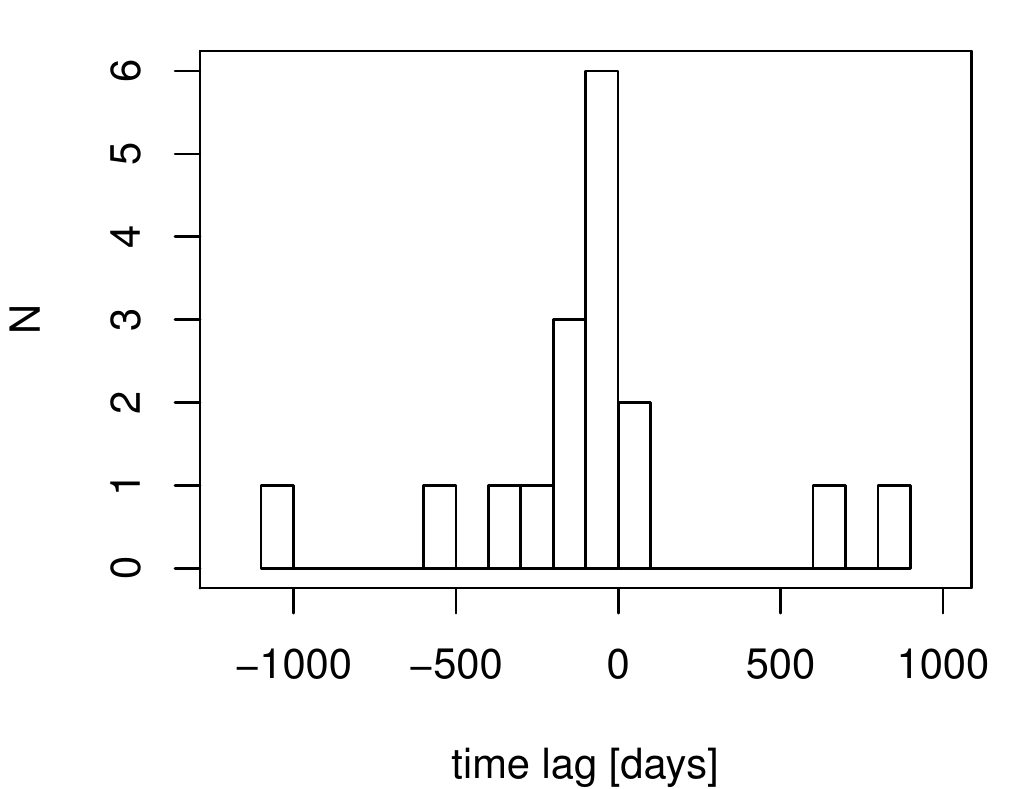}
\caption{Histogram of the time lags between optical and radio light
  curves. A negative lag means that optical leads the radio variations}\label{fig:corr}
\end{figure}
Of the 27 sources for which we calculated the DCF, 17 show
correlations at a $>3\sigma$ level. The distribution of the
time lags is shown in Fig.~\ref{fig:corr}. A negative time lag means
that optical leads the radio variations. 
The range of time lags is $-1030$ days (optical leading, 1ES~0647+250) to 860 days (radio leading, 3C~66A). In the case of 1ES~0647+250 the
correlation is barely significant and most likely due to a common
rising trend in the light curves, rather than correlated
flares (see also Appendix A). Similarly, in 3C~66A the correlation is barely significant and due to the large radio flare at the beginning of the monitoring period. The mean and median time lags are $-79$ and $-70$ days,
respectively, showing that typically the optical variations lead the
radio variations. 

\section{Discussion}

The results from the three methods applied to study the connection
between the emission at 15\,GHz and optical R-band in VHE $\gamma$-ray
detected BL~Lac object population are discussed below. The findings
for individual sources are discussed in Appendix A.

The modulation index analysis shows that the
variability in optical and radio bands differs significantly, with the
variability amplitudes in the optical being significantly larger than
in the radio. This is consistent with the findings for a much larger
blazar sample where the optical and 15\,GHz radio modulation indices
were compared \citep{hovatta14a}.
However, there seems to be a significant connection in these two bands when longer
time scales are studied. This has been previously found for single
sources of our sample (PG~1424+240 and MAGIC~J2001+439 \citep{Alek14a, Alek14b}), which partially
triggered the study presented here. The connection is evident both in
the simplistic approach of looking at overall trend of the light curve
as well as in DCF analysis. Comparing the results of these two analyses, we find that:
\begin{itemize}
\item For ten sources in our sample both analyses show connected variability, which is a strong indication
of common origin of the emission in these wavebands, both in very long
time scales (scoped by the linear regression) and shorter times scales
(correlated flares). 
\item For three sources (PG~1424+240, Mkn~501 and
1ES~1959+650) the linear regression analysis indicates common trends,
but the DCF plot shows no peak with significance of
$>3\sigma$. However, in visual inspection of light curves there seems to be ``correlated'' flares (less evident in the case of PG~1424+240 due to very strong increasing trend in the radio light curve). The DCF curves show some 2$\sigma$ peaks,
indicating that there is possibly multiple time scales involved, but
none of the peaks reaches 3$\sigma$ limit. 
\item For seven sources for which DCF finds significant correlation
(1ES~0033+595, 1ES~0414+009, 1ES~0806+524, 1ES~1218+304, PG~1553+113,
H~1722+119 and B3~2247+381), but no common trend is found. In case of
the first two, the linear regression analysis shows that there are no
significant trends in either optical or radio light curves. For three
sources (1ES~0806+524, 1ES~1218+304, PG~1553+113), there is no
significant trend in optical, but significant trend in radio. Finally, for
two (H~1722+119 and B3~2247+381) there is no trend in radio, but significant
trend in optical. These cases demonstrate a weakness in the linear
regression method: the significant trend is sometimes a result of a
single flare in the beginning or the end of the light curve, and a non-detection of a trend when the visually apparent trend changes direction within the time window we study.
\item Finally for 12 sources we find no indication of a connection
  between optical and radio variability. However, for five of these
  (RGB~0136+391, 1ES~0229+200, 1ES~0502+675, VER~J0648+152,
  RGB~0847+115), we did not even perform DCF analysis as there were too many low signal-to-noise points for estimating the modulation index. The remaining sample of 7 sources consists of
  two sources that are very weak in both optical and radio
  (RGB~1136+676 and 1ES~1426+428), two sources that are very weak in
  the optical (HB89~0317+185 and 1ES~1741+196) and three sources
  (S5~0716+714, ON~231 and ON~325) that show clear outbursts in both
  wave bands without apparent correlated behavior. For these weak sources,   there still might be connection, but as the
sources are weak, our measurements and methods fail to find them. 
  The remaining three sources are discussed below and
  individually in the Appendix.
\end{itemize}

The time lags we found between optical and radio variations are
similar to lags obtained using longer light curves of mainly bright
quasars and BL~Lac objects. \cite{tornikoski94} studied the
correlation between up to 15 years of radio and optical light curves
of 18 sources. They found several sources with correlated variations,
with the optical leading radio variations by 0 to few hundred
days. They attributed the lack of correlation to under-sampled light
curves. Similar results were also obtained by \cite{hanski} again
using long-term radio and optical data. In their study, seven out of
20 sources showed correlations in the DCF analysis with delays from 0
to several hundred days. In recent study by \cite{ramakrishnan} 2.5 years of data was used. In their study two out of nine (37\,GHz) and three out nine (95\,GHz) sources showed significant correlation, the lags varying from 78 to 272 days. \cite{tornikoski94}, \cite{hanski} and \cite{ramakrishnan}
used higher frequency radio data from 22 to 95\,GHz. At lower radio
frequencies a similar study was done by \cite{clements95} using
4.8, 8, and 14.5\,GHz radio data in comparison with optical light
curves of up to 26 years long. They also found correlated variations
in half of their sample of 18 sources, with optical variations leading
by 0 to 14 months. 

In our case the light curves are well sampled and at least in
some sources the lack of correlation seems to be due to lack of strong
variations in the radio light curves. Another alternative is that our
light curves are not long enough to detect variations, as variability
time scales in the radio light curves are typically long, on average
$4-6$ years \citep{hovatta07}. However, as discussed above, there are
also several sources with clear outbursts in both bands, but no
significant correlation (S5~0716+714, ON~231, ON~325, PKS~1424+240,
Mrk~501 and 1ES~1959+650). All but S5~0716+714 show at least two
2$\sigma$ peaks in the DCF, so it might be that for these sources
multiple time scales are involved. 


\subsection{Common Emission Component}

\begin{figure}
\includegraphics[scale=0.45]{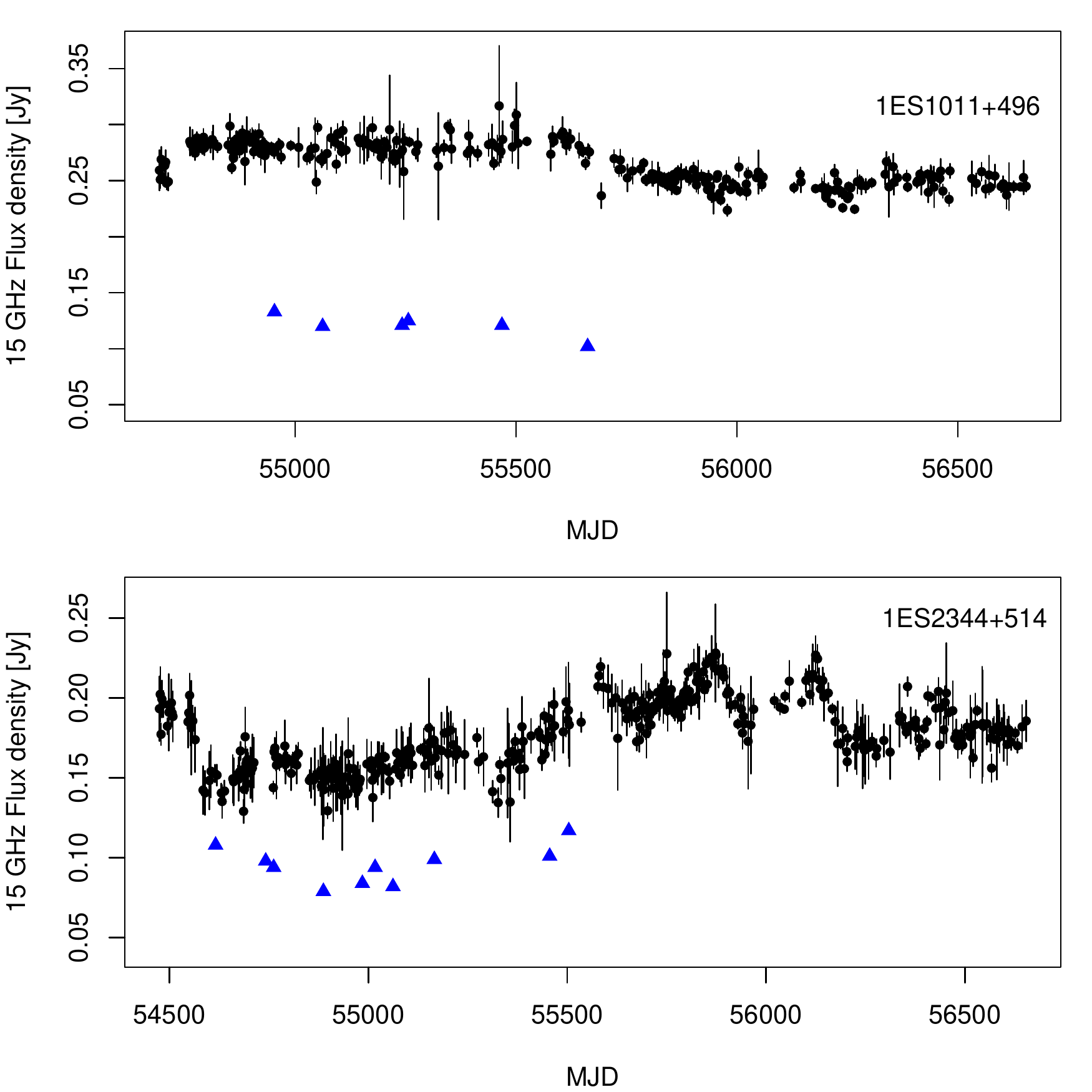}
\caption{Comparison of 15\,GHz light curve (black filled circles) and VLBA 15\,GHz core fluxes (blue triangles) from \cite{lister13} for 1ES~1011+496 (top) and 1ES~2344+514 (bottom).}\label{fig:corefluxes}
\end{figure}

In \cite{Alek14a}, studying PKS~1424+240, it was suggested that the
common trend seen in the radio and optical light curves is due to a
common emission component, which was suggested to be the 15\,GHz VLBA
core. In Fig.~\ref{fig:corefluxes} it is shown that indeed the brightness of the VLBA core closely follows the 15\,GHz light curve as has been previously found also at the higher frequencies \citep{savolainen02}.

\begin{figure*}
\includegraphics[scale=0.25, angle=-90]{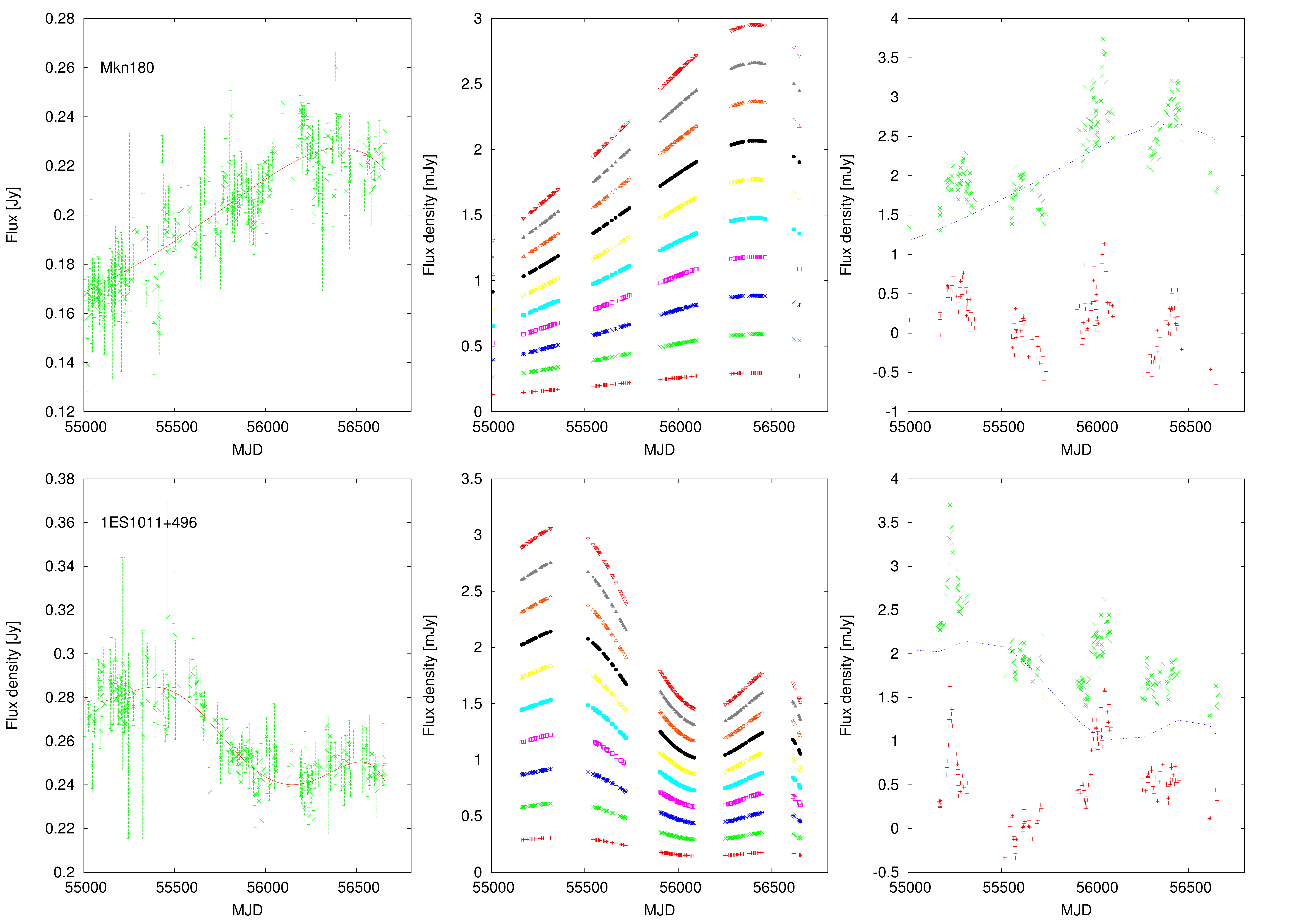}
\caption{Three steps to estimate the contribution from the slowly varying component to the optical light curves for Mkn~180 (top) 1ES~1011+496 (bottom).{\bf Left:} The polynomial (red solid line) is fit to the radio light curve (green symbols). {\bf Middle:} The polynomial is scaled to the same average flux density as the optical light curve and then multiplied with 0.1 (red crosses), 0.2, 0.3,...1.0 (red triangles). These polynomial-multiples are subtracted from the optical light curve. {\bf Right:} The optical light curve from which the polynomial-multiple that minimizes the rms (blue line, 0.9 for Mkn~180 and 0.7 for 1ES~1011+496) has been subtracted (red) and the original optical data (green).}\label{fig:example}
\end{figure*}

In order to study this further in our sample of objects, we used a simple approach to estimate the contribution from the slowly varying component to the optical light curves. We do this by fitting a polynomial to the radio light curve, which enables us to simply quantify the observed variations. We then subtract this polynomial from the optical light curve and calculate the rms of the polynomial-subtracted optical light curve and compare
it to the rms of the original optical light curve.
The three steps are shown in Fig.~\ref{fig:example} and include:
\begin{enumerate}
\item We fit a polynomial to the radio light curve. The order of the polynomial is defined by adding new orders until the $\chi^2$ of the fit does not improve any more. We define this by fitting first 40th order polynomials to determine the scatter in $\chi^2$ values and define that the fit did not improve when the improvement is smaller than this scatter. As the radio light curves of the sources are very different from each other, the number of orders differs from 1 to $\sim30$ for the best fits. The polynomial fits are overlaid on top of the radio light curve in Figs.~\ref{Fig:lc1}-\ref{Fig:lc32}.
\item We scale the polynomial fit to the same average flux density as the optical data{\footnote{We calculate the average and variance of the polynomial sampled with the dates of the optical data and subtract the average. We calculate the average and variance of the optical data and subtract the average. Finally we scale the polynomial such that the variances are equal and add the average of the optical data.}} and then the polynomial fit is multiplied with 0.1, 0.2, 0.3,...1.0 and the resulting curve is subtracted from the optical data. We calculate the rms of the resulting light curves and select as best-fit the one that minimizes the rms. We also tested if shifting the polynomial fit by the amount of the most significant lag would decrease the rms of the subtracted light curve, but on average this did not seem to be the case.
\item To estimate the fractional contribution of this slowly varying component to the optical flux density, we divide the rms of the best-fit-subtracted data with the rms of the original data:
Fraction$ = 1 -(\mathrm{rms_{\,subtracted\,optical\,data}}/\mathrm{\,rms_{optical\,data}})$.
\end{enumerate}

The polynomial fits to light curves have $\chi^2_{\mathrm{reduced}}>1$ meaning that it does not describe {\it all} radio variability. Furthermore this analysis cannot account for varying fraction or different time scales of the flares, i.e. for sources which show multiple flares in both bands, with varying amplitude ratio (e.g. 3C~66A, 1ES~0806+524), or different time scales (typically faster rise of the optical flare, e.g. PG~1553+113), the results are rather poor. 

The fractions we find vary from $0-63\%$, with in total nine sources showing no significant change in rms with the subtracted optical light curve compared to the non-subtracted optical light curve. There are 7 sources that have contribution $\sim50\%$: 1ES~0414+009, VER~0521+211, VER~J0648+152, Mkn~421, Mkn~180, 1ES~1218+304 and MAGIC~J2001+439. These are all sources\footnote{except VER~0648+152 for which correlation analysis was not performed} for which also the DCF analysis showed significant correlation and the majority also showed same direction in the linear regression analysis. The average fraction for the whole sample is 0.09 and for the sample, from which we have removed the 12 sources that showed no connection between optical and radio in our DCF and linear regression analysis, it is 0.27. However, due to the limitations of the analysis described above, these fractions should be considered as lower limits.
It is still clear that significant fraction of the variability of the optical light curve originates from another component and that the relative contribution of these two components vary from one source to another. 
In the future work we will study if this other component can be associated to the region that also emits the VHE $\gamma$-rays (Reinthal et al. in prep.). 
Moreover, it is evident that as some of the sources also show clear flares in their radio light curves (e.g., BL~Lac), there can be multiple emission regions contributing to the observed flux density also in the radio band. This needs to be accounted for in the SED modelling.

\section{Summary and Conclusions}

In this work we have presented the first study of the optical and
radio variability of VHE $\gamma$-ray detected BL~Lac objects. The
population consists of mainly HSPs, which are in general faint radio
sources and therefore rather little studied in this waveband. Still we
find that all studied sources, for which we can calculate the
modulation indices, are variable at 3$\sigma$ level. Using linear
regression analysis, we find significant increasing or decreasing
trends in the radio light curves of 21 of our 32 sources. In the case
of 13 sources, the trend is in same direction as the trend found in
optical light curves and our simulations show that chance coincidence
for this has $p=0.005$. We also found significant correlation between
radio and optical light curves for 17 sources. This clearly supports
the common origin for radio and optical emission for some sources, which is in conflict with the most commonly adopted SED model, the one-zone SSC model,
where the optical emission is assumed to originate from VHE
$\gamma$-ray emitting region and radio emission from separate outer
region and is excluded from the modelling.

We also study the amplitude of the variability of the radio and
optical light curves. We find that modulation indices found for
optical light curves are significantly larger than for radio light
curves. Inspection of the light curves shows that many sources show
fast sharp flares in the optical band, which are absent in most of the
radio light curves. It is therefore evident, that in addition to
common emission component, there is a second component contributing to
the optical emission, which can indeed be linked to VHE $\gamma$-ray
emission. We quantified this by estimating the slowly varying
component in the optical light curves using the trends in the radio
curves as a guidance, and found that on average, at least 27\% of the
optical emission is coming from the slowly varying component. This supports the two-zone models that have been suggested for e.g. Mrk~501 \citep{katarz, doert} and most recently for PKS~1424+240 \citep{Alek14a}, but also potentially provides a method to separate the emission from these components, by means of comparing the long term radio and optical light curves. In future work we will compare the method with the method suggested in \cite{barresalmeida} using the optical polarization to separate the SED components.     

As the sample studied in this work is VHE $\gamma$-ray selected, we
also checked if average flux densities or modulation indices correlate
with the VHE $\gamma$-ray flux given in Table~1. We found no
significant correlation, which is in agreement with the emission
scenario presented above. However, it should be noted that our VHE
$\gamma$-ray data is not coherent in a sense that it could present
different state for different sources (e.g. some sources have been
only observed once). We will address also this question in future
work (Fallah Ramazani et al. in preparation).

\begin{acknowledgements}
TH was supported by the Academy of Finland project number 267324.
The OVRO 40-m monitoring program is
supported in part by NASA grants NNX08AW31G 
and NNX11A043G, and NSF grants AST-0808050 
and AST-1109911.
This research has made use of the NASA/IPAC Extragalactic Database (NED) which is operated by the Jet Propulsion Laboratory, California Institute of Technology, under contract with the National Aeronautics and Space Administration.
\end{acknowledgements}

\appendix
\section{Comments on individual sources}

\begin{figure}
\includegraphics[width=0.45\textwidth]{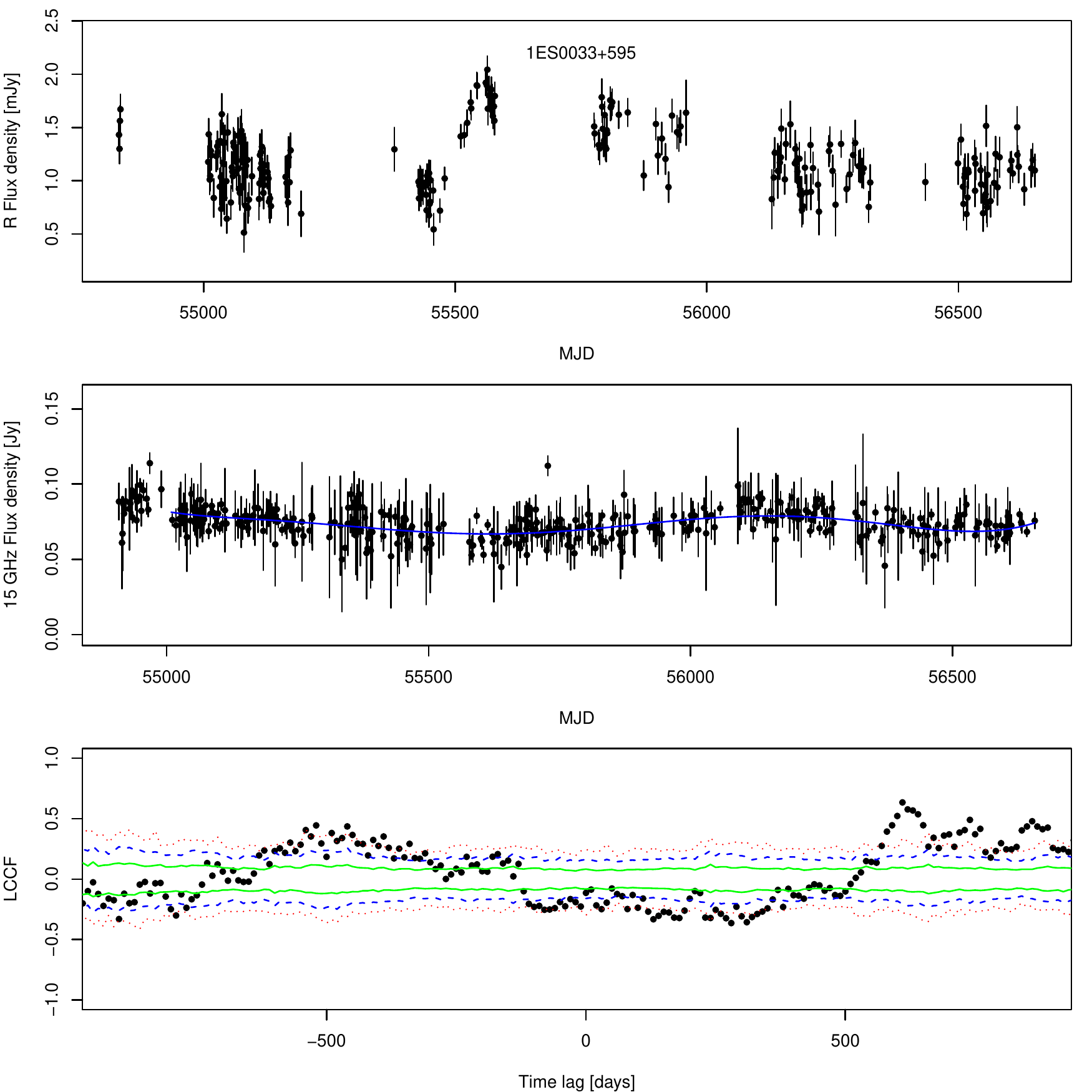}
\caption{The optical R-band light curve (top) and radio 15\,GHz light curve (middle) of 1ES~0033+595. The solid blue line in middle panel shows the polynomial fit to radio data, which is subtracted from the optical light curve to estimate the contribution of the slowly varying component to optical flux. The bottom panel shows the results of the DCF study; the green, blue and red line representing the $1\sigma$, $2\sigma$ and $3\sigma$ significance limits, respectively.}
\label{Fig:lc1}
\end{figure}

{\bf 1ES~0033+595}: The average flux density in the optical and radio bands is
below the median of the sample. The radio nor the optical light curve
show significant trends, but DCF shows significant correlation with
radio leading by 610 days (see Fig.~\ref{Fig:lc1}). This correlation is due to a few radio data
points in the beginning of the light curve with rather small uncertainties. Subtracting the polynomial fitted to radio light curve does not reduce the rms of the optical light curve at all, i.e. also this method does not find connection between the optical and radio bands. 

\begin{figure}
\includegraphics[width=0.45\textwidth]{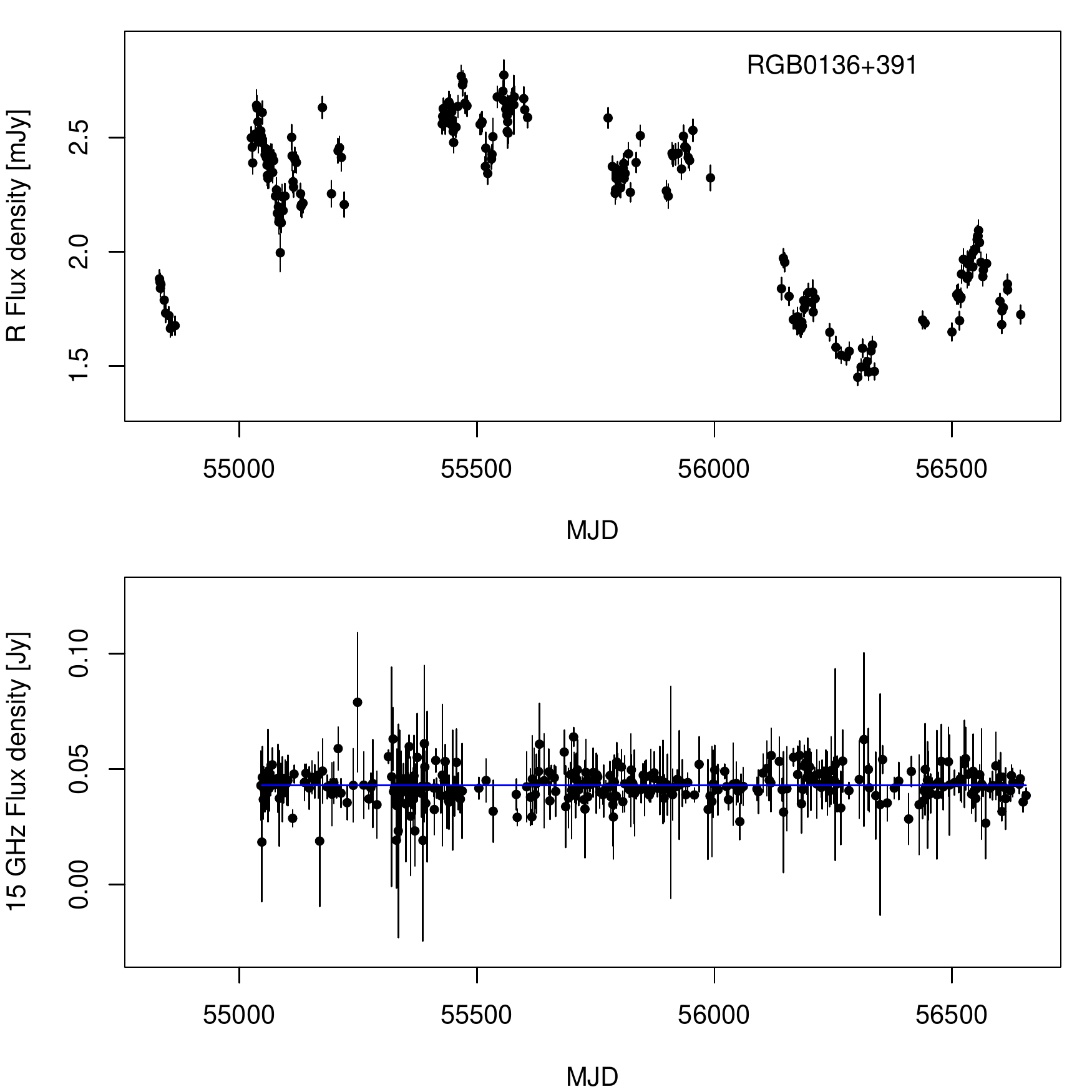}
\caption{The optical R-band light curve (top) and radio 15\,GHz light curve (bottom) of RGB~0136+391. The solid blue line in the bottom panel shows the polynomial fit to the radio data. As the modulation index for radio data could not be determined, no DCF analysis was performed.}
\label{Fig:lc2}
\end{figure}

{\bf RGB~0136+591}: One of the faintest radio sources, while in the
optical the average flux density is above the median. In the radio
band the modulation index calculation failed due to large
uncertainties and faintness of the source. The optical light curve has
a significant decreasing trend, while the radio light curve does not
show any significant trend. This is in agreement with the results from
simulated light curves, where in only 2 cases out of 1000 a significant trend occurs in the radio band. In the polynomial fitting, this was one of the few sources for which adding higher order polynomial did not improve the fit significantly, i.e. straight line was the best fit for the radio light curve (see Fig.~\ref{Fig:lc2}).  

\begin{figure}
\includegraphics[width=0.45\textwidth]{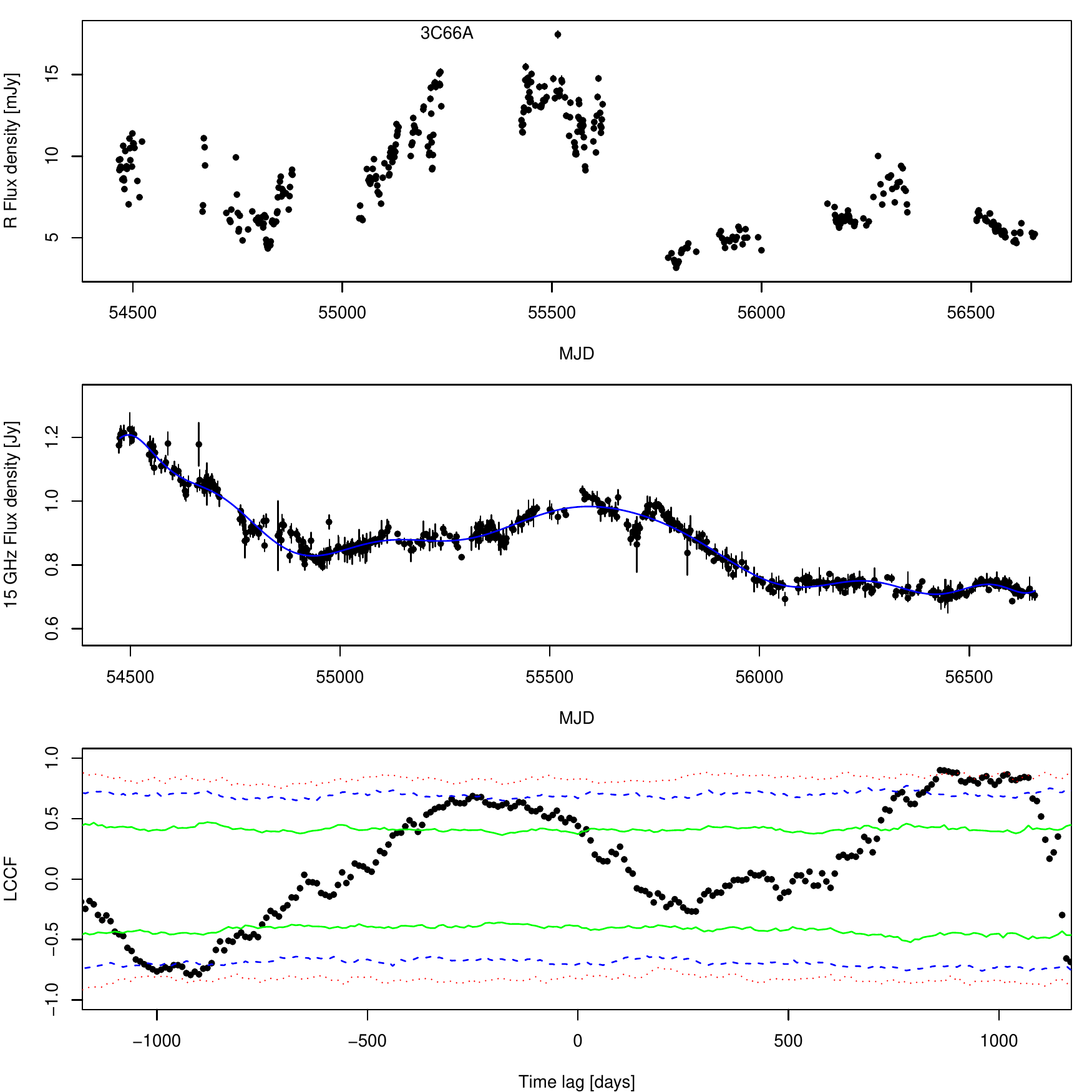}
\caption{The optical R-band light curve (top) and radio 15\,GHz light curve (middle) of 3C~66A. The solid blue line in middle panel shows the polynomial fit to radio data, which is subtracted from the optical light curve to estimate the contribution of the slowly varying component to optical flux. The bottom panel shows the results of the DCF study; the green, blue and red line representing the $1\sigma$, $2\sigma$ and $3\sigma$ significance limits, respectively.}
\label{Fig:lc3}
\end{figure}

{\bf 3C~66A}: One of the brightest sources in both bands. The optical
light curve is characterized by fast variability, while in the radio
band the variability is clearly slower. In both light curves there is
a highly significant decreasing trend. As discussed in Sect. 4.3, the time lag of 860 days (radio leading) is caused by the large radio flare in the beginning of the light curve (see Fig.~\ref{Fig:lc3}). However, visually it seems that the two bands correlate on long time scales with optical leading the radio, but this peak has significance of $\sim2\sigma$ only in the DCF plot. The result is in agreement with previous correlation studies \citep[e.g.][]{hanski}. It is also evident that our simplistic method to define the fraction of common radio-optical component to the optical light curve fails, because the ratio of the amplitudes of the flares in optical and radio is clearly variable.

\begin{figure}
\includegraphics[width=0.45\textwidth]{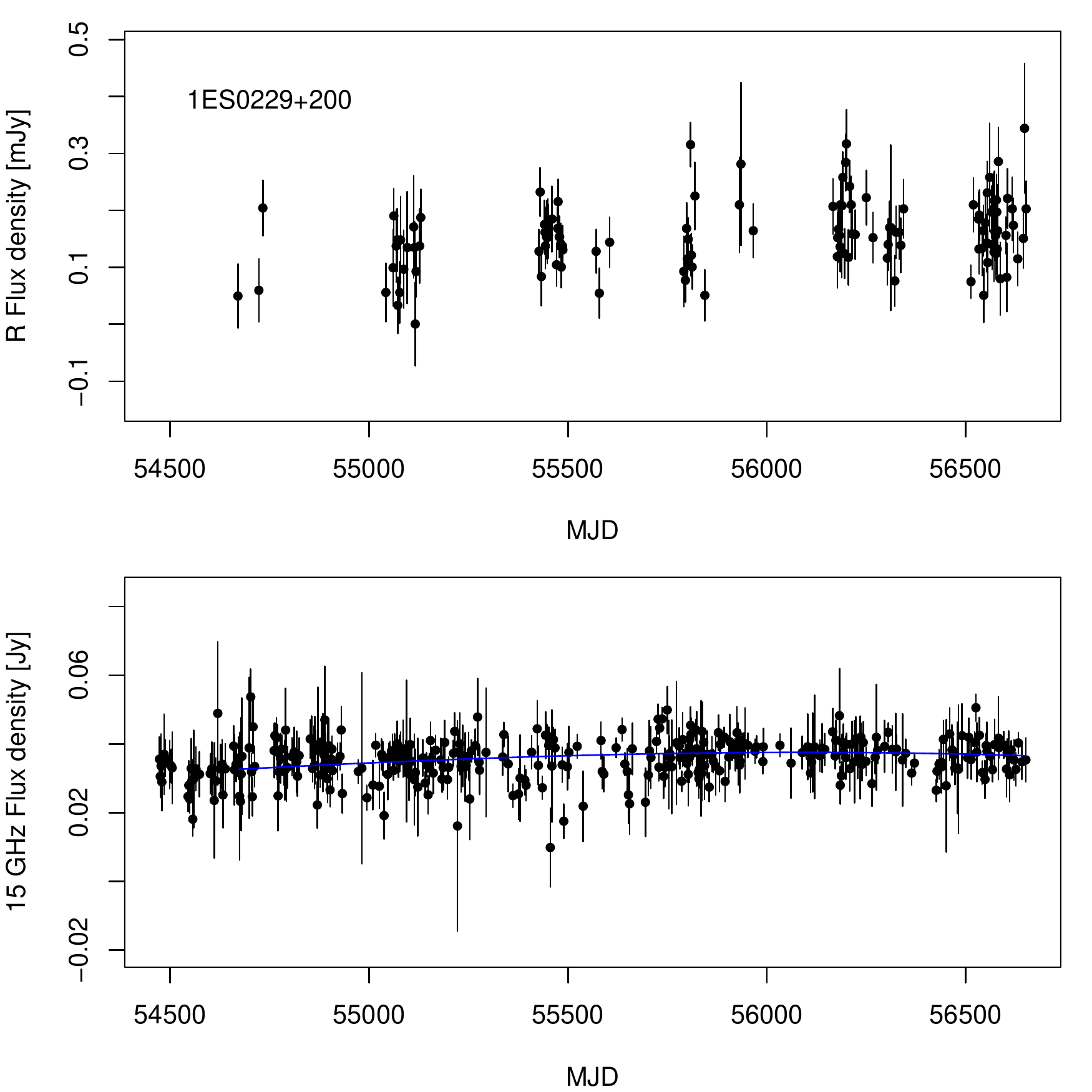}
\caption{The optical R-band light curve (top) and radio 15\,GHz light curve (bottom) of 1ES~0229+200. The solid blue line in the bottom panel shows the polynomial fit to the radio data. As the modulation index for radio data could not be determined, no DCF analysis was performed.}
\label{Fig:lc4}
\end{figure}

{\bf 1ES~0229+200}: The faintest source in our optical sample, resulting in failing calculation of the modulation index for the source. The source is also one of the weakest sources in the radio band in our sample. Still the radio light curve shows significant increasing trend, while in the optical light curve the increasing trend is significant only at 99.9\% significance level. Our polynomial subtraction method decreases the rms of the optical light curve very little, which is probably due to large uncertainty in the optical data. Visually the shape of the polynomial seems to trace the general shape of the optical light curve (see Fig.~\ref{Fig:lc4}). 

\begin{figure}
\includegraphics[width=0.45\textwidth]{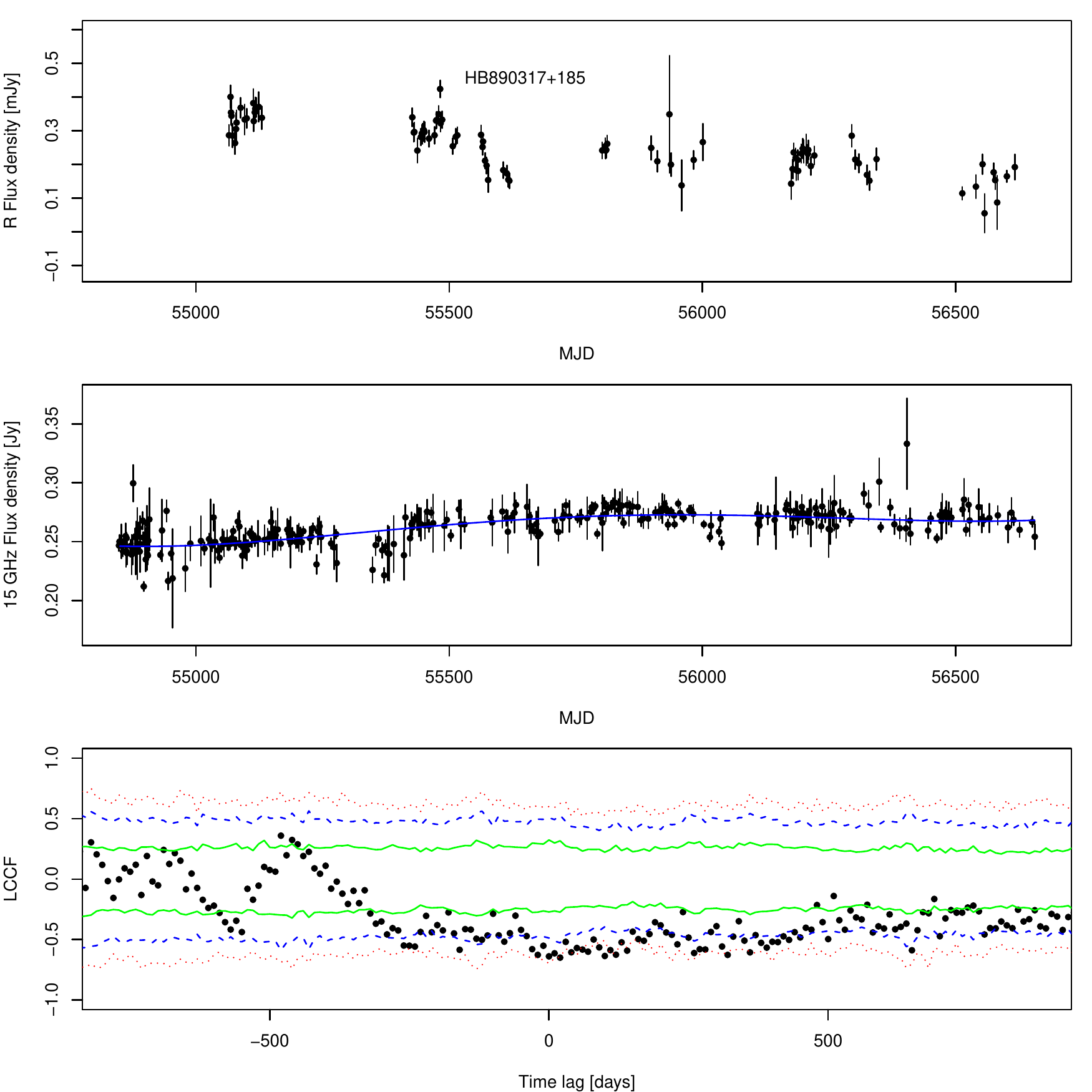}
\caption{The optical R-band light curve (top) and radio 15\,GHz light curve (middle) of HB89~0317+185. The solid blue line in middle panel shows the polynomial fit to radio data, which is subtracted from the optical light curve to estimate the contribution of the slowly varying component to optical flux. The bottom panel shows the results of the DCF study; the green, blue and red line representing the $1\sigma$, $2\sigma$ and $3\sigma$ significance limits, respectively.}
\label{Fig:lc5}
\end{figure}

{\bf HB89~0317+185}: One of the weakest sources of the sample in the optical, while the radio flux density is above the median of the sample. The
optical light curve shows significant decreasing trend, while the
radio light curve shows significant increasing trend. Also DCF finds
no significant correlation between the two bands, and therefore it is not surprising that our polynomial subtraction method does not decrease the rms much. We note that the
optical light curve is poorly sampled compared to most other sources in our sample (see Fig.~\ref{Fig:lc5}).

\begin{figure}
\includegraphics[width=0.45\textwidth]{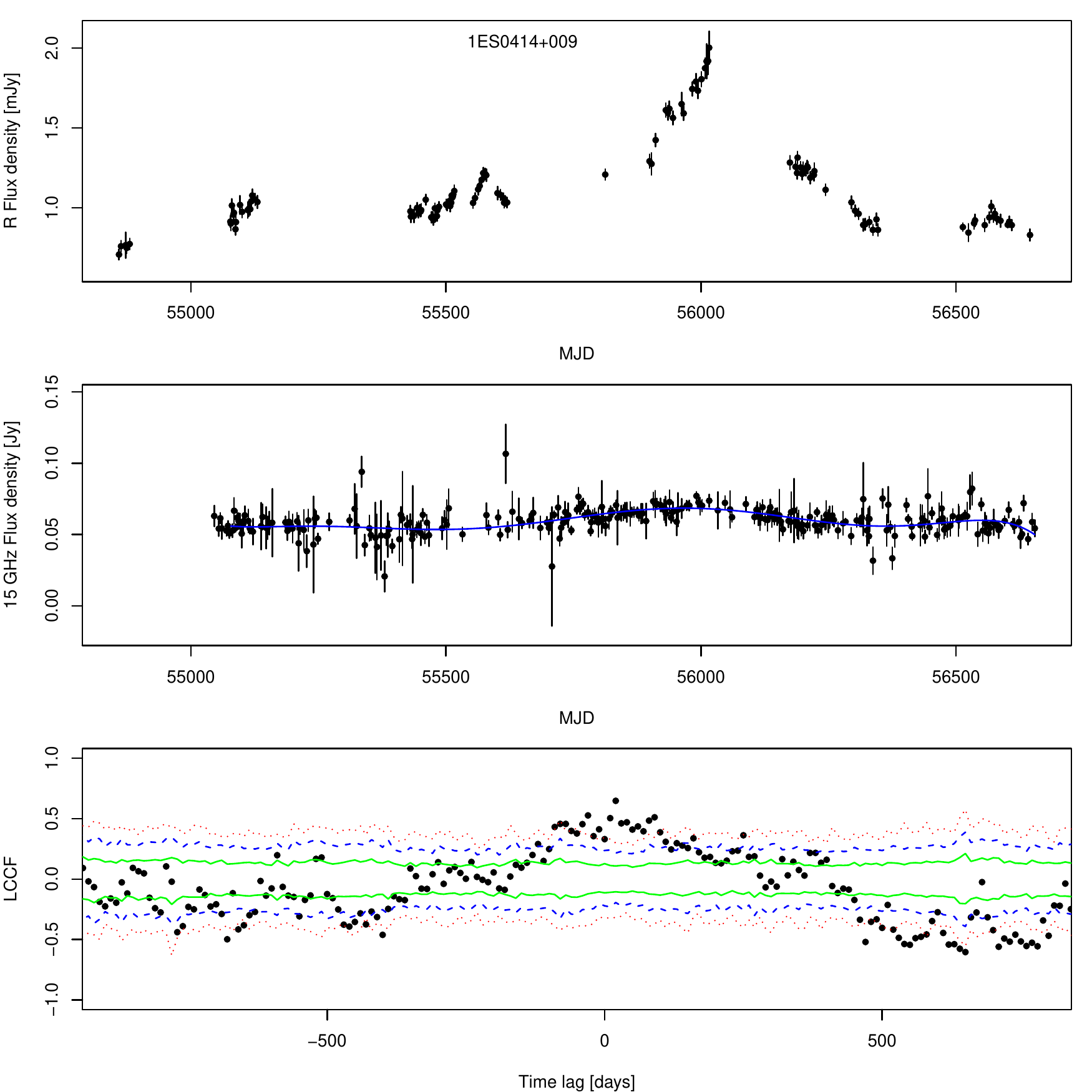}
\caption{The optical R-band light curve (top) and radio 15\,GHz light curve (middle) of 1ES~0414+009. The solid blue line in middle panel shows the polynomial fit to radio data, which is subtracted from the optical light curve to estimate the contribution of the slowly varying component to optical flux. The bottom panel shows the results of the DCF study; the green, blue and red line representing the $1\sigma$, $2\sigma$ and $3\sigma$ significance limits, respectively.}
\label{Fig:lc6}
\end{figure}

{\bf 1ES~0414+009}: The source is rather weak in both bands. Both
light curves show no trend, but DCF shows a significant correlation
with radio leading optical by 20 days. Visually it appears that the
correlation is a result of a common long term behaviour, rather than correlated flares. With the trend changing direction within the studied period, the trend analysis fails, but instead the polynomial subtraction is successful and we find that at least $\sim50\%$ of the flux in optical originates from the common slowly varying optical-radio component (see Fig.~\ref{Fig:lc6}).

\begin{figure}
\includegraphics[width=0.45\textwidth]{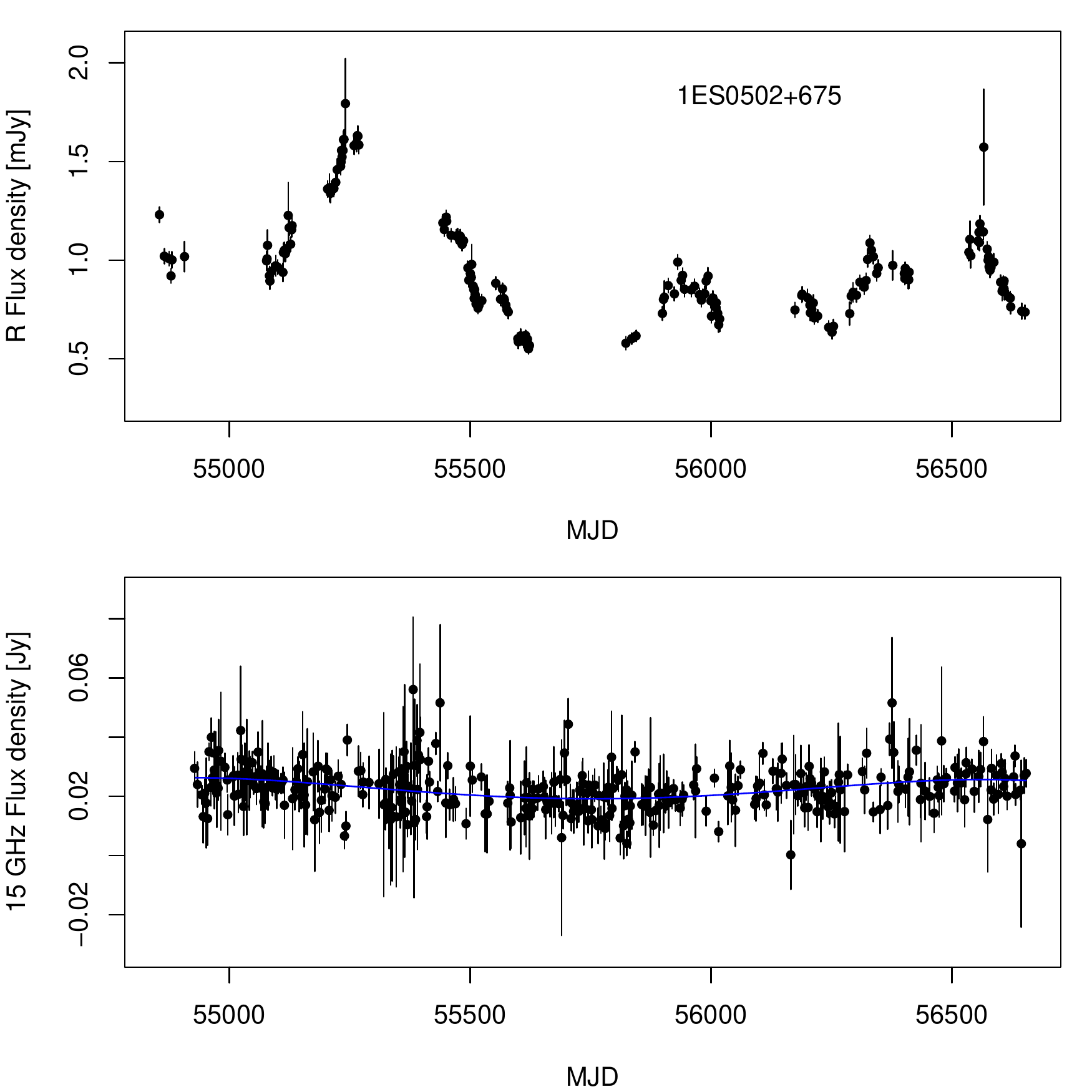}
\caption{The optical R-band light curve (top) and radio 15\,GHz light curve (bottom) of 1ES~0502+675. The solid blue line in the bottom panel shows the polynomial fit to the radio data. As the modulation index for radio data could not be determined, no DCF analysis was performed.}
\label{Fig:lc7}
\end{figure}

{\bf 1ES~0502+675}: The source is the weakest radio source in our
sample and therefore the calculation of the modulation index
fails. Thus, we perform no DCF analysis for this source. In the
optical the source shows clear outbursts and a highly significant decreasing trend (see Fig.~\ref{Fig:lc7}).

\begin{figure}
\includegraphics[width=0.45\textwidth]{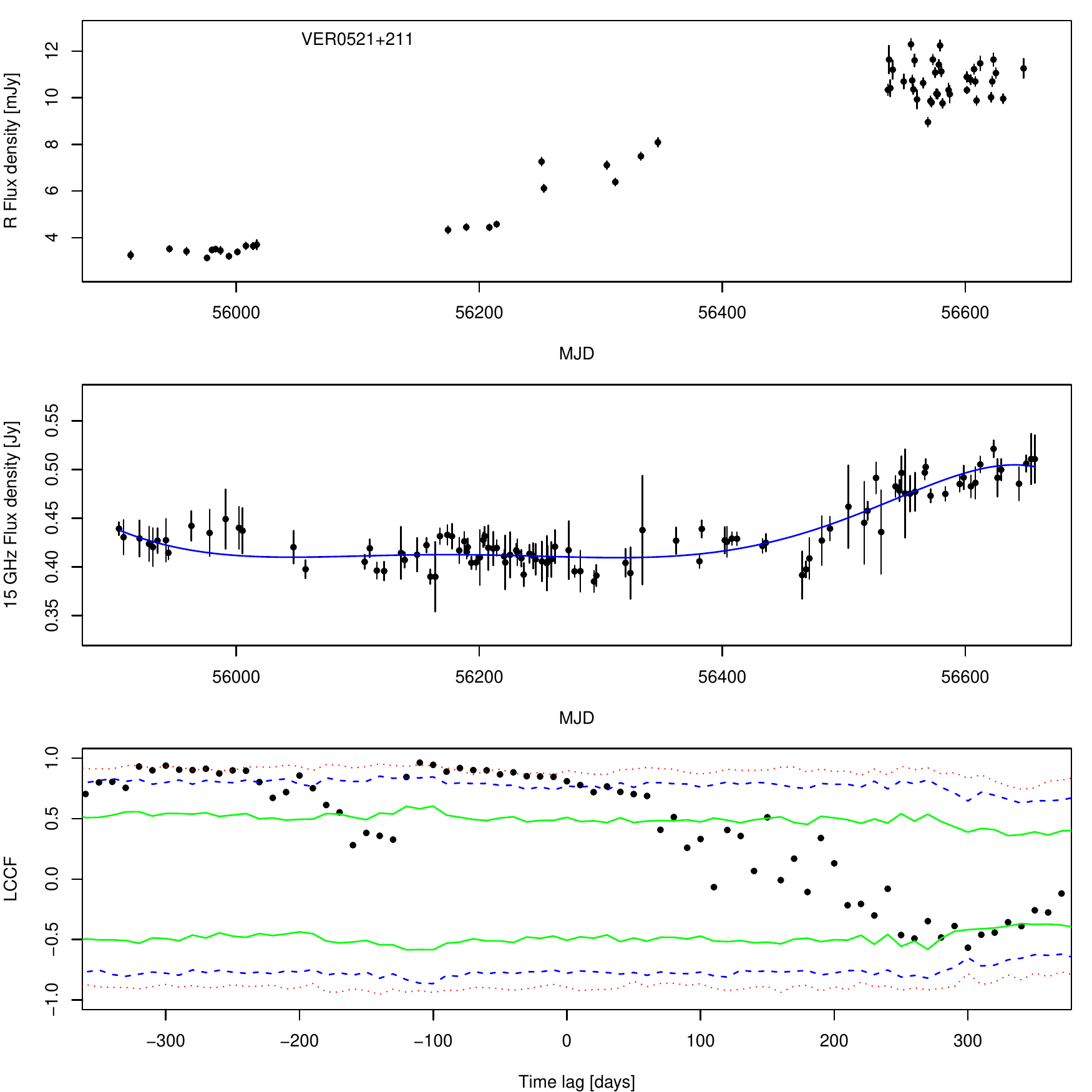}
\caption{The optical R-band light curve (top) and radio 15\,GHz light curve (middle) of VER~J0521+211. The solid blue line in middle panel shows the polynomial fit to radio data, which is subtracted from the optical light curve to estimate the contribution of the slowly varying component to optical flux. The bottom panel shows the results of the DCF study; the green, blue and red line representing the $1\sigma$, $2\sigma$ and $3\sigma$ significance limits, respectively.}
\label{Fig:lc8}
\end{figure}

{\bf VER~0521+211}:
The source is rather bright both in the optical and radio bands, and shows common increasing trend, as well as significant correlation with optical leading by 110 days. There are several 3$\sigma$ points in the correlation plot and actually these points form rather flat plateaus than single peaks, indicating that the correlation is probably due to common trend rather than correlated flares (see Fig.~\ref{Fig:lc8}). Subtracting the polynomial fit 
of the radio light curve from optical light curve shows that $\sim50\%$ of the optical flux originates from the common radio-optical component (which should be considered as lower limit, see main text).

\begin{figure}
\includegraphics[width=0.45\textwidth]{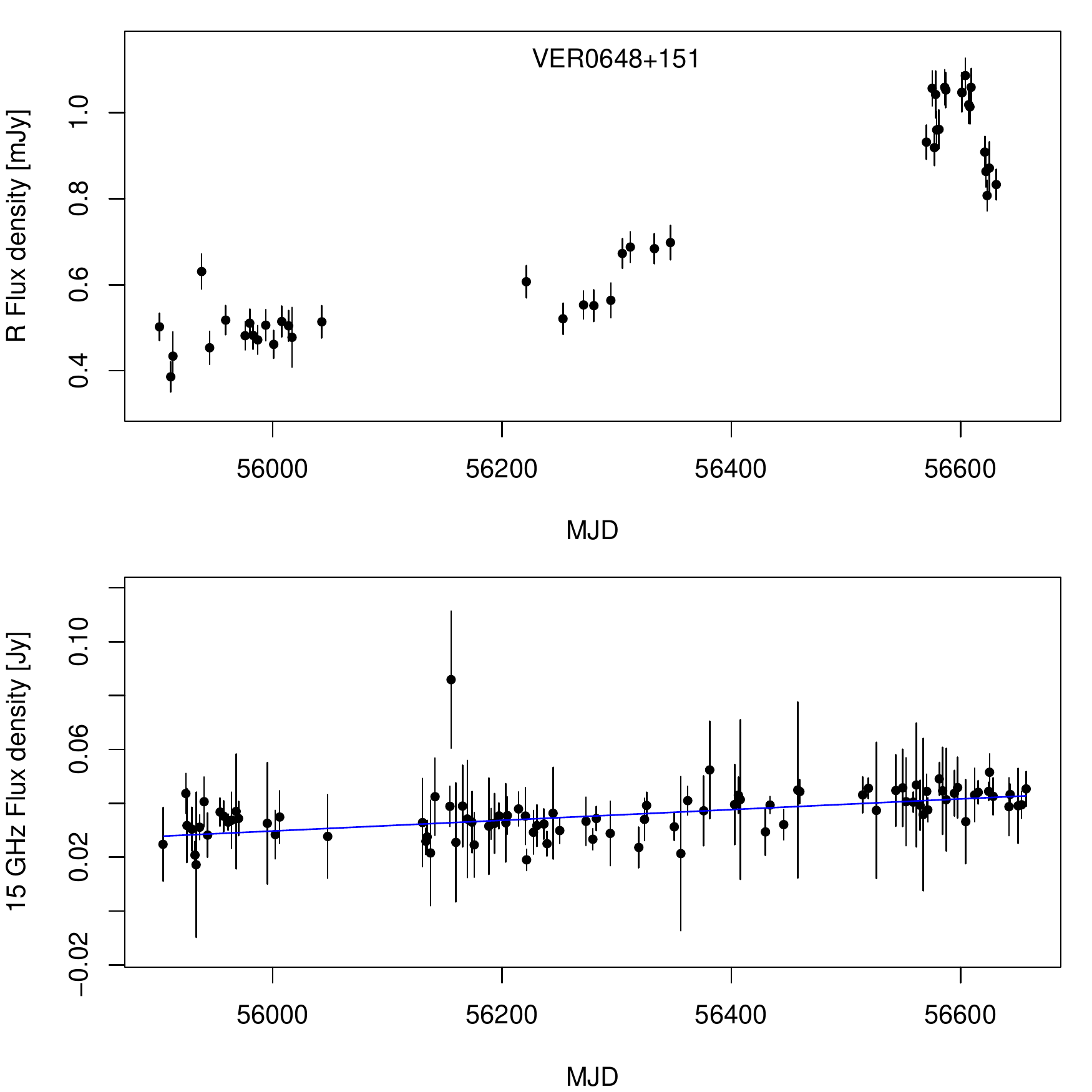}
\caption{The optical R-band light curve (top) and radio 15\,GHz light curve (bottom) of VER~J0648+152. The solid blue line in the bottom panel shows the polynomial fit to the radio data. As the modulation index for radio data could not be determined, no DCF analysis was performed.}
\label{Fig:lc9}
\end{figure}

{\bf VER~0648+152}: Weak optical and radio source (see Fig.~\ref{Fig:lc9}), and for the radio the modulation index calculation fails due to large uncertainties and
low measured flux density. However, we find significant increasing
trend in both bands. This is particularly puzzling as the simulations
indicate that no significant trends are expected in radio. Visual
inspection supports the result of the trend analysis as does the polynomial fitting to the radio data. 
We suggest that the problem with the modulation index calculation and simulated light
curves arises from the overestimation of the uncertainties in the
radio band. However, due to this disagreement between real data and
simulations, we exclude the source from further analysis. 

\begin{figure}
\includegraphics[width=0.45\textwidth]{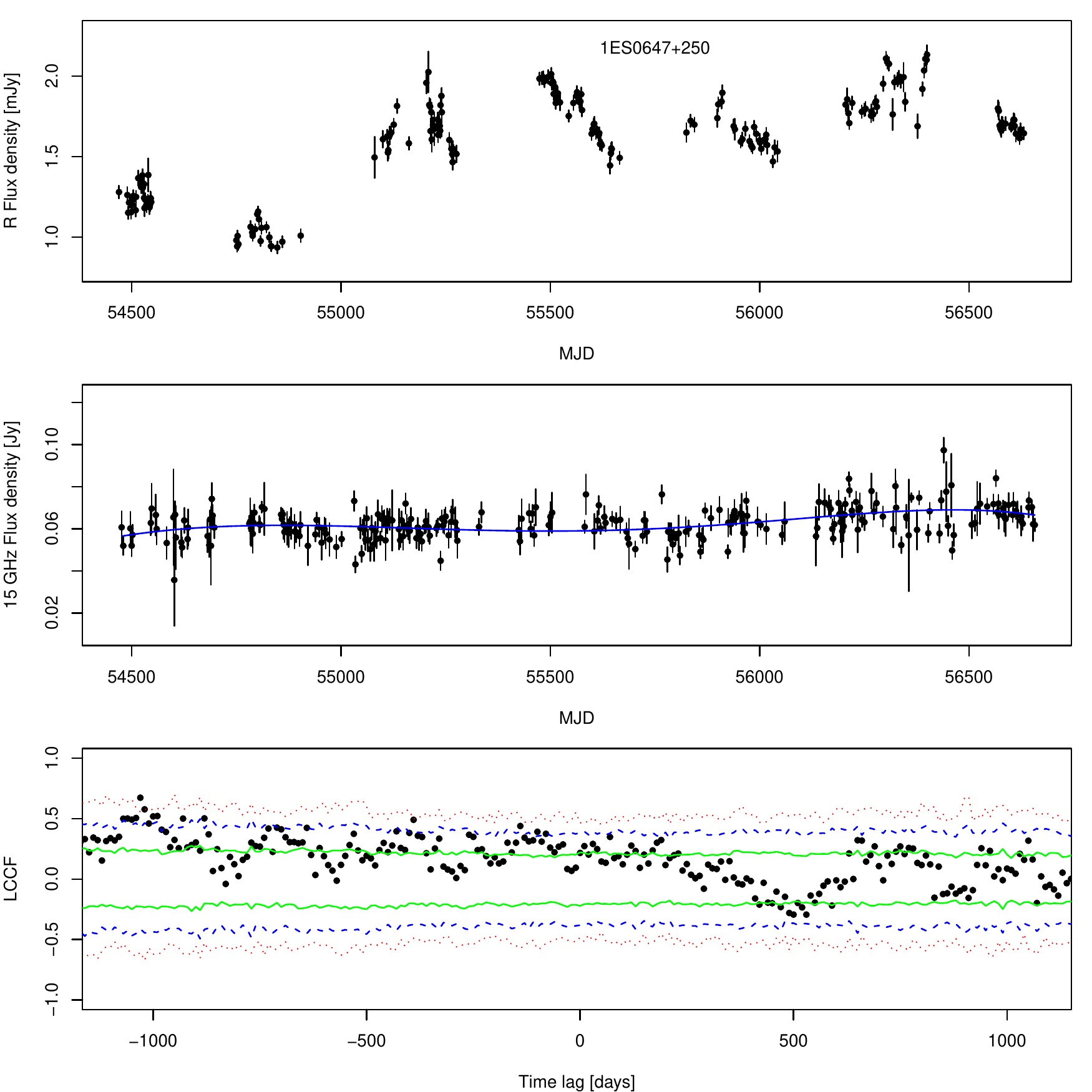}
\caption{The optical R-band light curve (top) and radio 15\,GHz light curve (middle) of 1ES~0647+250. The solid blue line in middle panel shows the polynomial fit to radio data, which is subtracted from the optical light curve to estimate the contribution of the slowly varying component to optical flux. The bottom panel shows the results of the DCF study; the green, blue and red line representing the $1\sigma$, $2\sigma$ and $3\sigma$ significance limits, respectively.}
\label{Fig:lc10}
\end{figure}

{\bf 1ES~0647+250}: The optical light curve shows clear outbursts and
significant increasing trend. In radio the source is rather weak, but
the increasing trend is significant also in this band. As discussed in
Sect. 4.3., the positive correlation found in DCF analysis, with time
lag of 1030 days, is due to this common trend rather than correlated
flares. The polynomial subtraction suggests that at least $\sim10\%$
of the optical flux would originate from common optical-radio
component (see Fig.~\ref{Fig:lc10}).

{\bf S5~0716+714}: The average flux densities in the optical and radio
are the second highest of the sample and the modulation indices are
the highest of the sample in both bands. The variability is fast and
visually there is no clear connection between the two bands, which is
also supported by our analysis methods finding no connection in
flaring activity nor long-term behavior. The study of \cite{raiteri03}
found varying long-term trend from the optical light curves with a
characteristic time scale of about 3.3 years, while a longer period of
$5.5-6$ years was found to characterize the radio long-term
variations. \cite{villata08} concludes that major optical outbursts
may have modest radio counterparts (at least in 37\,GHz) and thus, the
optically-emitting jet region is sometimes not completely opaque to
the high radio frequencies, while lower frequencies are at least
partially absorbed and a delay is observed. In recent study by \cite{ramakrishnan} significant correlation was found between 95\,GHz radio and optical data (but not 37\,GHz and optical), which further supports this conclusion. 

\begin{figure}
\includegraphics[width=0.45\textwidth]{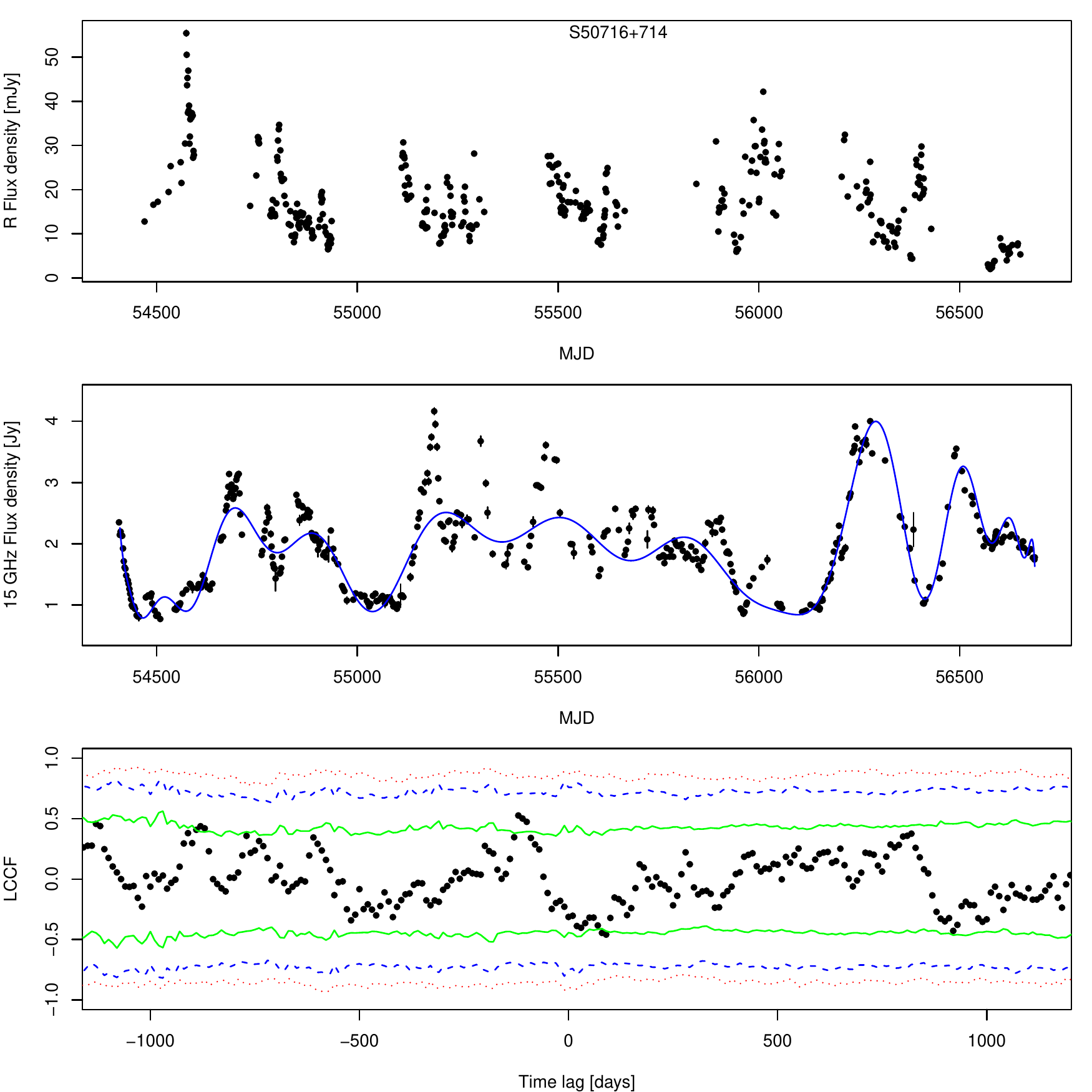}
\caption{The optical R-band light curve (top) and radio 15\,GHz light curve (middle) of S5~0716+714. The solid blue line in middle panel shows the polynomial fit to radio data, which is subtracted from the optical light curve to estimate the contribution of the slowly varying component to optical flux. The bottom panel shows the results of the DCF study; the green, blue and red line representing the $1\sigma$, $2\sigma$ and $3\sigma$ significance limits, respectively.}
\label{Fig:lc11}
\end{figure}

Our analysis is not optimal for finding periodicities, although one would naively expect them to result in trends, and we find no significant correlation. Therefore, we cannot confirm the results from these previous studies, but this might simply suggests that methods adopted here are too simplistic for the case of S5~0716+714 (see Fig.~\ref{Fig:lc11}).

\begin{figure}
\includegraphics[width=0.45\textwidth]{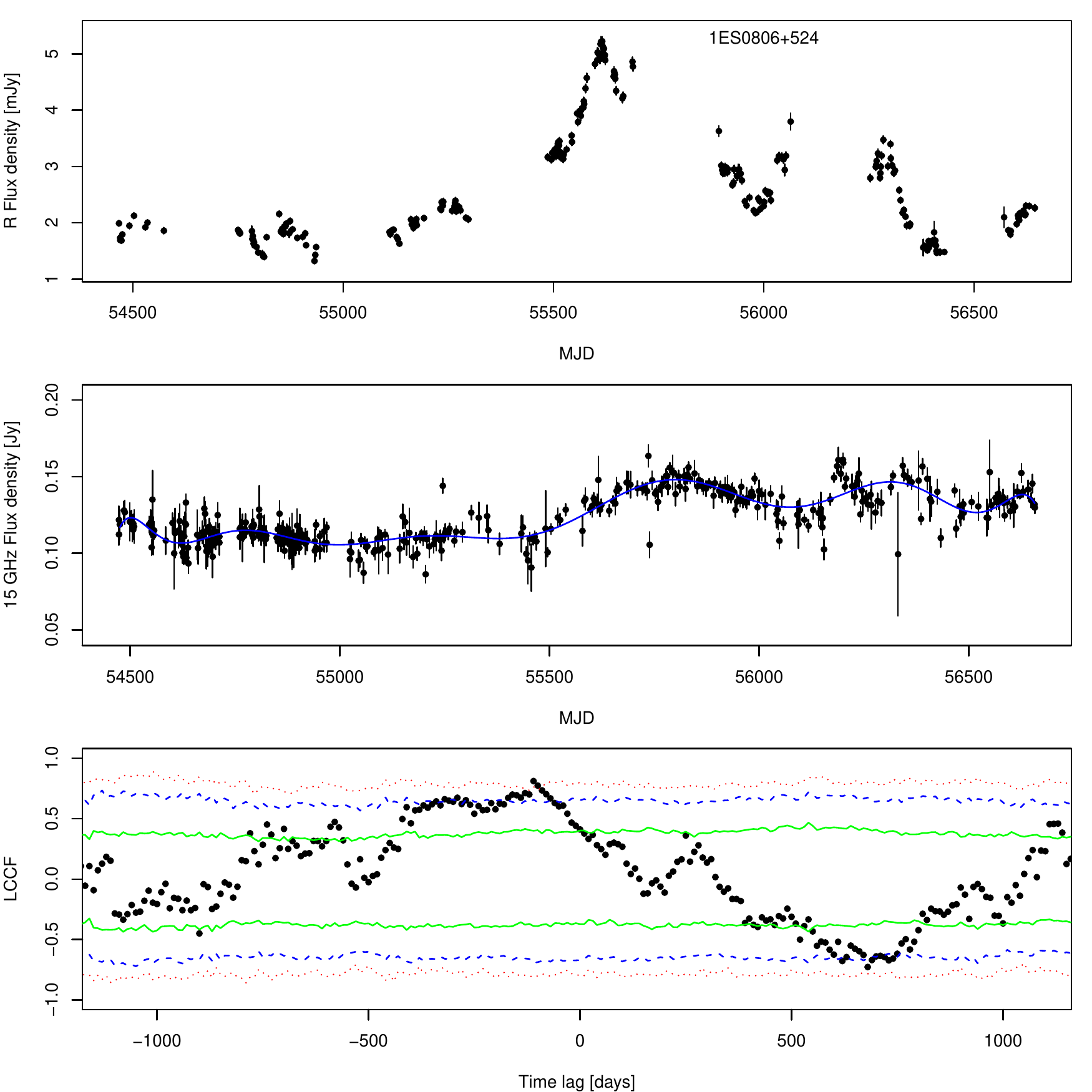}
\caption{The optical R-band light curve (top) and radio 15\,GHz light curve (middle) of 1ES~0806+524. The solid blue line in middle panel shows the polynomial fit to radio data, which is subtracted from the optical light curve to estimate the contribution of the slowly varying component to optical flux. The bottom panel shows the results of the DCF study; the green, blue and red line representing the $1\sigma$, $2\sigma$ and $3\sigma$ significance limits, respectively.}
\label{Fig:lc12}
\end{figure}

{\bf 1ES~0806+524}: This source is close to the median flux density of
the sample in both wavebands. Both light curves show very little
variability before MJD 55400, after which there is
a clear outburst in both wavebands. The peak in the optical is reached
$\sim$200 days before the peak in radio, but due to a gap in the
optical light curve we cannot exclude the possibility of a second peak
at the time of the radio maxima. After the outburst the optical flux density steadily decreases and no significant trend is found in our trend analysis. DCF finds a significant correlation with optical leading radio by 110 days. In \cite{Alek15b} short period around the optical flare was studied and no correlation found. The shape of the polynomial fitted to the radio light curve visually resembles the shape of the optical light curve very well, but as for 3C~66A, the ratio of the amplitudes of optical and radio outbursts is variable ((see Fig.~\ref{Fig:lc12}) and therefore the subtraction of the polynomial does not decrease the rms of the optical light curve significantly.  

\begin{figure}
\includegraphics[width=0.45\textwidth]{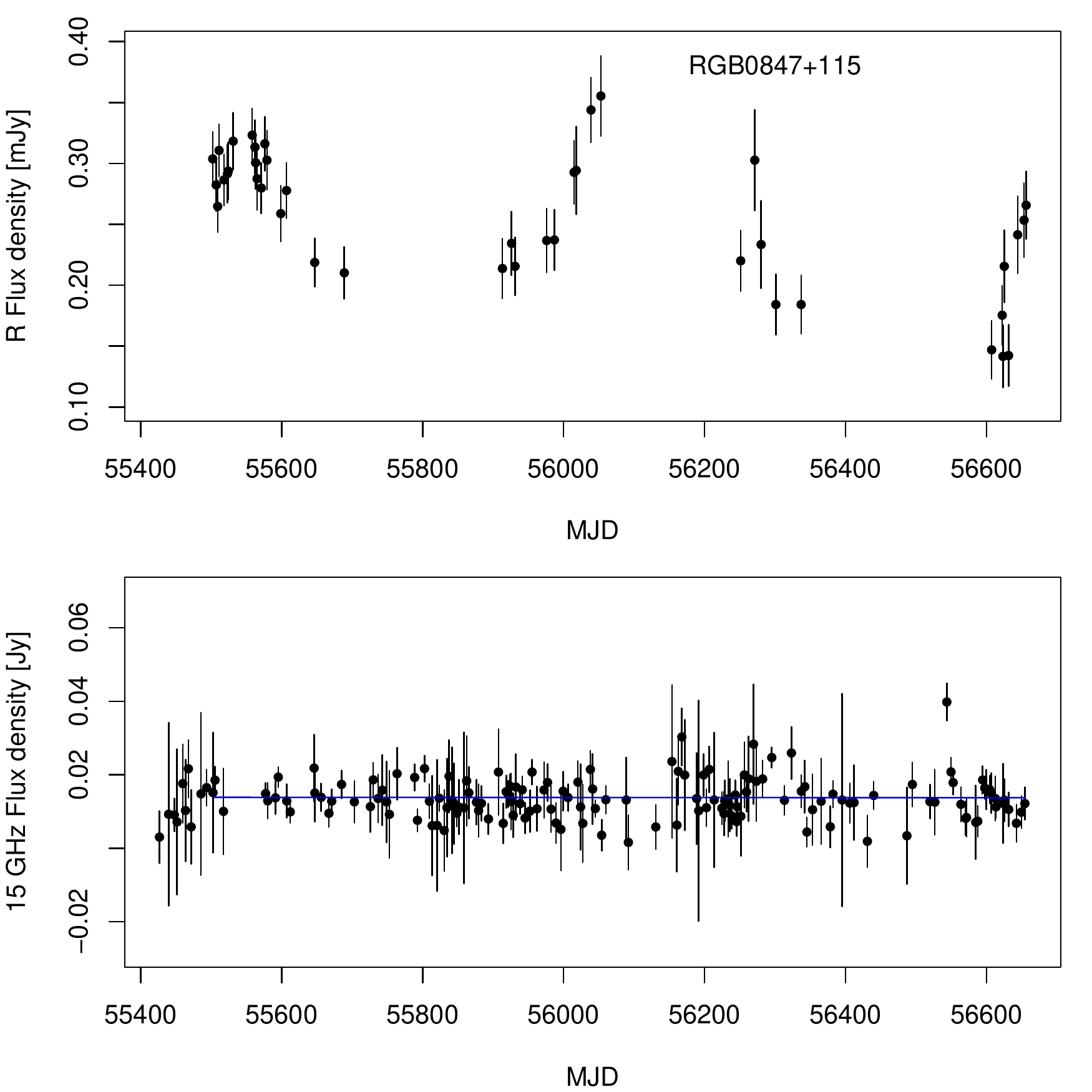}
\caption{The optical R-band light curve (top) and radio 15\,GHz light curve (bottom) of RGB~0847+115. The solid blue line in the bottom panel shows the polynomial fit to the radio data. As the modulation index for radio data could not be determined, no DCF analysis was performed.}
\label{Fig:lc13}
\end{figure}

{\bf RGB~0847+115}: Among the weakest sources in our sample in both
optical and radio. The calculation of the radio modulation index fails. The linear regression does not find significant trend in the radio light curve, but the polynomial fit does favour second order polynomial with decreasing trend. As such trend is clearly present in optical light curve, subtracting the ``polynomial'' (which in this case is just a line) decreases the rms of the optical light curve significantly. 
We note that the light curves of this source
have a few data points compared to other sources, as it was added to
monitoring programs only after the {\it Fermi}-LAT detection (see Fig.~\ref{Fig:lc13}).  

\begin{figure}
\includegraphics[width=0.45\textwidth]{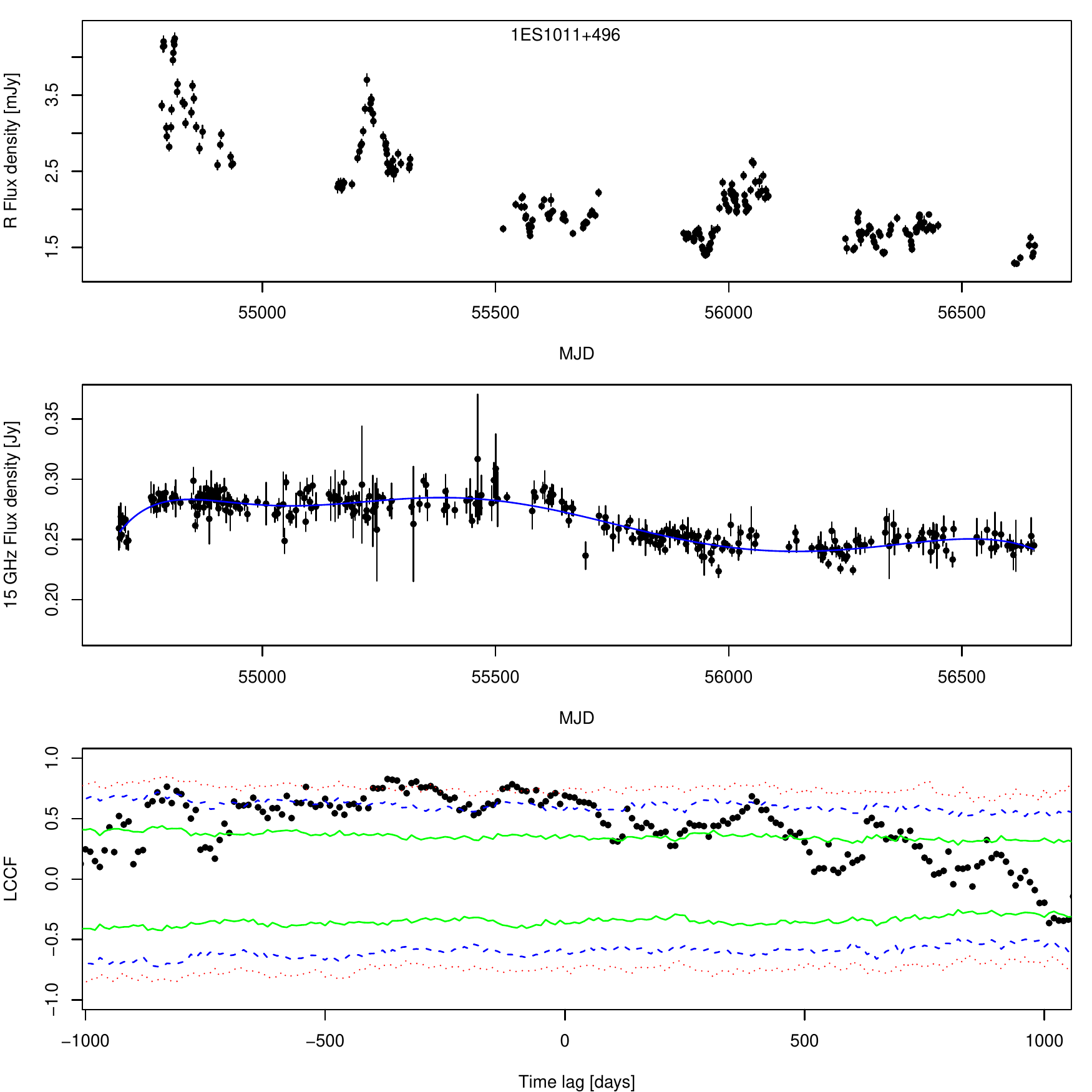}
\caption{The optical R-band light curve (top) and radio 15\,GHz light curve (middle) of 1ES~1011+496. The solid blue line in middle panel shows the polynomial fit to radio data, which is subtracted from the optical light curve to estimate the contribution of the slowly varying component to optical flux. The bottom panel shows the results of the DCF study; the green, blue and red line representing the $1\sigma$, $2\sigma$ and $3\sigma$ significance limits, respectively.}
\label{Fig:lc14}
\end{figure}

{\bf 1ES~1011+496}: This source is close to the median flux density of the sample in both wavebands. The radio light curve shows little variability, but a clear decreasing trend. The modulation index of the optical light curve is much larger, but in addition to the short-term variability, the source also shows highly significant decreasing trend. The DCF analysis shows several peaks and is actually rather flat, suggesting that the correlation is probably due to common decreasing trend of the light curves rather than correlated flares. As shown in Fig.~6 and Fig.~\ref{Fig:lc14}, for this source the polynomial subtraction method works rather well and suggest that at least $\sim25\%$ of the flux comes from common optical-radio emission component.

\begin{figure}
\includegraphics[width=0.45\textwidth]{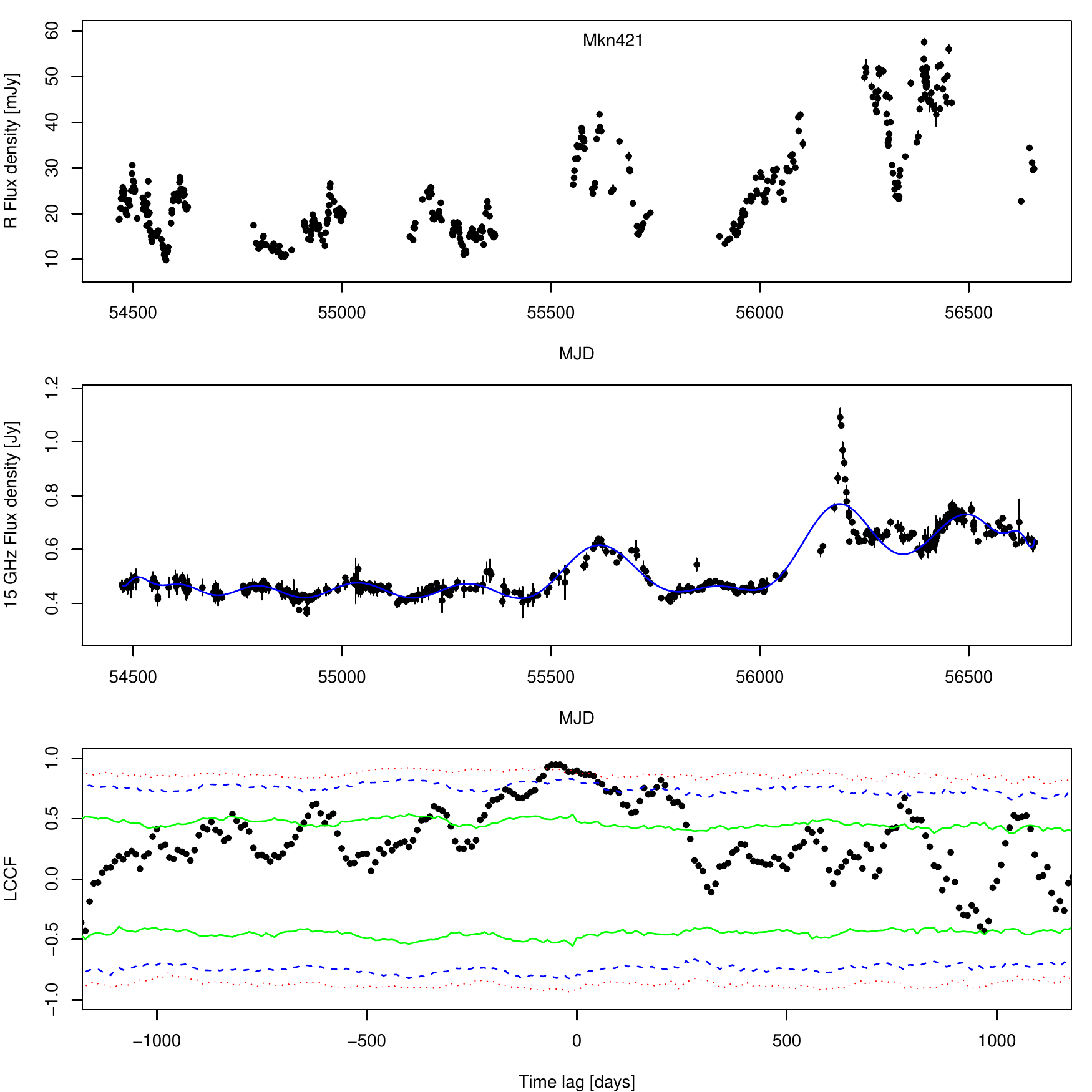}
\caption{The optical R-band light curve (top) and radio 15\,GHz light curve (middle) of Mkn~421. The solid blue line in middle panel shows the polynomial fit to radio data, which is subtracted from the optical light curve to estimate the contribution of the slowly varying component to optical flux. The bottom panel shows the results of the DCF study; the green, blue and red line representing the $1\sigma$, $2\sigma$ and $3\sigma$ significance limits, respectively.}
\label{Fig:lc15}
\end{figure}

{\bf Mrk~421}: The brightest optical source in our sample with rather
large modulation index. Both light curves show clear outbursts that
visually seem correlated which is confirmed by the DCF analysis (see Fig.~\ref{Fig:lc15}). Both
light curves also show significant increasing trend and polynomial subtraction suggests that at least $\sim50\%$ of the optical flux originates from common optical-radio component. For this source a significant correlation is also found between the radio and $\gamma$-ray light curves, when the largest extreme flare in 2012 is studied \citep{hovatta15}. The $\gamma$-ray variations lead the radio by $40-70$ days, which is consistent with the delay of $-60$ days obtained between our
radio and optical light curves. The visibility gap in the optical
light curve hinders a more detailed study of the 2012 flare in the
optical band.

\begin{figure}
\includegraphics[width=0.45\textwidth]{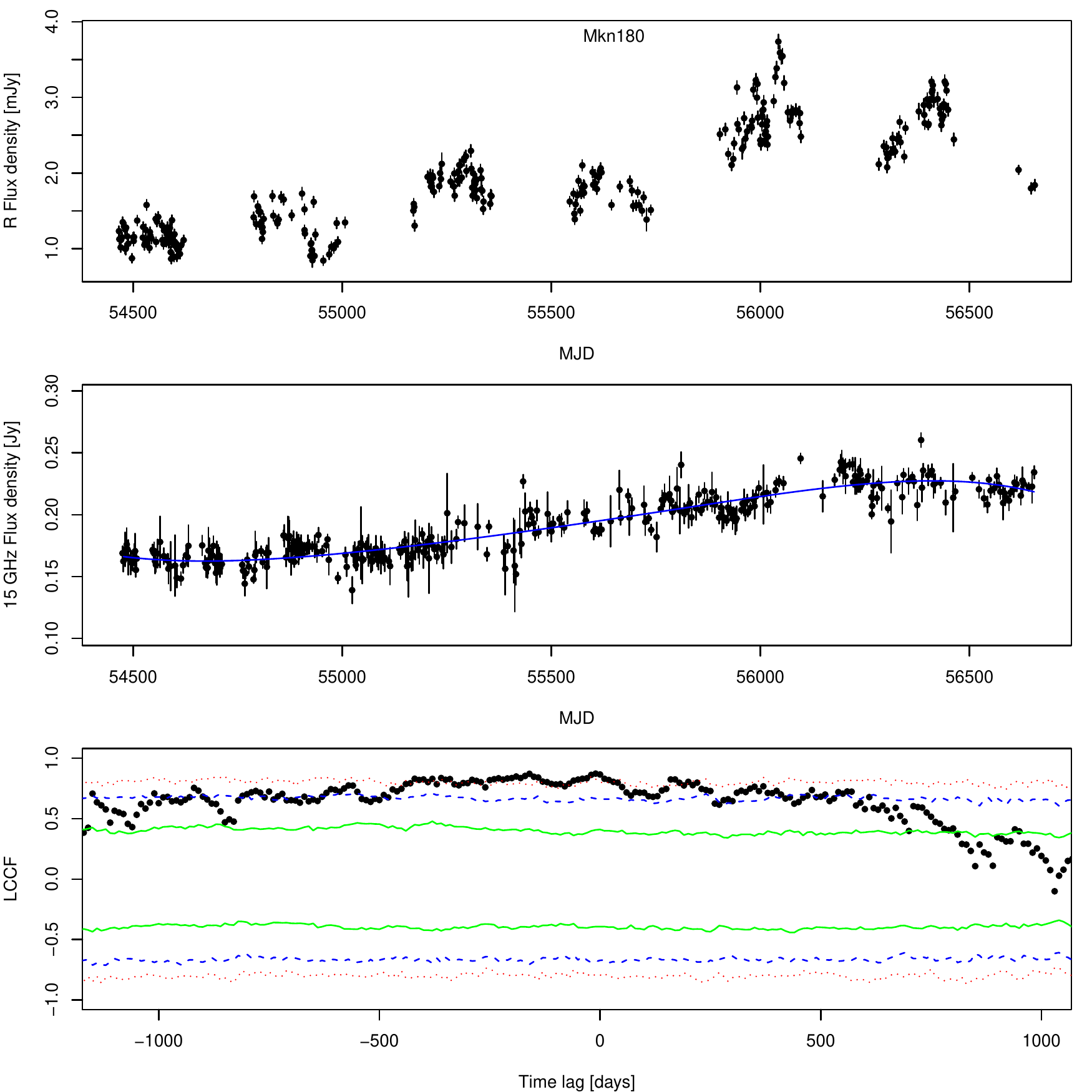}
\caption{The optical R-band light curve (top) and radio 15\,GHz light curve (middle) of Mkn~180. The solid blue line in middle panel shows the polynomial fit to radio data, which is subtracted from the optical light curve to estimate the contribution of the slowly varying component to optical flux. The bottom panel shows the results of the DCF study; the green, blue and red line representing the $1\sigma$, $2\sigma$ and $3\sigma$ significance limits, respectively.}
\label{Fig:lc16}
\end{figure}

{\bf Mrk~180}: The source is close to the median flux density of the
sample in both optical and radio bands. Very significant (highest
significance within our sample) increasing trends extending through
several years are found in both optical and radio light
curves. Besides this, both light curves show flares. However, the DCF curve is really flat, with several subsequent points above 3$\sigma$, suggesting that the significant correlation we find is mainly due to common trend seen in the optical and radio light curves.
It seems that all four visually identified optical flares have also counterpart in the radio light curve (see Fig.~\ref{Fig:lc16}), which might indicate that all of the optical emission (also the ``fast'' flares) in the studied period originates from the same emission region as the radio emission. Our polynomial subtraction indicates that at least $\sim50\%$ of the optical emission originates from this region.  

\begin{figure}
\includegraphics[width=0.45\textwidth]{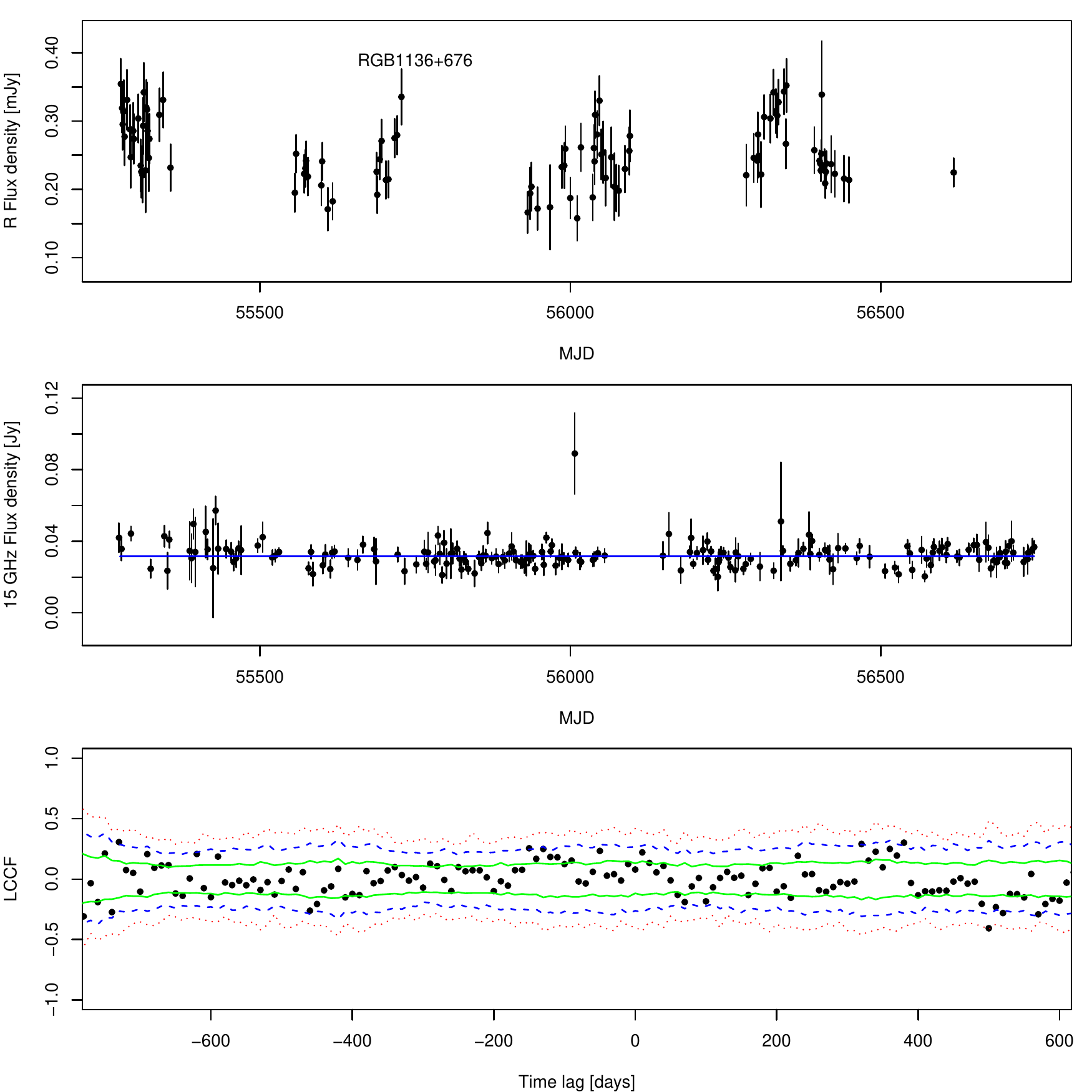}
\caption{The optical R-band light curve (top) and radio 15\,GHz light curve (middle) of RGB~1136+676. The solid blue line in middle panel shows the polynomial fit to radio data, which is subtracted from the optical light curve to estimate the contribution of the slowly varying component to optical flux. The bottom panel shows the results of the DCF study; the green, blue and red line representing the $1\sigma$, $2\sigma$ and $3\sigma$ significance limits, respectively.}
\label{Fig:lc17}
\end{figure}

{\bf RGB~1136+676}: One of the faintest sources in our sample both in
the radio and optical. The light curves show no trends and no correlated variability (see Fig.~\ref{Fig:lc17}) .
Accordingly the rms of the optical light curve does not improve with the subtraction of the polynomial.

\begin{figure}
\includegraphics[width=0.45\textwidth]{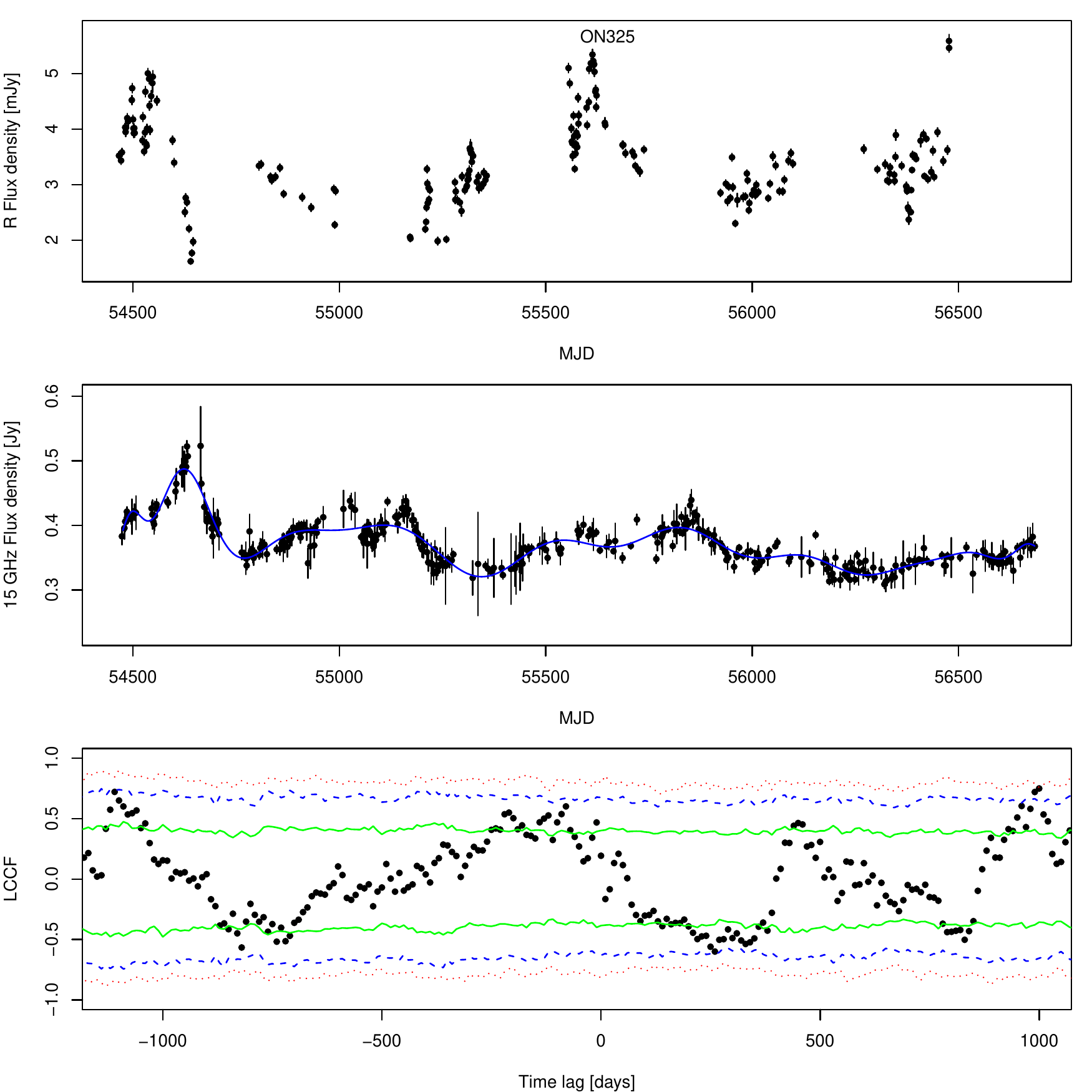}
\caption{The optical R-band light curve (top) and radio 15\,GHz light curve (middle) of ON~325. The solid blue line in middle panel shows the polynomial fit to radio data, which is subtracted from the optical light curve to estimate the contribution of the slowly varying component to optical flux. The bottom panel shows the results of the DCF study; the green, blue and red line representing the $1\sigma$, $2\sigma$ and $3\sigma$ significance limits, respectively.}
\label{Fig:lc18}
\end{figure}

{\bf ON~325}: The average brightness in both bands are above the median
and light curves show clear outbursts. Visually, the light curves seem
to show an anti-correlation (see Fig.~\ref{Fig:lc18}) instead of a correlation with the radio flux density increasing when the optical is decreasing. Our analysis finds no common trend and DCF reveals no significant peaks and is
therefore in agreement with this visual impression. Interestingly, at
the time of the $\gamma$-ray flare observed by {\it Fermi}-LAT
\citep[around MJD 54700][]{fermibright} there was a major decaying
outburst in the radio (gap in the optical light curve), while during
the Very High Energy $\gamma$-ray detection by MAGIC \citep[around MJD
55570][]{Alek12a}, there was an major outburst in the optical, but not
in the radio. In \cite{Alek12a} it was suggested, based on optical
polarization degree dropping during the optical flare, that there are
two components (one variable and one presenting a standing shock)
contributing to the optical emission. However, in the present work, we
do not find signatures that would link one of these regions with the
radio core, the polynomial subtraction does not decrease the rms of the optical light curve. It could be that also in the radio band there are multiple emission regions contributing and our simple method fails to discriminate them.

\begin{figure}
\includegraphics[width=0.45\textwidth]{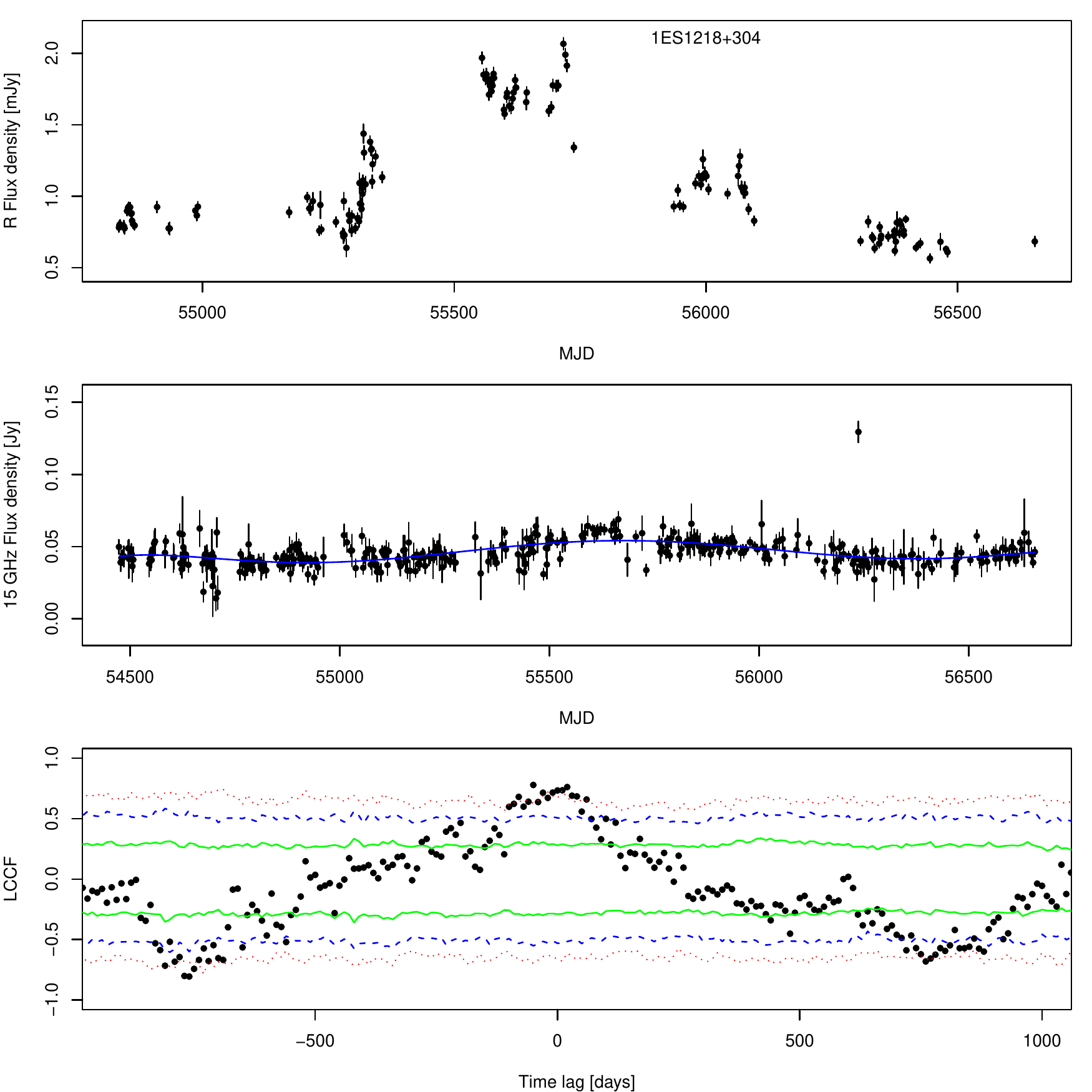}
\caption{The optical R-band light curve (top) and radio 15\,GHz light curve (middle) of 1ES~1218+304. The solid blue line in middle panel shows the polynomial fit to radio data, which is subtracted from the optical light curve to estimate the contribution of the slowly varying component to optical flux. The bottom panel shows the results of the DCF study; the green, blue and red line representing the $1\sigma$, $2\sigma$ and $3\sigma$ significance limits, respectively.}
\label{Fig:lc19}
\end{figure}

{\bf 1ES~1218+304}: One of the faintest radio sources and also in the
optical band the average flux density is below the median. Visually
both light curves show increasing trends until MJD 55800, after which
the flux density begins to decrease in both bands (see Fig.~\ref{Fig:lc19}). Due to this change
of direction, the trend analysis shows no significant trend in the
optical. However, the DCF analysis shows a clear correlation between
the two with the optical leading by 50 days. The correlation is a result of these common trends in the light curves, rather than correlated flares. The polynomial subtraction suggests that at least $\sim50\%$ of the optical flux originates from common radio-optical component.

\begin{figure}
\includegraphics[width=0.45\textwidth]{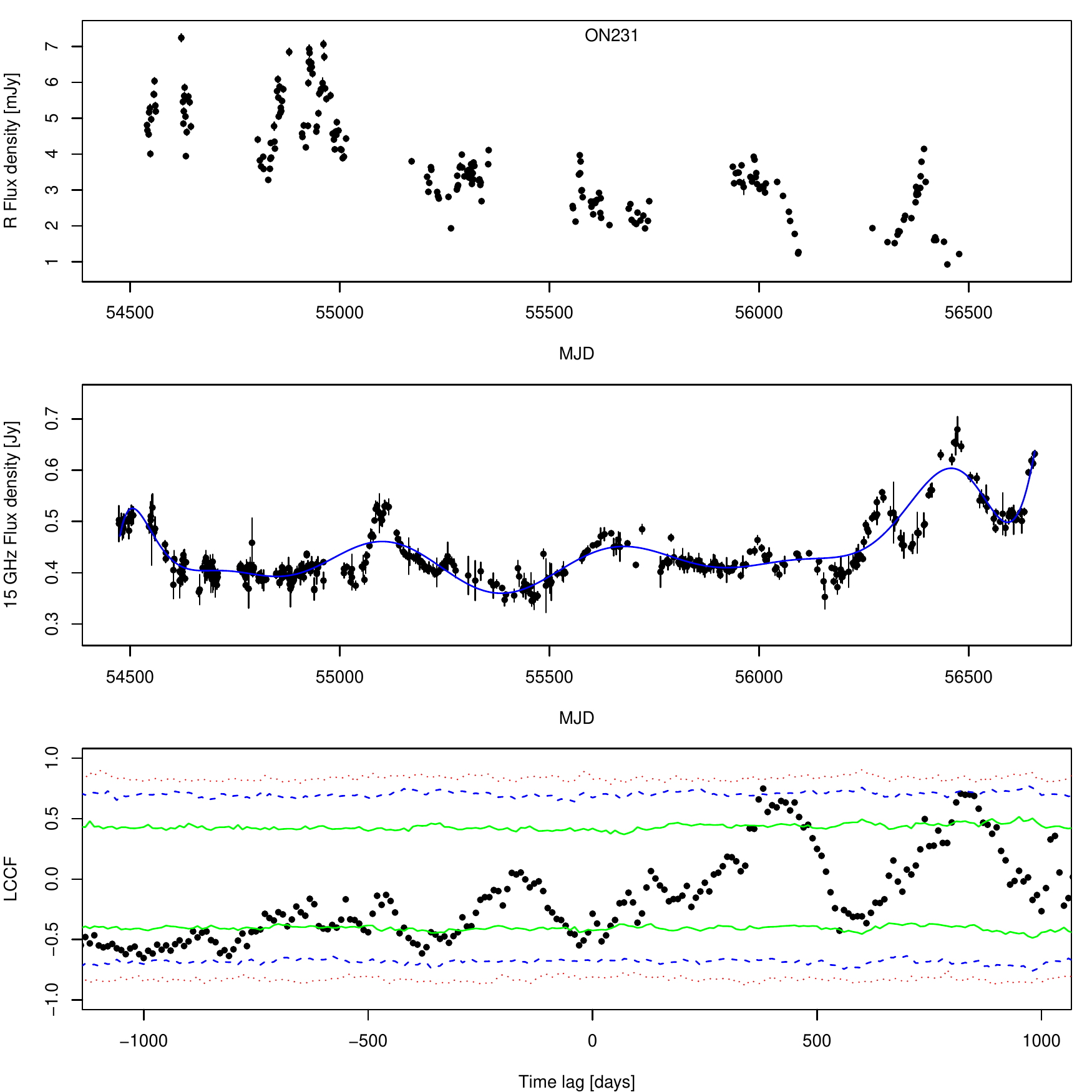}
\caption{The optical R-band light curve (top) and radio 15\,GHz light curve (middle) of ON~231. The solid blue line in middle panel shows the polynomial fit to radio data, which is subtracted from the optical light curve to estimate the contribution of the slowly varying component to optical flux. The bottom panel shows the results of the DCF study; the green, blue and red line representing the $1\sigma$, $2\sigma$ and $3\sigma$ significance limits, respectively.}
\label{Fig:lc20}
\end{figure}

{\bf ON~231}: The average brightness in both bands is above the median
of the sample, and the light curves show clear outbursts. Visually the
outbursts do not appear correlated and while the optical light curve
shows highly significant decreasing trend, the radio light curve shows
an increasing trend. DCF analysis finds no significant correlation (see Fig.~\ref{Fig:lc20}). In summary, the visual appearance and obtained results are very similar as for ON~325. 
According to our light curves, the detection of
a strong VHE $\gamma$-ray flare \citep[MJD 54625][]{Accia09} took
place just after the peak of a major optical flare (brightest in our
light curve covering six years of data). There is no obvious radio
counterpart for this flare. Recently, in \cite{sorcia}, it was
suggested based on optical polarization data, that there are two
emission components contributing to optical emission. As for ON~325,
we do not find signatures (and the polynomial subtraction does not decrease the rms of the optical light curve), that would link one of these components with the radio emission and we suggest that this might be due to a more complex emission pattern.  

\begin{figure}
\includegraphics[width=0.45\textwidth]{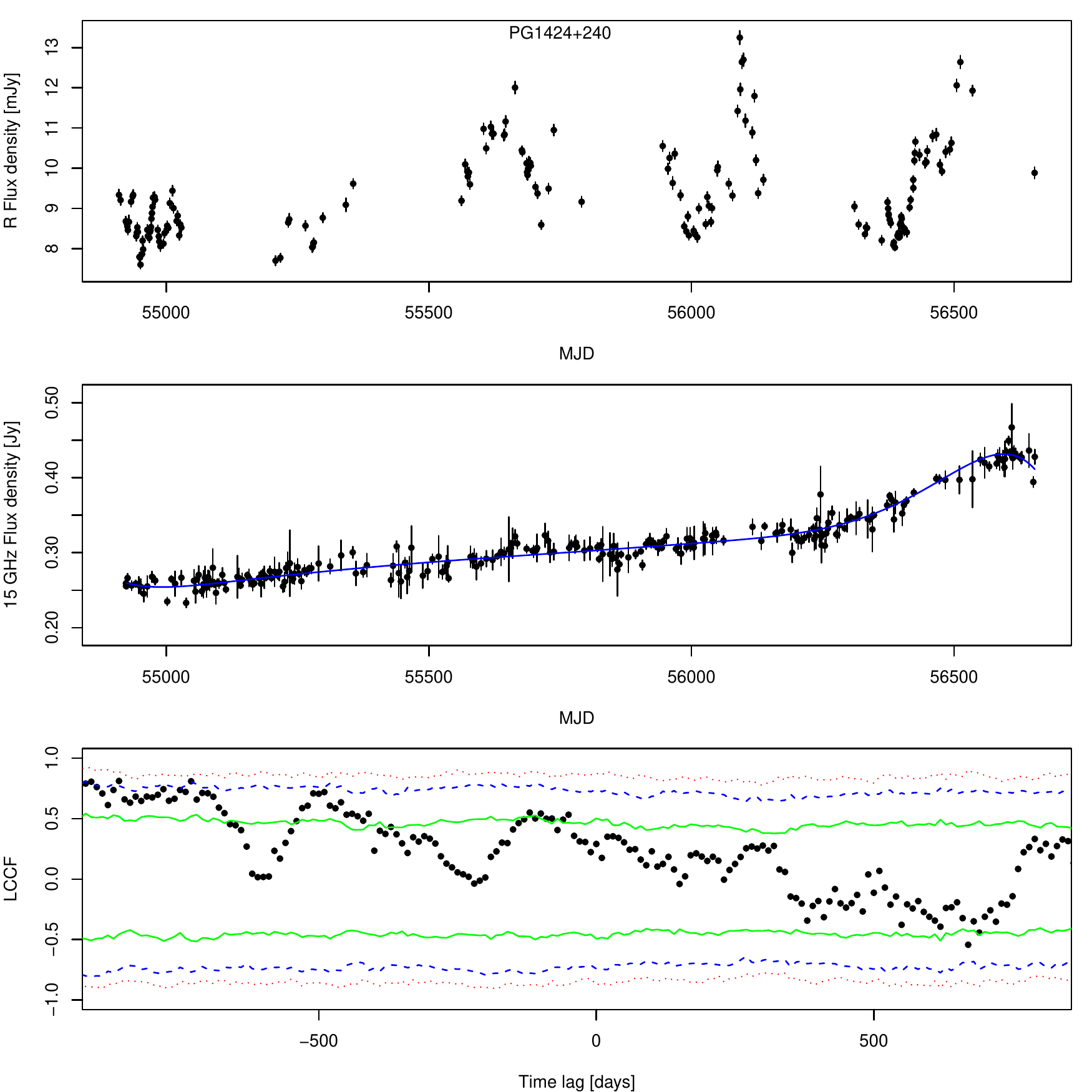}
\caption{The optical R-band light curve (top) and radio 15\,GHz light curve (middle) of PG~1424+240. The solid blue line in middle panel shows the polynomial fit to radio data, which is subtracted from the optical light curve to estimate the contribution of the slowly varying component to optical flux. The bottom panel shows the results of the DCF study; the green, blue and red line representing the $1\sigma$, $2\sigma$ and $3\sigma$ significance limits, respectively.}
\label{Fig:lc21}
\end{figure}

{\bf PG~1424+240}: A significant increasing trends extending through
several years is found in both optical and radio light curves. In
addition the optical light curve shows several fast outburst, the one
starting around MJD 55600 is marginally visible in the radio light
curve. In \cite{Alek14a} it was concluded that optical emission
originates in two components, one connected to variable high energy
emission and one to 15GHz VLBA core. For the study presented here, we
have added two more years of data, which increases the significance of
the radio trend and decreases that of the optical, but both remain
significant. We find no significant correlation in the DCF analysis and also rather surprisingly the subtraction of the polynomial fit does not decrease the rms of the optical light curve (see Fig.~\ref{Fig:lc21}).  

\begin{figure}
\includegraphics[width=0.45\textwidth]{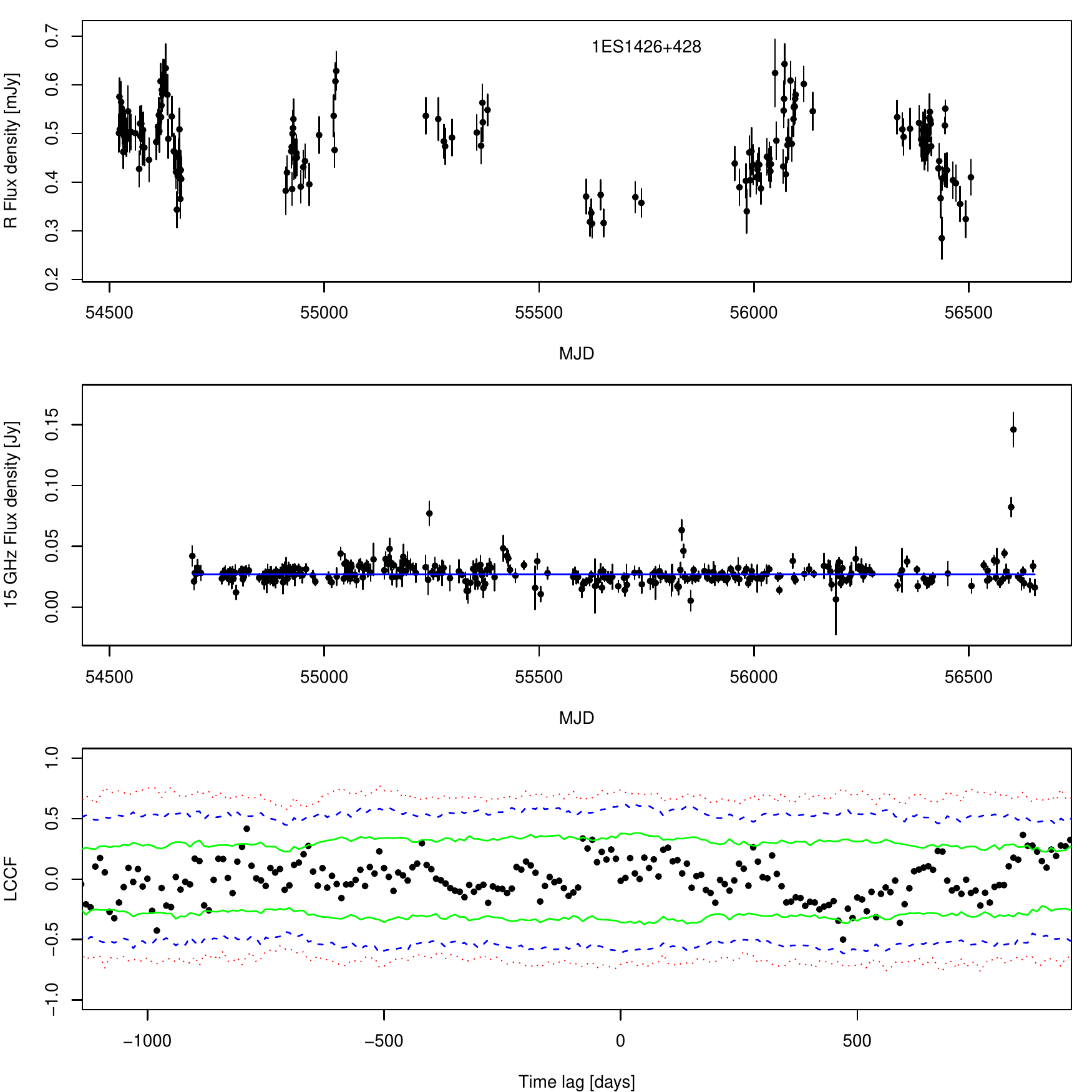}
\caption{The optical R-band light curve (top) and radio 15\,GHz light curve (middle) of 1ES~1426+428. The solid blue line in middle panel shows the polynomial fit to radio data, which is subtracted from the optical light curve to estimate the contribution of the slowly varying component to optical flux. The bottom panel shows the results of the DCF study; the green, blue and red line representing the $1\sigma$, $2\sigma$ and $3\sigma$ significance limits, respectively.}
\label{Fig:lc22}
\end{figure}

{\bf 1ES~1426+428}: One of the faintest sources in both bands showing
moderate variability in both bands. The surprisingly large modulation index in the radio band is possibly artefact of the few outliers. There is no significant trend in the radio data, we find no significant correlation between optical and radio light curves and polynomial subtraction does not improve the rms of the optical light curve (see Fig.~\ref{Fig:lc22}).

\begin{figure}
\includegraphics[width=0.45\textwidth]{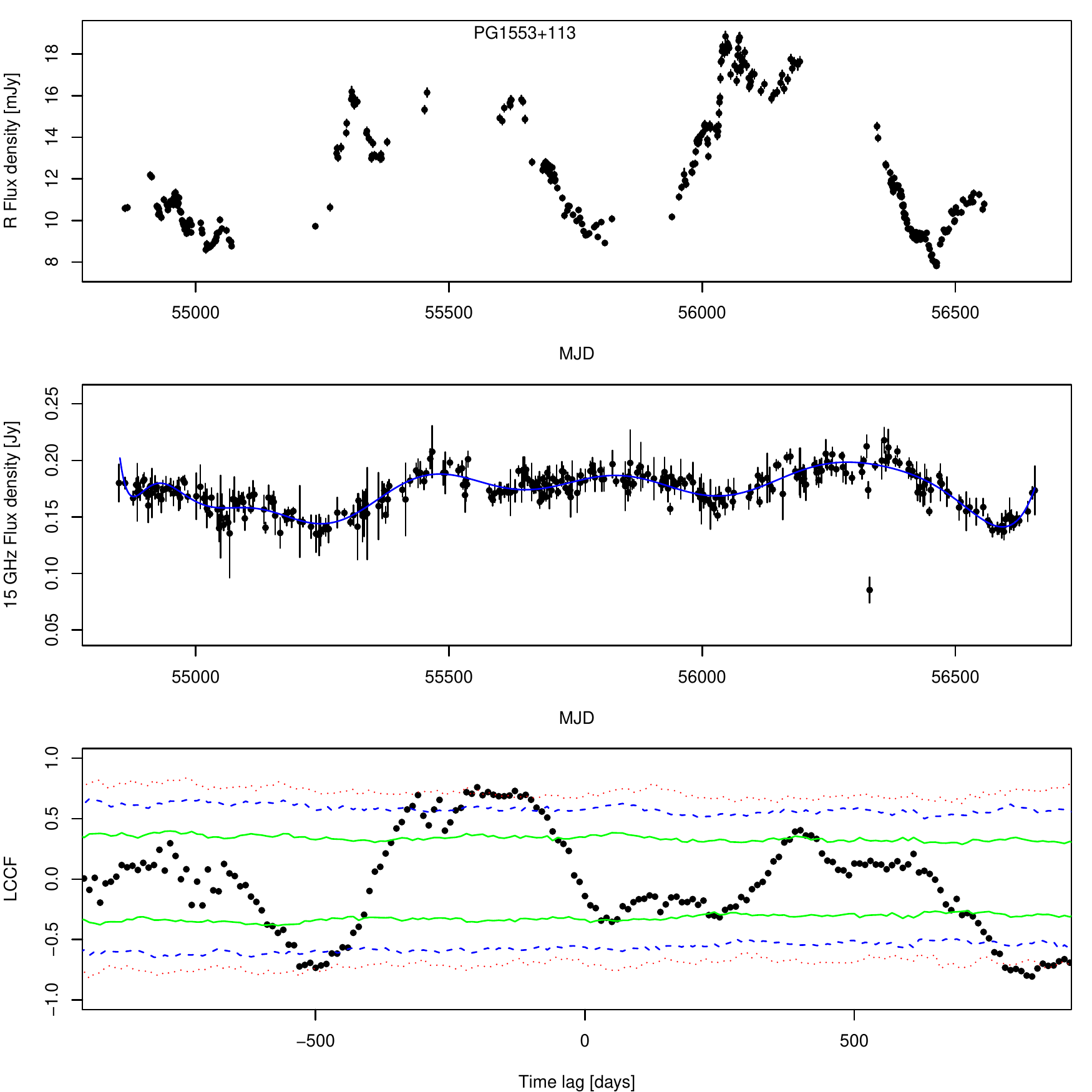}
\caption{The optical R-band light curve (top) and radio 15\,GHz light curve (middle) of PG~1553+113. The solid blue line in middle panel shows the polynomial fit to radio data, which is subtracted from the optical light curve to estimate the contribution of the slowly varying component to optical flux. The bottom panel shows the results of the DCF study; the green, blue and red line representing the $1\sigma$, $2\sigma$ and $3\sigma$ significance limits, respectively.}
\label{Fig:lc23}
\end{figure}

{\bf PG~1553+113}: The source is one of the brightest ones in optical
band while in radio its average flux density is around the median of the sample. The optical light curve shows several clear outbursts and the modulation index is large, while in radio the appearance of the light curve is smoother and modulation index small.  
In the beginning of both light curves there seems to be a decreasing
trend ending around MJD 55200 (see Fig.~\ref{Fig:lc23}). Analysis of \cite{2015ApJ...813L..41A} revealed a two-year periodicity in the $\gamma$-ray
light curve and visually this periodicity is apparent also in our
light curves. As discussed in the case of S5~0716+714, our analysis
methods are not optimal for looking at periodicities. However, our results (no significant trend, significant correlation) are in agreement with the short period of the suggested periodicity.
During the latest optical outburst (starting MJD 55900), which showed
high Very High Energy $\gamma$-ray flux \citep{cortina12a,cortina12b,Aliu15,Abram15,Alek15c}, the radio
outburst is clearly delayed compared to the optical one, which is in
agreement with the scenario we suggest, where some optical outbursts
occur closer to the central engine (where 15\,GHz emission is still self-absorbed) and are associated with VHE $\gamma$-ray emission. In this case, the delayed radio outburst is caused when the emission region propagates down the jet and becomes transparent to radio emission. We suggest that this wider shape of the radio outbursts is also the reason why the polynomial subtraction did not decrease the rms of the optical light curve, even when we shifted it with 200 days like suggested by the DCF analysis.  

\begin{figure}
\includegraphics[width=0.45\textwidth]{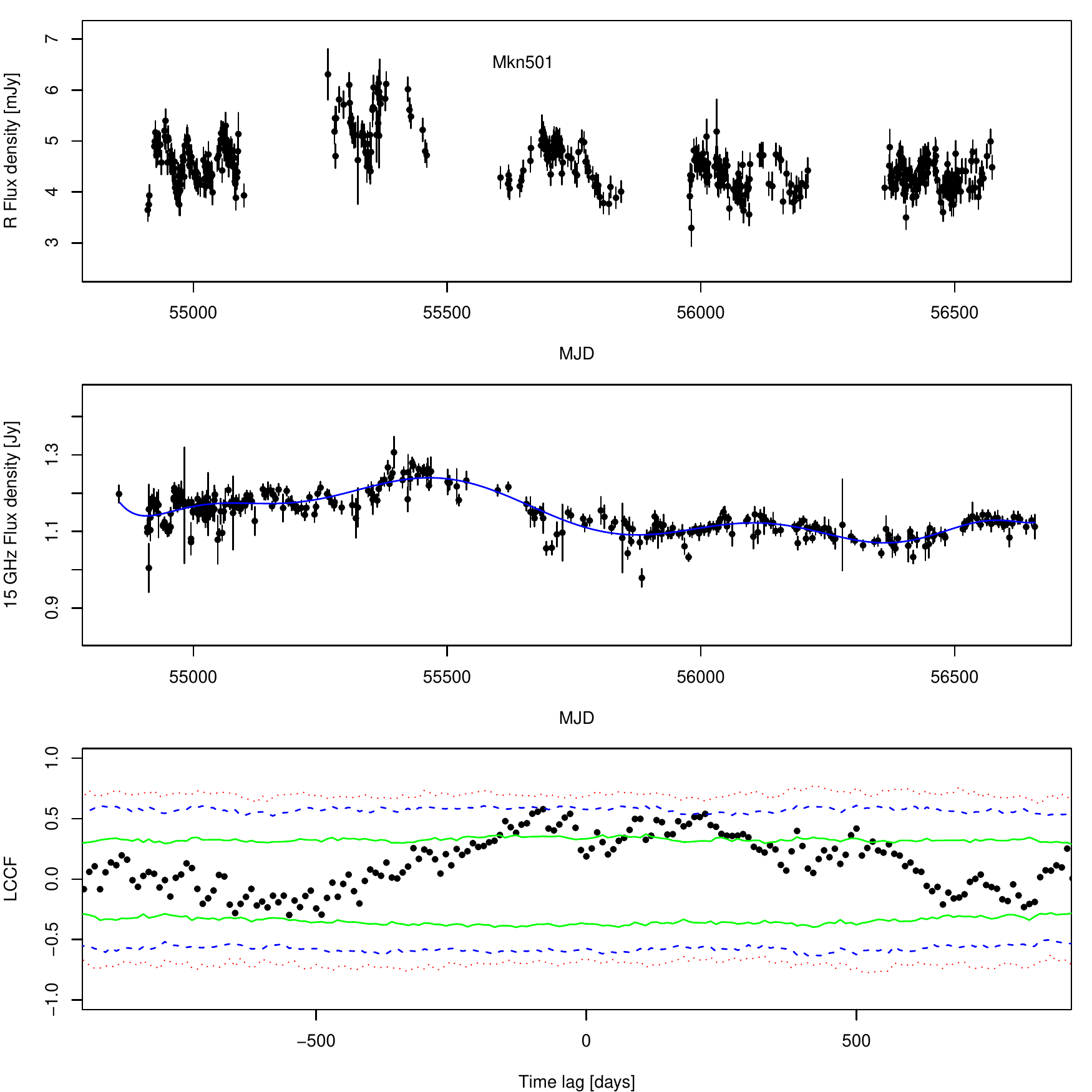}
\caption{The optical R-band light curve (top) and radio 15\,GHz light curve (middle) of Mkn~501. The solid blue line in middle panel shows the polynomial fit to radio data, which is subtracted from the optical light curve to estimate the contribution of the slowly varying component to optical flux. The bottom panel shows the results of the DCF study; the green, blue and red line representing the $1\sigma$, $2\sigma$ and $3\sigma$ significance limits, respectively.}
\label{Fig:lc24}
\end{figure}

{\bf Mrk~501}: The source is one of the brightest ones in the radio
and optical bands, but shows only modest variability resulting in the
lowest modulation indices of the sample. Visual inspection of the
radio and optical light curves shows a decreasing trend starting
$\sim$MJD 55500, which is confirmed by our trend analysis showing
significant decreasing trends for both bands. The outburst around MJD
55400 is visible both in the radio and optical, and in general the two
light curves follow the same patterns. However, the DCF analysis does
not reveal a significant correlation. As discussed earlier, this might
be due to multiple time scales in the light curve, producing two
2$\sigma$ peaks (see Fig.~\ref{Fig:lc24}). This seems to be also the reason why the polynomial subtraction is only mildly successful in reducing the rms of the optical light curve. It suggests that at least $\sim20\%$ of the optical flux would originate from common radio-optical component. Still, it is apparent that the emission in these two
bands originates largely from the same emission region. 
Being one of the brightest VHE $\gamma$-ray sources, it has been extensively
monitored in the VHE $\gamma$ band, and correlation analyses have not
revealed any correlation between the optical and VHE $\gamma$-rays
\citep[e.g.][]{furniss15}. However, \cite{doert} found a rotation of the optical
polarization angle associated with the VHE $\gamma$-ray flare,
revealing that small fraction of the optical emission does originate
from the VHE $\gamma$-ray emitting region, but the flux density from
this region is very small compared to the other components contributing in the optical. 

\begin{figure}
\includegraphics[width=0.45\textwidth]{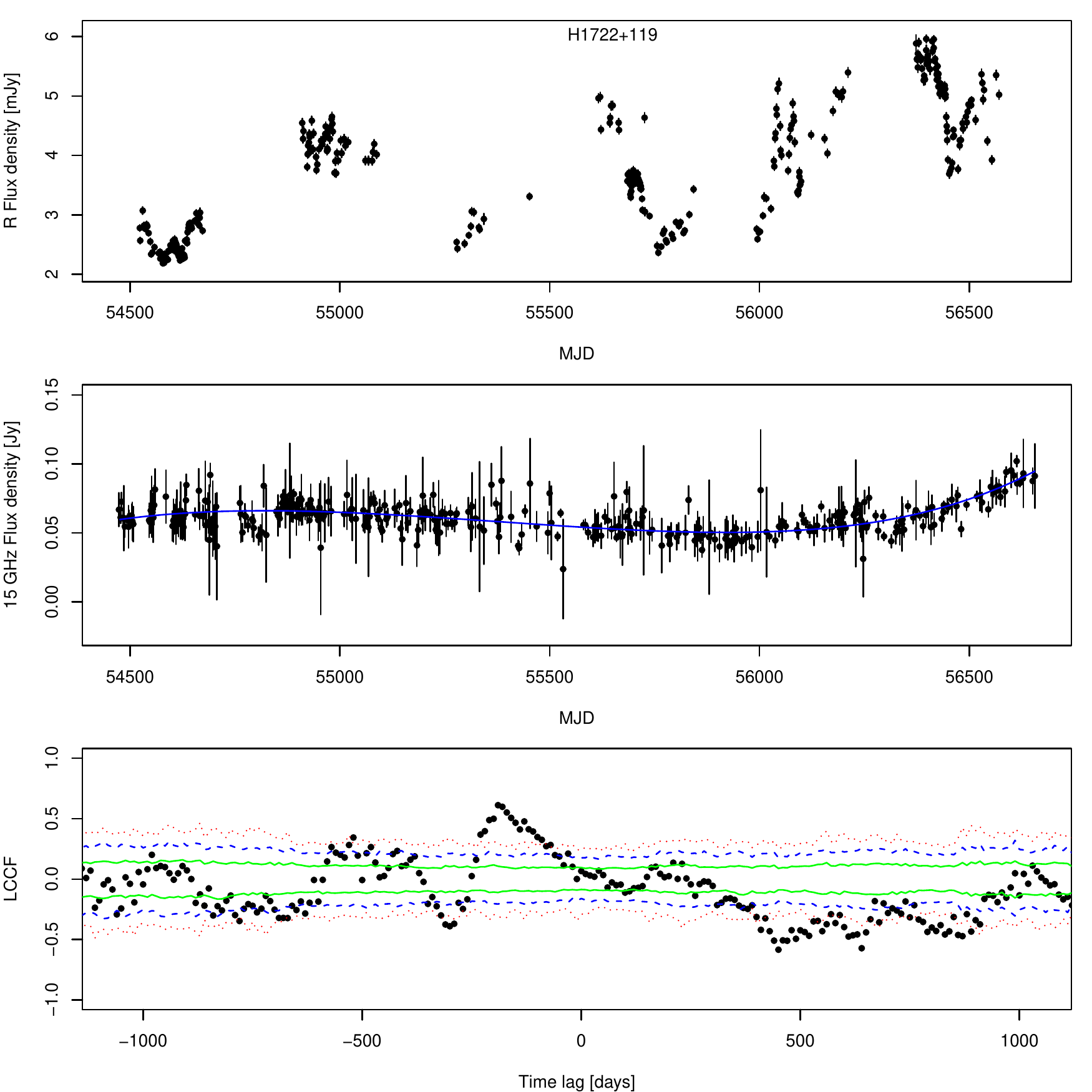}
\caption{The optical R-band light curve (top) and radio 15\,GHz light curve (middle) of H~1722+119. The solid blue line in middle panel shows the polynomial fit to radio data, which is subtracted from the optical light curve to estimate the contribution of the slowly varying component to optical flux. The bottom panel shows the results of the DCF study; the green, blue and red line representing the $1\sigma$, $2\sigma$ and $3\sigma$ significance limits, respectively.}
\label{Fig:lc25}
\end{figure}

{\bf H~1722+118}: Faint radio source with large amplitude optical
outbursts. Visual inspection suggests increasing trend in the radio
light curve starting around MJD~56000 and decreasing before it,
resulting in no significant trend in our trend analysis (see Fig.~\ref{Fig:lc25}). There is a
significant correlation between the two bands, and in the absence of
flares in the radio light curve, it seems to be a result of slow
variability common in the two bands. However, the polynomial subtraction does not decrease the rms of the optical light curve significantly. In addition to the increasing trend, the optical
light curve shows very sharp optical flares, which are probably
associated with much more compact emission region and therefore
possibly also with the VHE $\gamma$-ray emission. This is supported by the discovery of VHE $\gamma$-ray emission during an optical flare in spring 2013 \citep{Ahnen16}.

\begin{figure}
\includegraphics[width=0.45\textwidth]{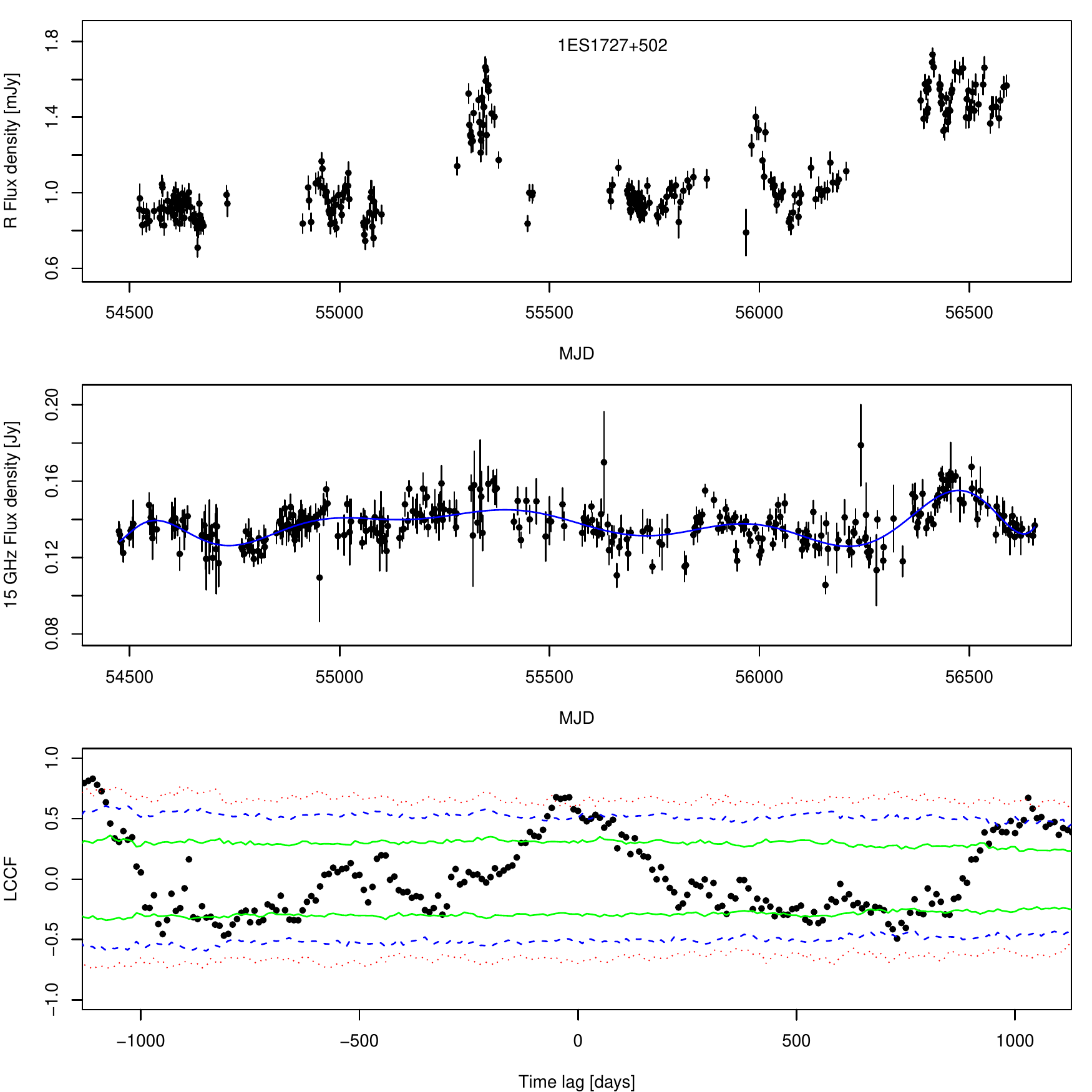}
\caption{The optical R-band light curve (top) and radio 15\,GHz light curve (middle) of 1ES~1727+502. The solid blue line in middle panel shows the polynomial fit to radio data, which is subtracted from the optical light curve to estimate the contribution of the slowly varying component to optical flux. The bottom panel shows the results of the DCF study; the green, blue and red line representing the $1\sigma$, $2\sigma$ and $3\sigma$ significance limits, respectively.}
\label{Fig:lc26}
\end{figure}

{\bf 1ES~1727+502}: The source is rather weak with mean flux density
below the median flux density of the sample in both optical and radio
bands. Both light curves show significant increasing flux density
trend throughout the period. There are two major optical outburst in
the source, around MJD 55350 and MJD 56300, which are  visible also in
the radio light curve, and DCF finds a significant correlation between the two bands (see Fig.~\ref{Fig:lc26}). At the time of the detection of VHE $\gamma$-rays from the source \citep{Alek14c}, there is no flare in optical or radio.
Therefore, it seems that for this source a major part of the optical emission originates from same region as the radio emission. The polynomial subtraction suggests that at least $\sim33\%$ originates from common region. 

\begin{figure}
\includegraphics[width=0.45\textwidth]{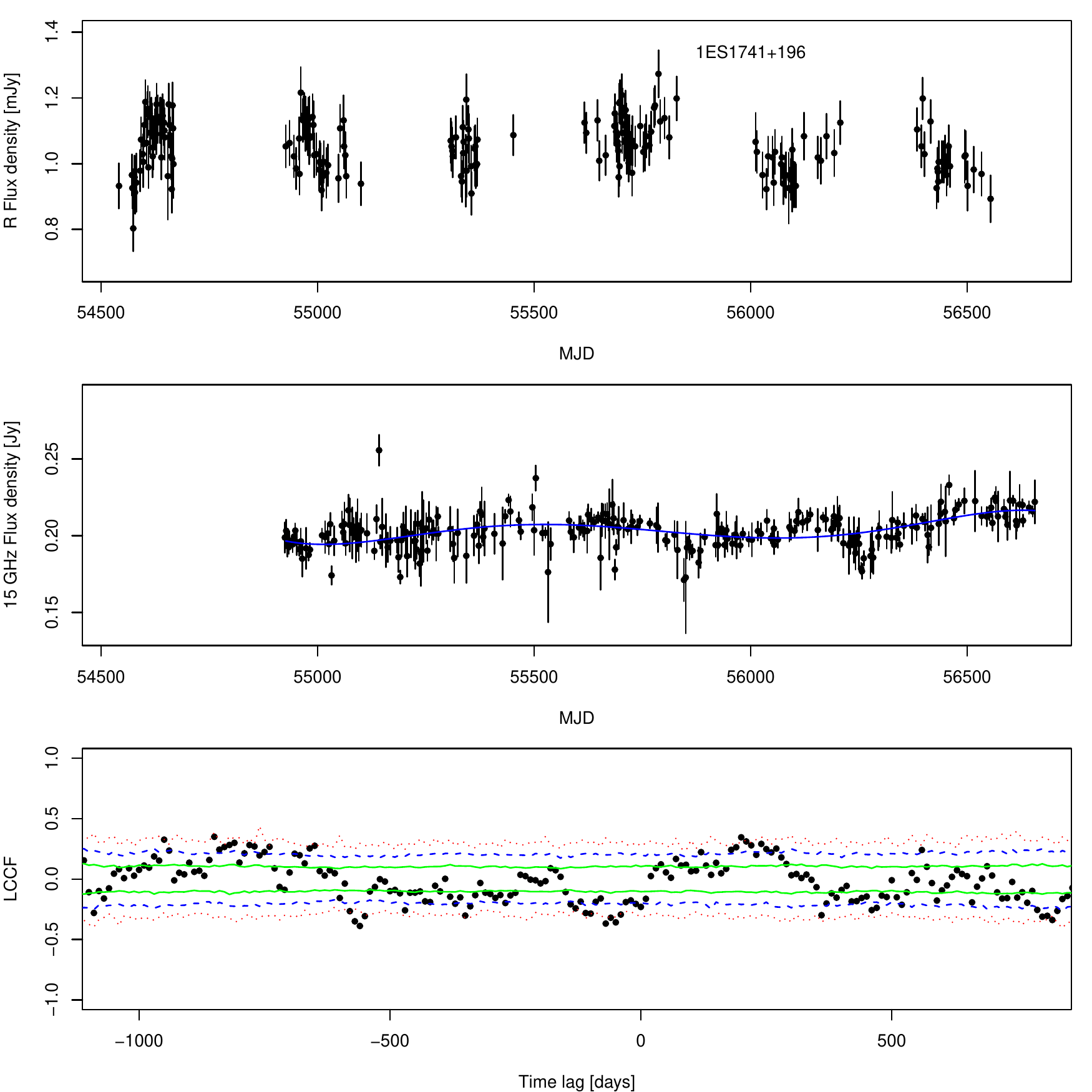}
\caption{The optical R-band light curve (top) and radio 15\,GHz light curve (middle) of 1ES~1741+196. The solid blue line in middle panel shows the polynomial fit to radio data, which is subtracted from the optical light curve to estimate the contribution of the slowly varying component to optical flux. The bottom panel shows the results of the DCF study; the green, blue and red line representing the $1\sigma$, $2\sigma$ and $3\sigma$ significance limits, respectively.}
\label{Fig:lc27}
\end{figure}

{\bf 1ES~1741+196}: The source is close to the median flux density of
the sample in both optical and radio band, but shows very little
variability. In the optical the modulation index is the smallest of
the sample and in radio among the smallest of the sample together with
Mrk~501 and HB89~0317+185. Visually, there seems to be an increasing
trend in the end of the radio light curve, and almost a 3$\sigma$ peak
in the DCF with a delay of $\sim$250 days (see Fig.~\ref{Fig:lc27}). polynomial subtraction does not decrease the rms of the optical light curve.  

\begin{figure}
\includegraphics[width=0.45\textwidth]{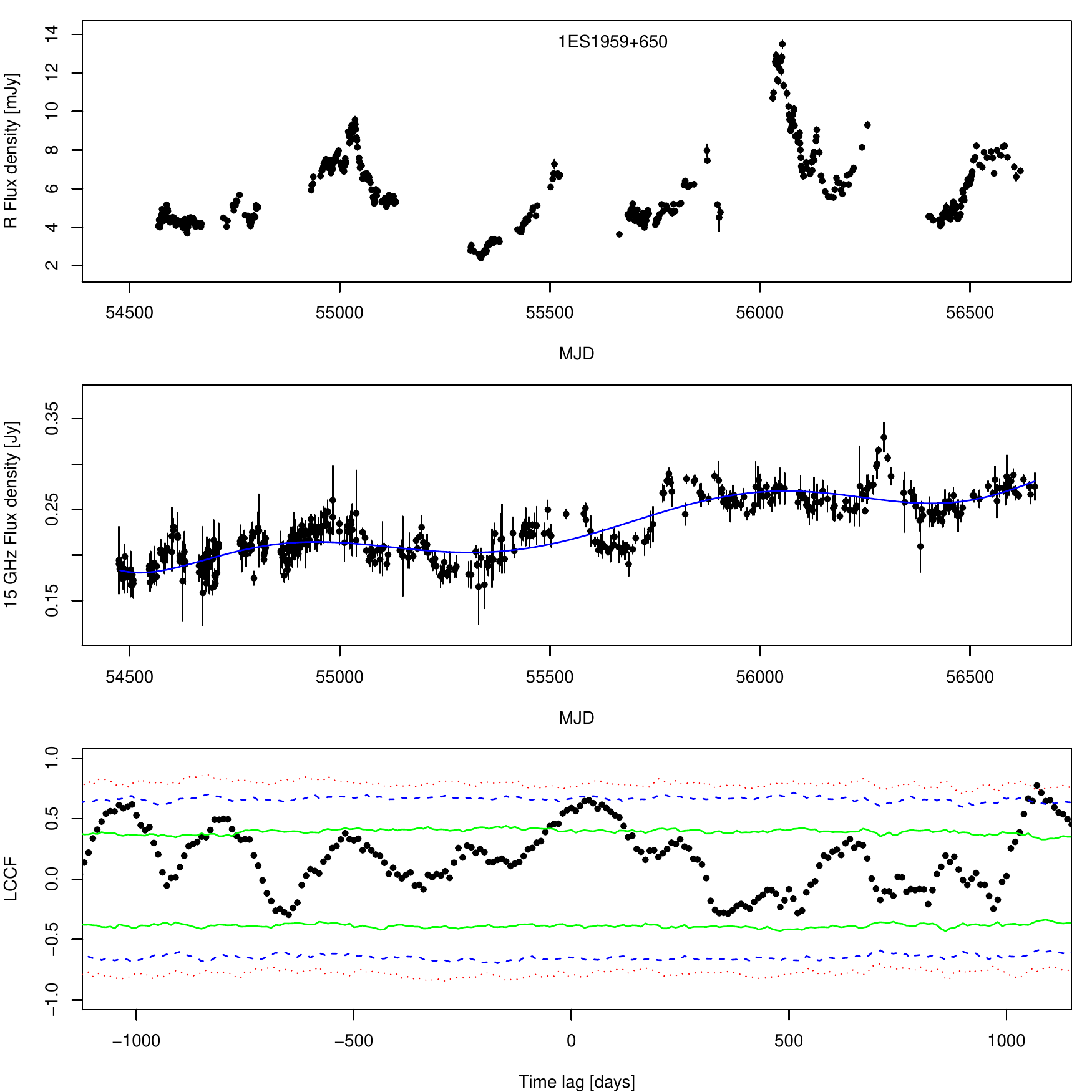}
\caption{The optical R-band light curve (top) and radio 15\,GHz light curve (middle) of 1ES~1959+650. The solid blue line in middle panel shows the polynomial fit to radio data, which is subtracted from the optical light curve to estimate the contribution of the slowly varying component to optical flux. The bottom panel shows the results of the DCF study; the green, blue and red line representing the $1\sigma$, $2\sigma$ and $3\sigma$ significance limits, respectively.}
\label{Fig:lc28}
\end{figure}

{\bf 1ES~1959+650}: The visual inspection of the light curves reveals
similar flaring behaviour in the radio and optical band with quasi-simultaneous outbursts starting around MJD 54900, 55350, 55700, 56200 and 56500 (see Fig.~\ref{Fig:lc28}).  
The linear regression analysis shows a highly significant flux density
increase in both radio and optical bands throughout the observing
period, but similarly to Mrk~501 the DCF does not find a significant correlation. This is probably due to multiple time scales as well as varying ratio of the amplitude of the outbursts. The polynomial subtraction suggests that at least $\sim15\%$ the optical flux would originate from common emission component.

\begin{figure}
\includegraphics[width=0.45\textwidth]{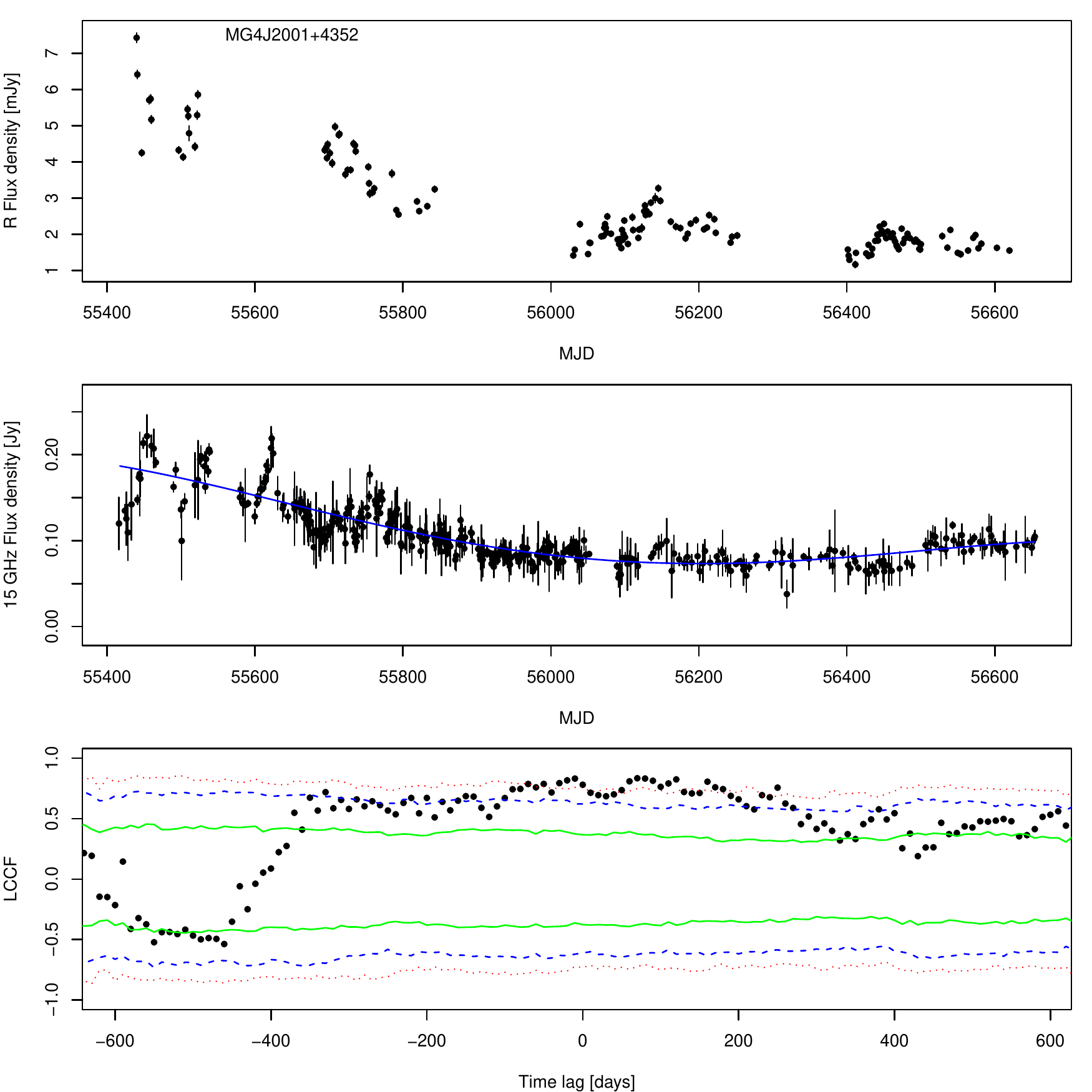}
\caption{The optical R-band light curve (top) and radio 15\,GHz light curve (middle) of MAGIC J2001+439. The solid blue line in middle panel shows the polynomial fit to radio data, which is subtracted from the optical light curve to estimate the contribution of the slowly varying component to optical flux. The bottom panel shows the results of the DCF study; the green, blue and red line representing the $1\sigma$, $2\sigma$ and $3\sigma$ significance limits, respectively.}
\label{Fig:lc29}
\end{figure}

{\bf MAGIC J2001+439}: The common decreasing trend and significant
correlation between the radio and optical bands found in \citep{Alek14b} are confirmed by our results using slightly more data (see Fig.~\ref{Fig:lc29}).
The flat DCF curve, with many subsequent data points above 3$\sigma$ limit, suggests that the correlation is due to common decreasing trend.
Also the polynomial subtraction suggests that at least $\sim50\%$ of the optical flux would originate from common radio-optical emission region.

\begin{figure}
\includegraphics[width=0.45\textwidth]{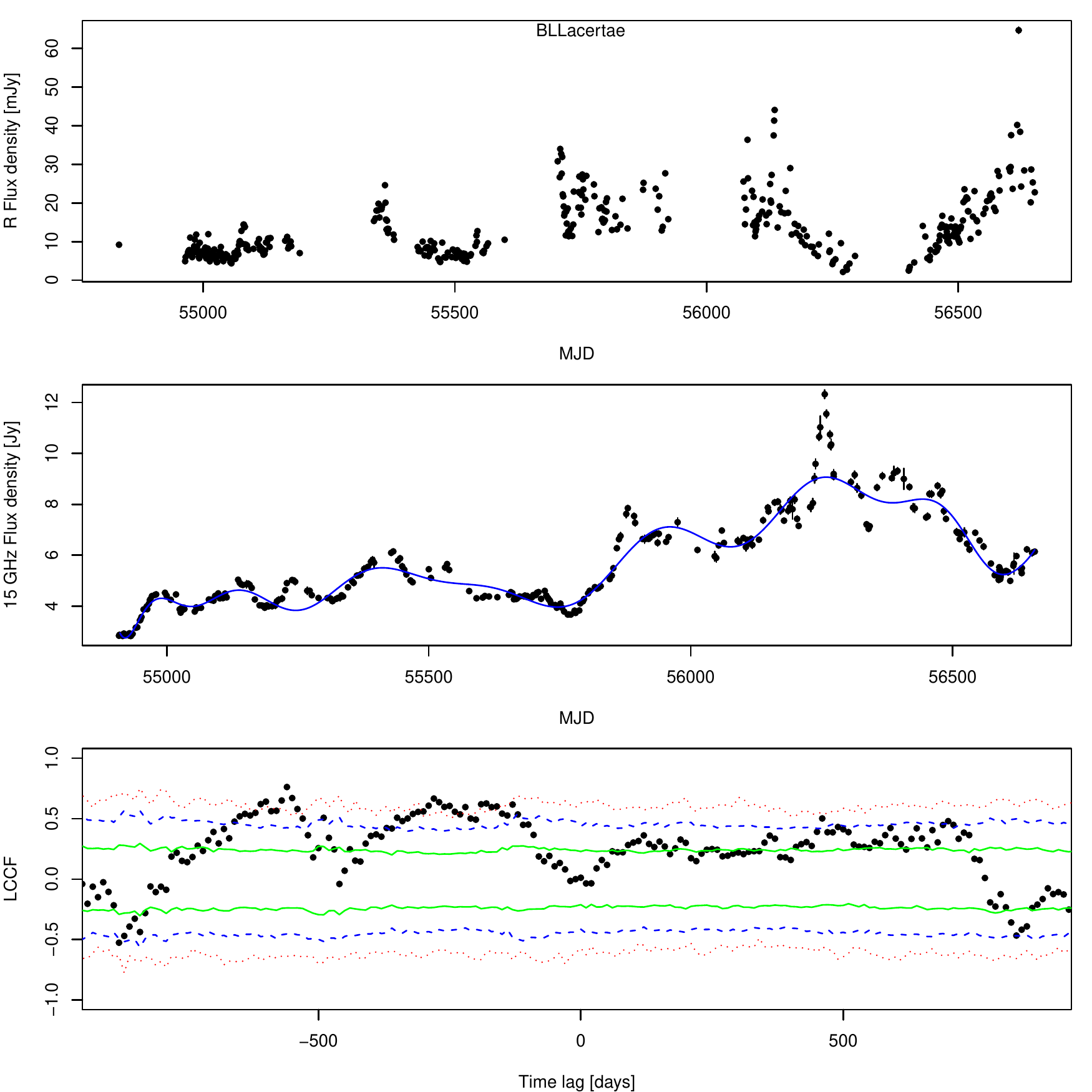}
\caption{The optical R-band light curve (top) and radio 15\,GHz light curve (middle) of BL~Lac. The solid blue line in middle panel shows the polynomial fit to radio data, which is subtracted from the optical light curve to estimate the contribution of the slowly varying component to optical flux. The bottom panel shows the results of the DCF study; the green, blue and red line representing the $1\sigma$, $2\sigma$ and $3\sigma$ significance limits, respectively.}
\label{Fig:lc30}
\end{figure}

{\bf BL~Lac}: This source has been studied previously in numerous
large multiwavelength campaigns, the longest of which cover data from 1968
to 2003 \citep{villata04}, revealing significant $\sim$8 years
periodicity in the radio light curves. The average radio flux density of this
source is an order of magnitude larger than the mean average flux
density of the sample and is by far the brightest radio source in our
sample. Also the optical flux density is one of the highest in the
sample. Both light curves show fast and large amplitude variability
and correspondingly the modulation indices are among the largest for
the sample. Both light curves reveal highly significant increasing
trends, which could be a result of a slowly varying component. In
agreement with \cite{villata04}, the visual inspection does not reveal
evident connection between the radio and optical outbursts, but the
correlation analysis suggests a significant correlation with time lag
of 540 days (optical leading). \cite{villata04} concluded that there are two radio
components soft and hard, such that only the hard components have
optical counterparts. This conclusion is further supported by recent study, where significant correlation is found between 95\,GHz and optical, but not 37\,GHz and optical \citep{ramakrishnan}.
As one would expect that at 15\,GHz the
outbursts are mainly "soft" bursts, and we find a significant
correlation, our result may conflict with this model. However, we
note, as in the case of S5~0716+714, that our methods might be too
simplistic for these two extremely variable sources that clearly differ from the other sources in our sample (see Fig.~\ref{Fig:lc30}).    

\begin{figure}
\includegraphics[width=0.45\textwidth]{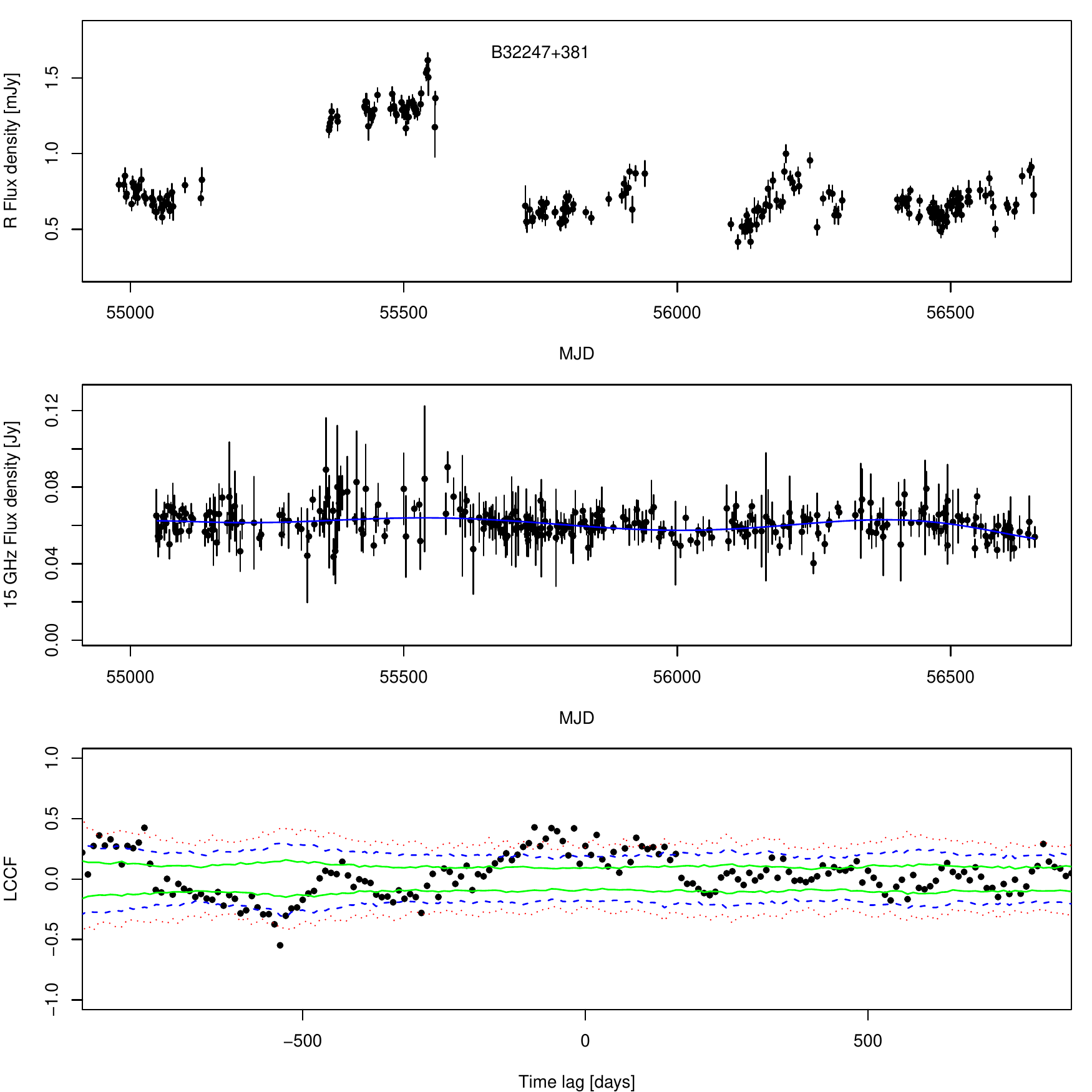}
\caption{The optical R-band light curve (top) and radio 15\,GHz light curve (middle) of B3~2247+381. The solid blue line in middle panel shows the polynomial fit to radio data, which is subtracted from the optical light curve to estimate the contribution of the slowly varying component to optical flux. The bottom panel shows the results of the DCF study; the green, blue and red line representing the $1\sigma$, $2\sigma$ and $3\sigma$ significance limits, respectively.}
\label{Fig:lc31}
\end{figure}

{\bf B3~2247+381}: One of the faintest sources in the sample both in
radio and optical band. There is a clear optical high state between
MJD 55380 and 55600, during which the source was discovered in VHE $\gamma$-rays \citep{Alek12b}, but there is no clear high
state in radio during this period. Polynomial subtraction suggests that at least $\sim15\%$ of the optical emission originates from the common optical-radio emitting region. The DCF suggests a correlation with optical leading by 90 days (see Fig.~\ref{Fig:lc31}). 

\begin{figure}
\includegraphics[width=0.45\textwidth]{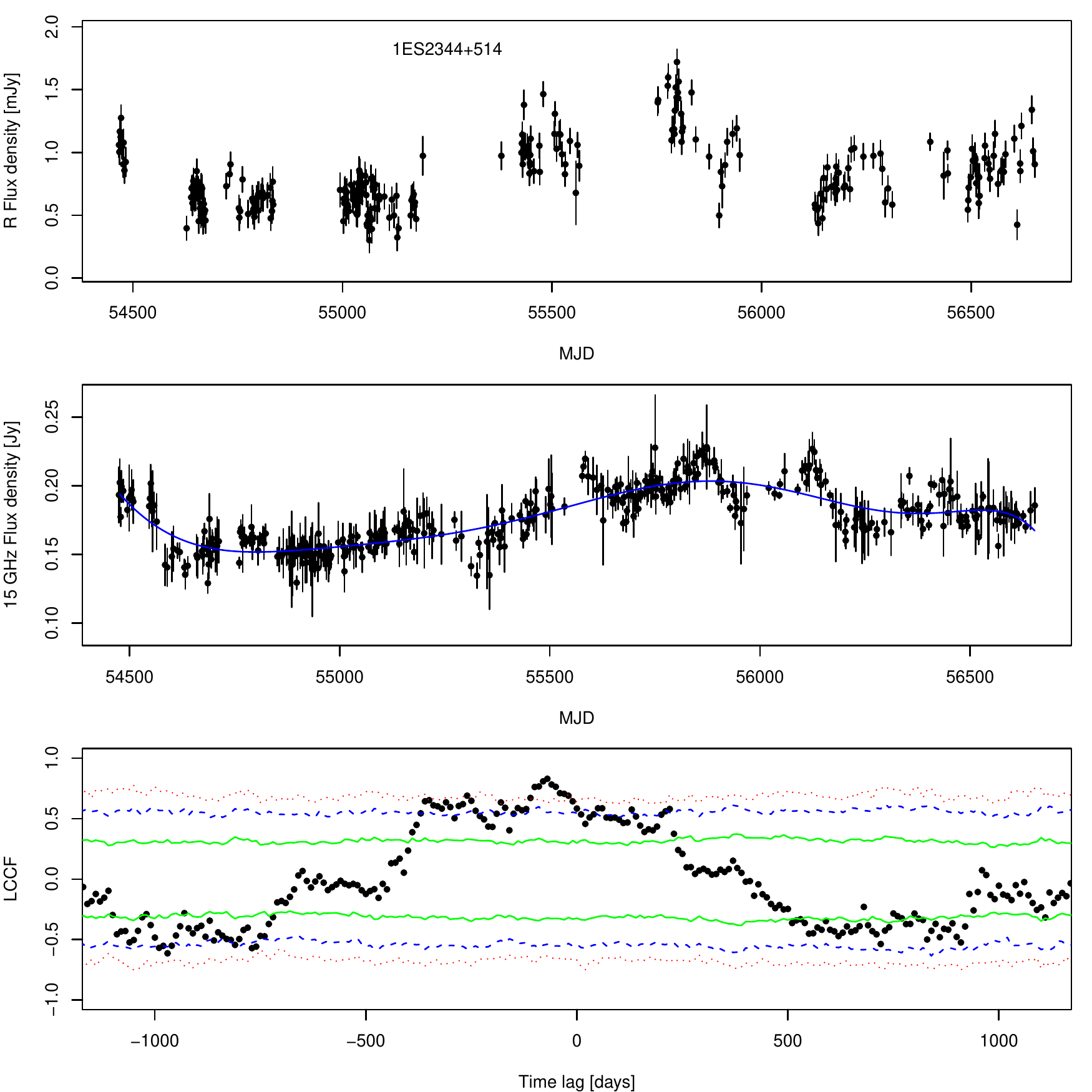}
\caption{The optical R-band light curve (top) and radio 15\,GHz light curve (middle) of 1ES~2344+514. The solid blue line in middle panel shows the polynomial fit to radio data, which is subtracted from the optical light curve to estimate the contribution of the slowly varying component to optical flux. The bottom panel shows the results of the DCF study; the green, blue and red line representing the $1\sigma$, $2\sigma$ and $3\sigma$ significance limits, respectively.}
\label{Fig:lc32}
\end{figure}

{\bf 1ES~2344+514}: The radio flux density is close to median for this
sample, while the optical flux density is one of the faintest in our
sample. The visual inspection of the radio light curve shows slowly
increasing flux density starting $\sim$ MJD 54850 and continuing for
$\sim$1000 days. The same increasing trend is visually present also in
the optical light curve (see Fig.~\ref{Fig:lc32}). The trend analysis shows a highly significant
trend in both light curves and the DCF shows a significant correlation
with optical leading by 70 days. Similarly, as for Mrk~501 and 1ES~1959+650, we suggest that for this source, the optical emission originates from the same regions as the radio emission (polynomial subtraction gives lower limit of $\sim25\%$), with no clear association to the VHE $\gamma$-ray emitting region. 

\clearpage

\clearpage

\clearpage

\clearpage

\section{Finding charts and Calibrated magnitudes of comparison stars}

For five sources the magnitudes of the comparison star were calibrated using the observations of stars with known magnitude from the same night.  The stars are marked in the finding charts in Fig.~B1 and B2. The R-band magnitudes are given in Tables~B1 and B2. The estimation of the host galaxy fluxes contributing to the aperture of 5'' are given in Table B~3.

\begin{table}[!h]
\centering
\caption{\small{Calibrated magnitudes of comparison stars in R-band.}}
\begin{tabular}{cccc}
\hline
\hline
Star & VER~J0521+211 & VER~J0648+152 & RGB~0847+115\\
\hline
1 & $15.26\pm 0.03$ & $14.10\pm0.03$ & $14.12\pm 0.03$\\
2 & $12.85\pm 0.03$ & $15.29\pm0.03$ & $14.97\pm 0.03$\\
3 & $15.31\pm 0.03$ & $14.80\pm0.03$ & $15.61\pm 0.03$\\
\hline
\end{tabular}
\label{standards}
\end{table}

\begin{table}[!h]
\centering
\caption{\small{Calibrated magnitudes of comparison stars in R-band.}}
\begin{tabular}{ccc}
\hline
\hline
Star & MAGIC~J2001+439& B3~2247+381\\
\hline
1 &$11.22\pm0.03$&$12.64\pm0.03$\\
2 &$11.31\pm0.03$&$13.10\pm0.03$\\
3 &$11.84\pm0.03$&$13.98\pm0.03$\\
4 &$14.18\pm0.03$&$12.65\pm0.03$\\
5 &&$15.46\pm0.03$\\
\hline
\end{tabular}
\label{standards2}
\end{table}

\begin{table}[!h]
\caption{\small{Host galaxy fluxes}}
\begin{tabular}{ccc}
\hline
\hline
source&host flux[mJy]& host flux error[mJy]\\ 
\hline
VER~J0521+211 &0.25&0.05\\
VER~J0648+152 &0.25&0.05\\
RGB~0847+115 &0.26&0.05\\
\hline
\end{tabular}
\label{hosts}
\end{table}

\begin{figure}
\includegraphics[width=0.4\textwidth]{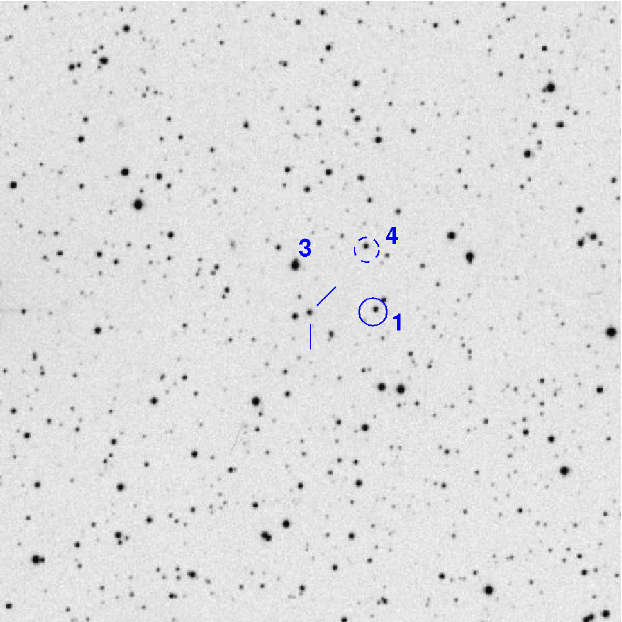}
\hspace{1.0mm}
\includegraphics[width=0.4\textwidth]{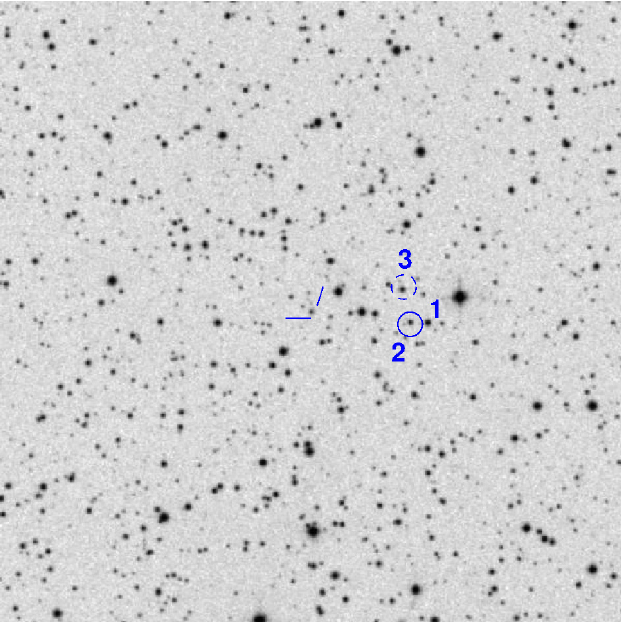}
\hspace{1.0mm}
\includegraphics[width=0.4\textwidth]{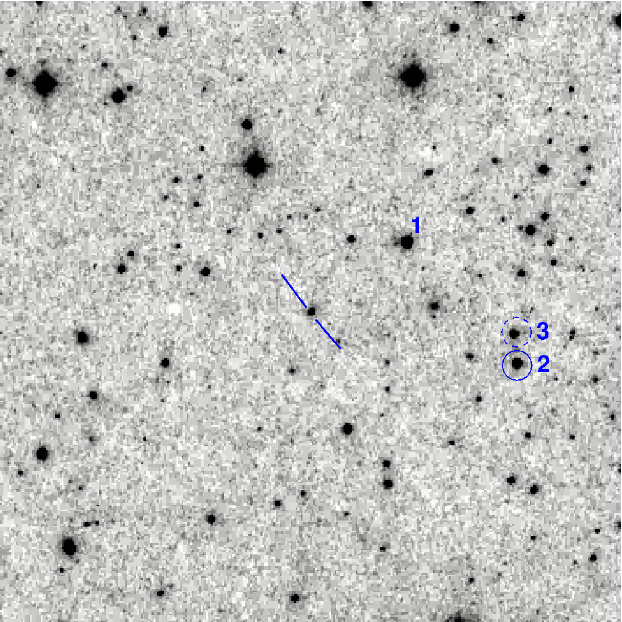}
\caption{Finding charts for VER~0521+211 (top), VER~J0648+152 (middle) and RGB0847+115 (bottom).The field of view of the finding charts is 12 arcmin. The stars 1-3 have been calibrated in this work (see Table B1). The stars circled are used as comparison (continuous line) and control (dashed line) stars for the light curves in this work.}
\label{fc1}
\end{figure}

\begin{figure*}
\includegraphics[width=0.45\textwidth]{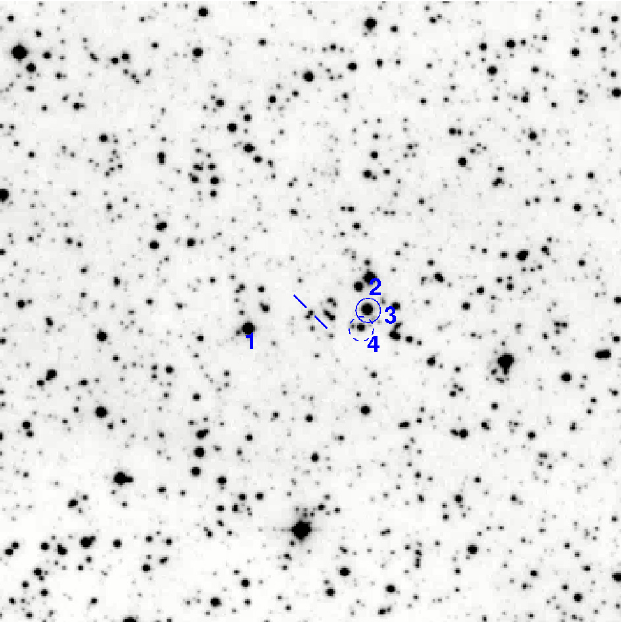}
\vspace{1.0mm}
\includegraphics[width=0.45\textwidth]{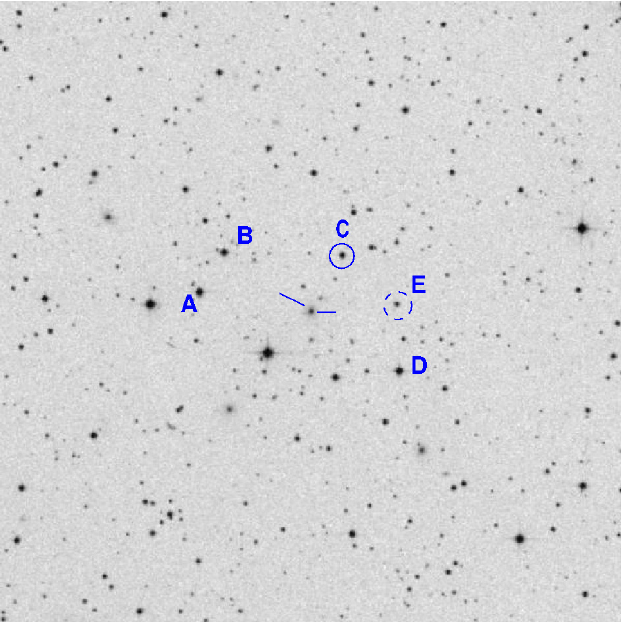}
\caption{Finding charts for MAGIC~J2001+439 (left) and B3~2247+381 (right).The field of view of the finding charts is 12 arcmin. The stars 1-4 and A-E have been calibrated in this work (see Table B2). The stars circled are used as comparison (continuous line) and control (dashed line) stars for the light curves in this work.}
\label{fc2}
\end{figure*}

\end{document}